\documentclass[letterpaper,11pt]{article}

\usepackage{jheppub}
\usepackage{xspace}

\usepackage{graphicx,color}  %
\usepackage{bm}  %
\usepackage{amssymb, graphics, amsmath}

\newcommand{\beq}{\begin{equation}}
\newcommand{\eeq}{\end{equation}}
\newcommand{\bea}{\begin{eqnarray}}
\newcommand{\eea}{\end{eqnarray}}
\newcommand{\ben}{\begin{eqnarray*}}
\newcommand{\een}{\end{eqnarray*}}

\newcommand{\Li}{\textrm{Li}}
\newcommand{\df}{\textrm{d}}

\def\lixo#1{}

\def\slashchar#1{\setbox0=\hbox{$#1$}           
  \dimen0=\wd0                                    
  \setbox1=\hbox{/} \dimen1=\wd1                  
  \ifdim\dimen0>\dimen1                           
    \rlap{\hbox to \dimen0{\hfil/\hfil}}            
    #1                                             
  \else                                          
    \rlap{\hbox to \dimen1{\hfil$#1$\hfil}}        
    /                                           
 \fi}                                           %

\begin{document}

\title{Heavy Quark Fragmenting Jet Functions}

\author{Christian W. Bauer, }
\author{Emanuele Mereghetti}

\affiliation{Ernest Orlando Lawrence Berkeley National Laboratory,
University of California, Berkeley, CA 94720, U.S.A.}

\emailAdd{cwbauer@lbl.gov, EMereghetti@lbl.gov}

\abstract{

Heavy quark fragmenting jet functions describe the fragmentation of a parton  into a jet containing a heavy quark, carrying a fraction of the jet momentum.
They are two-scale objects, sensitive to the heavy quark mass, $m_Q$, and to a jet resolution variable, $\tau_N$.
We discuss how cross sections for heavy flavor production at high transverse momentum  can be expressed in terms of heavy quark fragmenting jet functions, and 
how the properties of these functions can be used to achieve a simultaneous resummation of logarithms of the jet resolution variable,  and  logarithms of the quark mass.
We calculate the heavy quark fragmenting jet function $\mathcal G_Q^Q$  at $\mathcal O(\alpha_s)$, and  the gluon and light quark fragmenting jet functions  into a heavy quark, $\mathcal G_g^Q$ and $\mathcal G_l^Q$,  at $\mathcal O(\alpha_s^2)$. We verify that, in the limit in which the jet invariant mass is much larger than $m_Q$,  the logarithmic dependence of the fragmenting jet functions on the quark mass is reproduced by the heavy quark fragmentation functions.
The fragmenting jet functions can thus be written as convolutions of the fragmentation functions with the matching coefficients $\mathcal J_{i j}$, which depend only on dynamics at the jet scale.  
We reproduce the known matching coefficients $\mathcal J_{i j}$ at $\mathcal O(\alpha_s)$, and we obtain the expressions of 
the coefficients $\mathcal J_{g Q}$ and $\mathcal J_{l Q}$ at $\mathcal O(\alpha_s^2)$.
Our calculation provides all the perturbative ingredients for the simultaneous resummation of logarithms of $m_Q$ and $\tau_N$. 

}


\maketitle

\section{Introduction}
The production of heavy flavors and heavy flavored jets, where by heavy flavor we here mean charm or bottom, plays an important role in collider experiments.
These processes are interesting in themselves, as a probe of QCD dynamics, since for heavy quarks one expects the closest correspondence between calculations at the parton level and experimentally measured hadrons. Furthermore, $b$ jets are found in interesting electroweak processes; for instance,
one of the most important channel to probe fermionic couplings of the Higgs boson is $ H \rightarrow b \bar b$.
Finally, a quick look at the ATLAS and CMS public results pages shows the almost-omnipresence of $b$ jets in searches of Beyond Standard Model physics. It is therefore important to have a good theoretical understanding of heavy flavor production (where a heavy flavored hadron is directly observed) and heavy flavored jets (where a jet is tagged by demanding that it contains at least one heavy flavored hadron) in collider experiments. 

The current state of the art of fixed order calculations for heavy flavor hadroproduction  is next-to-leading order (NLO) accuracy, and such NLO calculations have a long history \cite{Nason:1987xz,Nason:1989zy,Beenakker:1988bq,Beenakker:1990maa}.
By now, several processes, including heavy quark pair  production, associated production of weak bosons and heavy quarks, and Higgs production with decay into $b \bar b$, are implemented in the program MCFM  \cite{Campbell:2000bg} at NLO accuracy, and any distribution can be obtained for these processes.

In such fixed order calculations, the dependence on the heavy quark mass typically enters through the ratio 
\begin{equation}
r_Q = m_Q / p_T
\,,
\end{equation}
were $p_T$ is the transverse momentum of the heavy quark. Many of the available NLO calculations include the full dependence of the heavy quark mass, which is important at small to moderate values of $p_T$. For large values of $p_T$, the ratio $r_Q$ becomes negligible, and one might want to perform calculations with massless heavy quarks, for which NLO calculations are significantly simpler. However, besides a dependence on powers of $r_Q$, there is also a logarithmic dependence on that ratio, which arises from infrared divergences in the massless calculation which are regulated by the heavy quark mass. Thus, at higher orders in perturbation theory more  powers of these logarithms appear, requiring  resummation. This is accomplished by introducing a heavy quark fragmentation function~\cite{Mele:1990cw}. As was discussed for hadroproduction in Ref.~\cite{Cacciari:1993mq}, the heavy quark fragmentation function can be calculated perturbatively at scales $\mu \sim m_Q$ without encountering any large logarithms. Running the fragmentation function from this low scale to $\mu \sim p_T$ using the familiar DGLAP evolution then resums all logarithms of $m_Q / p_T$. 

A combination of both approaches is 
needed to describe heavy flavor production for both large and small values of $p_T$.
Such a combination, named ``Fixed Order plus Next-to-Leading-Log'' (FONLL), has been proposed in Ref.~\cite{Cacciari:1998it}, and applied to single inclusive  production of heavy flavored hadrons. The general idea of FONLL is 
to add the massive fixed order calculation to the resummed calculation, and then subtract the overlap of the two. The overlap can be calculated either as the massless limit of the fixed order calculation (keeping the logarithms), or as the expansion of the resummed calculation to the appropriate order. 
The FONLL approach has been successfully compared to Tevatron and LHC data, for a recent discussion see Ref.~\cite{Cacciari:2012ny}.

Over the past decade we have learned how to combine NLO calculations with parton shower algorithms. This provides final states which are fully showered and hadronized, but which still provide NLO accuracy for predicted observables. Since such calculations can be compared much more directly to experimental data, this is used a great deal in analyses. The most popular available methods are MC@NLO~\cite{Frixione:2002ik} and POWHEG~\cite{Nason:2004rx, Frixione:2007vw}, with several other approaches being pursued. 
 Both MC@NLO and POWHEG include heavy flavor production in their list of available processes~\cite{Frixione:2003ei,Frixione:2007nw}. 
Since parton showers resum leading logarithms in their evolution variable $t$,
any calculation that is interfaced with such a shower needs to provide at least the same amount of resummation. In fact, as was discussed in detail in Ref.~\cite{Alioli:2012fc}, any combination of a perturbative calculation with a parton shower algorithm requires at least LL resummation of the dependence on an infrared safe jet resolution variable, however this jet resolution variable does not necessarily have to be equal to the evolution variable of the parton shower. Thus, one can choose a resolution variable for which one has good theoretical control, such as $N$-jettiness. We will denote a general dimensionful $N$-jet resolution variable by $\tau_N$ and define the dimensionless ratio 
\begin{equation}
r_\tau = \tau_N / p_T
\,,
\end{equation}
 with $p_T$ denoting the transverse momentum of the hadron, or of the jet in which the hadron is found, which we assume to be not too different.

It follows from the above discussion that combining heavy quark production at large $p_T$ with parton shower algorithms requires the simultaneous resummation of logarithms of $r_Q$ and $r_\tau$.\footnote{It should be noted that the leading log resummation in the shower resums a subset of logarithms of the heavy quark mass, for example all terms originating from emissions from the heavy quark or antiquark in the final state. 
However, not all $\log r_Q$ are included  at leading logarithmic accuracy. In particular, gluon splitting in $Q \bar Q$, with almost collinear quark and antiquark, are included only at fixed order. 
For a full discussion, see~\cite{Frixione:2003ei,Frixione:2007nw}.}
Logarithmic dependence on a second ratio $r_\tau$ can also arise from explicit experimental cuts restricting the size of the jet resolution variable $\tau_N$. For example, observables that explicitly restrict extra jet activity through jet vetoes will have logarithmic dependence on the jet veto scale $\tau_N^{\rm cut}$. 

Extending the discussion to heavy-flavor tagged jets, one might expect jet observables to be  less sensitive to the heavy quark fragmentation function, and to logarithms of  $m_Q$~\cite{Frixione:1996nh}. This is because heavy-flavor tagged jets are essentially agnostic to the flavor of the heavy hadron and its energy fraction. 
Indeed heavy quark jets initiated by  heavy quarks produced directly in the hard interaction do not have a  logarithmic dependence on $m_Q$ . 
In this case it is not necessary  to introduce a fragmentation function, and to resum $\log r_Q$. 
The  resummation of $\log r_\tau$ can be achieved with methods similar to those used for light quark jets.

However, heavy-flavor tagging algorithms also tag  jets initiated by gluons or light quarks, where heavy quarks are produced through $g \to Q \bar Q$.
In this case, $Q$-tagging introduces an infrared dependence  on logarithms of $m_Q$, and large uncertainties \cite{Banfi:2007gu}. 
A possible way to deal with final state logarithms is not to label jets with gluon or light quark splittings into $Q \bar Q$ as heavy-flavor jets. 
Banfi, Salam and Zanderighi in Ref.~\cite{Banfi:2007gu} explored the interesting  possibility of using an IR safe jet flavor algorithm \cite{Banfi:2006hf}, 
which would label jets with no net heavy flavor as gluon or light quark jets.
Alternatively, one can improve the theoretical description of $Q$-tagged jet cross sections by resumming  $\log r_Q$, and,  in  the presence of another small ratio $r_\tau$, by simultaneously resumming $\log r_\tau$.

In this paper we develop a formalism that allows to simultaneously resum the logarithmic dependence on the heavy quark mass as well as on the additional small ratio $r_\tau $. This opens the door to deal with vetoed heavy flavor production in a systematic way, and perhaps more importantly to interface FONLL-type calculations with a parton shower. 
Our formalism is based on the idea of fragmenting jet functions (FJFs), introduced 
in Refs.~\cite{Procura:2009vm,Jain:2011xz}, and first applied to heavy quarks in Ref.~\cite{Liu:2010ng}.
The FJFs $\mathcal G_i^j$ describe the fragmentation of a parton $j$ inside a jet initiated by the parton $i$, and contain information both on the jet dynamics, and on the parton fragmentation function. Therefore, FJFs encode the dependence on both $m_Q$, as well as $\tau_N$. An important property of the heavy flavor FJFs is that the renormalization group evolution is independent of the heavy quark mass, with the anomalous dimension being identical to that of an inclusive jet function. Thus, the dependence on the jet resolution variable $\tau_N$ can be resummed in the same way as for processes with only light jets. 

To perform a simultaneous resummation of $r_Q$ and $r_\tau$ requires to separate these scales in the factorization theorem, and therefore factorize the FJFs themselves. For $\tau_N \gg m^2_Q/Q$ this is accomplished by integrating out the degrees of freedom responsible for the $\tau_N$ scale, with the remaining long-distance physics (and therefore the entire $m_Q$ dependence) determined by the heavy-quark fragmentation function. 

The main part of this paper is an explicit calculation of the heavy flavor FJFs in fixed order perturbation theory. We calculate the 
heavy quark initiated FJF at $\mathcal O(\alpha_s)$, and the gluon and light-quark initiated FJFs at $\mathcal O(\alpha_s^2)$. 
Besides being an important ingredient to obtain the resummed expressions, it also allows us to check explicitly various properties of the heavy flavor FJFs. In particular, we verify that the heavy quark fragmentation functions reproduce the logarithmic dependence on the heavy quark mass and that the anomalous dimension of the heavy-quark FJFs are independent of $m_Q$. Our calculations are performed using Soft Collinear Effective Theory (SCET)~\cite{Bauer:2000ew,Bauer:2000yr,Bauer:2001yt,Bauer:2002nz,Leibovich:2003jd}. 

The paper is organized as follows. In Section \ref{SCET} we recall the main SCET ingredients needed in the rest of the paper. We define and state important properties of heavy quark fragmentation functions in Section \ref{HQFF}, and of inclusive jet functions in Section \ref{JF}. In Section \ref{HQFJF} we introduce the fragmenting jet functions, extending the definition of Refs.~\cite{Procura:2009vm,Jain:2011xz} to heavy quarks. After reviewing the resummation of $\log r_Q$ in single inclusive observables, and of $\log r_\tau$ in jet observables in Section \ref{ResumRb}, we describe how to achieve the simultaneous resummation of logarithms of the quark mass and the jet resolution variable $\tau_N$ in Section \ref{ResumBoth}.
In Section~\ref{OneLoop} we calculate the FJFs $\mathcal G_Q^Q$ and $\mathcal G_g^Q$ at $\mathcal O(\alpha_s)$  in the massless limit, we give the expressions with full mass dependence in Appendix \ref{AppC}. In Sections \ref{GluonAlpha2} and \ref{LightQuark} we carry out the calculation of $\mathcal G_g^Q$
and $\mathcal G_l^Q$
at $\mathcal O(\alpha_s^2)$. We draw our conclusions in Section \ref{Conclusion}.
In Appendix \ref{AppA} we discuss some additional details on how to take the massless limit $m_Q\rightarrow 0$. In Appendix \ref{AppE} we give the analytic expression of the function $g^{C_F T_R}$, defined in Section \ref{GluonAlpha2}.

\section{Soft Collinear Effective Theory }\label{SCET}

In this paper we use the formalism of Soft Collinear Effective Theory (SCET) \cite{Bauer:2000ew,Bauer:2000yr,Bauer:2001yt,Bauer:2002nz},
generalized to massive quarks    \cite{Leibovich:2003jd}.
SCET is an effective theory for fast moving, almost light-like, quarks and gluons, and their interactions with soft degrees of freedom. 
It has been successfully applied to a variety of processes, from $B$ physics to quarkonium, and there is a growing body of application to the study of jet physics and collider observables. 

We are interested  in processes sensitive to three scales, $Q^2$, $Q \tau_N$ and $m^2_Q$.
$Q$ is the hard scattering scale, represented by the  $p_T$ of the hardest jet in the event. $\tau_N$ defines the jet scale, so the  typical  size of $Q \tau_N$ is the jet invariant mass, while $m_Q$ is the heavy quark mass. We are interested in the situation $Q^2 \gg  Q \tau_N, m^2_Q$.
In this case, degrees of freedom with virtuality of order  $Q^2$ can be integrated out by matching QCD onto SCET. 
The degrees of freedom of  SCET are collinear  quarks and gluons, with virtuality $p^2  \sim Q^2 \lambda^2$, and ultrasoft (usoft) quarks and gluons, 
with even smaller virtuality
$p^2 \sim Q^2 \lambda^4$. $\lambda$ is the SCET expansion parameter, $\lambda \sim m/Q \ll 1$, with $m$ the next relevant scale in the problem, e.g. $m^2 = Q\tau_N$.
In SCET different collinear sectors can only interact by exchanging  usoft degrees of freedom.
An important property of SCET is that usoft-collinear interactions can be moved from the SCET Lagrangian to matrix elements of external operators \cite{Bauer:2001yt},  greatly simplifying the proof of factorization theorems. Since the dynamics of different collinear sectors and of  usoft degrees of freedom factorize, we can focus in this paper on jets in a single collinear sector.

If there is a large hierarchy between the remaining two scales, $Q \tau_N \gg m^2_Q$, we can further lower the virtuality of the degrees of freedom in the effective theory  by integrating out  particles with virtuality  $Q \tau_N$ at the jet scale.
This second version of SCET has  collinear fields with  $p^2 \sim m^2_Q$. The additional matching step allows to factorize the dynamics of the two scales $m_Q$ and $\tau_N$, and to resum large logarithms of their ratio $m_Q^2/(Q \tau_N)$.

We now summarize some SCET ingredients needed in the rest of the paper. For more details, we refer to the original papers \cite{Bauer:2000ew,Bauer:2000yr,Bauer:2001yt,Bauer:2002nz,Leibovich:2003jd}.
We introduce two lightcone vectors   $n^{\mu}$ and $\bar n^{\mu}$, satisfying $n^2 =\bar n^2 = 0$, and $\bar n \cdot n  =2$.
The momentum of a particle can be decomposed in lightcone coordinates according  to
\begin{equation}
p^{\mu} = p^- \frac{n^{\mu}}{2} + p^+ \frac{\bar n^{\mu}}{2} + p_{\perp}^{\mu}\,.
\end{equation}
Particles collinear to the jet axis have $(p^+,p^-,p_{\perp}) \sim Q ( \lambda^2, 1,\lambda)$, where $\lambda \ll 1$ is the SCET expansion parameter.
Usoft quarks and gluons have all components of the momentum roughly of the same size $(p^+,p^-,p_{\perp}) \sim Q (\lambda^2,\lambda^2,\lambda^2)$.

The SCET Lagrangian can be written as
\begin{equation}
\mathcal L_{\textrm{SCET}} = \sum_i \mathcal L_{n_i} + \mathcal L_{\textrm{us}}\, .
\end{equation}
Each collinear sector  is described by a copy of the collinear Lagrangian $\mathcal L_n$.
For  massless quarks,  $\mathcal L_n$ is
\begin{equation}\label{SCET1}
\mathcal L_{n} = \bar \xi_{n} \left( i n \cdot D_n + g  n \cdot A_{us} + \left( \slashchar{\mathcal P}_{ \perp} + g \slashchar{A}_{n\perp} \right) W_n \frac{1}{\bar n \cdot \mathcal P} W_n^{\dagger} 
(\slashchar{\mathcal P}_{\perp} + g \slashchar{A}_{n\perp}) \right )\frac{\slashchar{\bar n}}{2}\xi_n\, . 
\end{equation}
$\xi_n$ and $ A_{n}$ are collinear quark and gluon fields, labeled by the lightcone direction $n$ and by the large components of their momentum $(p^-, p_{\perp})$.
We leave the momentum  label mostly implicit, unless explicitly needed.
The label momentum operator  $\mathcal P^{\mu}$ acting on collinear fields returns the value of the label, for example
\begin{equation}
\mathcal P^{\mu} \xi_{n,p}  = \left(  p^- \frac{n^{\mu}}{2} + p_{\perp}^{\mu}  \right) \xi_{n, p}\, .
\end{equation}
The collinear covariant derivative $D_n$ is defined as
\begin{equation}
i D_n^{\mu} = \left( \bar n \cdot  \mathcal P + g \bar n \cdot A_{n} \right)\frac{n^{\mu}}{2} 
+ \left( i n \cdot \partial + g n \cdot A_n \right) \frac{\bar n^{\mu}}{2} +   \mathcal P^{\mu}_{\perp} + g A^{\mu}_{n\perp} \, .
\end{equation}
$W_n$ are Wilson lines, constructed with collinear gluon fields, 
\begin{equation}
W_n (x) = \sum_{\textrm{perms}} \exp \left( -\frac{g}{\bar n \cdot \mathcal P} \bar n \cdot A_n (x)\right)\, .
\end{equation}
$A^{\mu}_{us}$ is an usoft gluon field.
At leading order in $\lambda$, usoft gluons couple to collinear quarks only through $n \cdot A_{us}$. This coupling can be eliminated from the Lagrangian via the BPS field redefinition \cite{Bauer:2001yt}:
\begin{eqnarray}
\xi_n^{(0)}(x) &=& Y_n^{\dagger}(x) \xi_{n}(x)\, , \\
A_n^{(0)}(x) &=& Y_n^{\dagger}(x) A_{n}(x) Y_{n}(x)\, . 
\end{eqnarray}
 $Y_n$ is a usoft Wilson line in the $n$ direction
\begin{equation}
Y_n (x) = \textrm{P} \exp \left[i g \int_{-\infty}^{0} \df s \, n \cdot A_{us} (x + s n)\right]\,, 
\end{equation}
with P denoting path ordering.
The effect of the field redefinition is to eliminate the usoft gluon in Eq.~\eqref{SCET1}, and to replace the collinear quark and gluon fields $\xi_n$ and $A_{n}$ with their non-interacting counterparts.
The same field redefinition also decouples usoft  from collinear gluons \cite{Bauer:2001yt}. From here on we always use decoupled collinear fields, and drop the superscript $(0)$.

For fast moving massive particles  there are additional mass terms \cite{Leibovich:2003jd},
\begin{equation}
\mathcal L_{m} = m_Q \bar \xi_n \left[ \left(\slashchar{\mathcal P}_{\perp} + g \slashchar A_{n\perp}\right) , W_n \frac{1}{\bar n \cdot \mathcal P} W_{n}^{\dagger} \right]\frac{\slashchar{\bar n}}{2}\xi_n - m^2_Q \bar \xi_n W_{n} \frac{1}{\bar n \cdot \mathcal P} W_{n}^{\dagger} \frac{\slashchar{\bar n}}{2}\xi_n\,.
\end{equation}
We work with  one massive quark  with mass $m_Q$,  $n_f - 1$ massless quarks and assume that quarks heavier than $m_Q$ have been integrated out.  
In this paper  we use $q$ to denote both heavy and light quarks, when it is not necessary to specify the quark mass.
$Q$ ($\bar Q$) is used exclusively for heavy quarks (antiquarks), while $l$ ($\bar l$) denotes the $n_l =  n_f - 1$ light quarks (antiquarks).

Using the Wilson line $W_n$ it is possible to construct gauge invariant combinations of collinear fields
\begin{equation}\label{chin}
\chi_n = W^{\dagger}_n \xi_n, \qquad \mathcal B^{\mu }_{n \perp} = \frac{1}{g} W^{\dagger}_n i D^{\mu}_{n \perp} W_n\, .
\end{equation}
Collinear gauge invariant operators are expressed in terms of matrix elements of these building blocks \cite{Bauer:2002nz}. In the next subsections, we discuss three such operators,  heavy quark fragmentation functions, inclusive quark and gluon jet functions, and heavy quark fragmenting jet functions.

\subsection{Heavy Quark Fragmentation Functions}\label{HQFF}

Fragmentation functions describe the fragmentation of a parton into a hadron $H$, which carries a fraction $z$ of the parton  momentum. 
In SCET, the operator definitions of the fragmentation function of a quark or a gluon into a hadron $H$ are given by  \cite{Procura:2009vm}
\begin{eqnarray}\label{qFF}
D^H_{q}(z) &=&  \frac{1}{2 N_c } \frac{1}{z}   \int \df^{d-2} p_{\perp  h} \sum_X  \nonumber\\ & &
\textrm{tr} \left[\frac{\slashchar{\bar n}}{2}  \langle 0 | \delta \left( \omega - \bar n \cdot \mathcal P \right) \delta^{(d-2)}(  \mathcal P_{\perp}) \chi_{n}(0) | H(p_h) \, X \rangle \langle H(p_h) X | \bar \chi_{n} (0)| 0 \rangle  \right]\, , \\
\label{gFF}
 D^H_{g}(z) & = &   - \frac{\omega}{d-2} \frac{1}{N_c^2 -1}  \frac{1}{z} \int \df^{d-2} p_{\perp h} \sum_X  \nonumber
\\
& &  \langle 0 | \delta\left(\omega- \bar n \cdot \mathcal P\right) \delta^{(d-2)}\left( \mathcal P_{\perp} \right) \mathcal B^{\mu, a}_{n \perp}(0) | H(p_h) \, X \rangle \langle H(p_h) X | \mathcal B^a_{n \perp,\, \mu}(0)| 0 \rangle\, .
\end{eqnarray}
The trace in Eq.~\eqref{qFF} is over Dirac and color indices.  The sum over $X$ denotes the integration over the phase space of all possible collinear final states. 
In Eqs.~\eqref{qFF} and \eqref{gFF}, $\omega$ denotes the large component of the  momentum of the fragmenting parton, and the frame in which the fragmenting parton has zero $p_{\perp}$ has been chosen. The hadron $H$ has momentum $p_h$. The perpendicular component $p_{\perp, h}$ is integrated over, while $p^-_h$, or, equivalently,  the momentum  fraction $z = p^-_h/\omega$ is measured. $N_c$ is the number of colors, $N_c = 3$.
The definitions in Eqs.~\eqref{qFF} and \eqref{gFF} are equivalent to the classical definition of fragmentation function in QCD, in Ref.~\cite{Collins:1981uw}.

The evolution of the fragmentation functions is governed by the DGLAP equation \cite{DGLAP}
\begin{equation}\label{DGLAP}
\frac{\df}{\df \log \mu^2} D_i(z,\mu^2) = \int \frac{\df \xi}{\xi} P_{j i}(\xi) D_j \left(\frac{z}{\xi}, \mu^2 \right)
\,,
\end{equation}
where the $P_{j i}(\xi)$ are the time-like splitting functions. The splitting functions are computed in perturbation theory 
\begin{equation}
P_{j i}(z) =  \frac{\alpha_s}{2\pi} \sum_{n = 0}^{\infty} \left( \frac{\alpha_s}{2\pi} \right)^{n} P_{j i}^{(n)}(z)\,,
\end{equation}
with, at one loop, \cite{DGLAP}
\begin{eqnarray}
P^{(0)}_{q_j q_i}(z) &=& \delta_{ij} C_F  \left[ \frac{1+z^2}{1-z}\right]_+ \, , \label{Pqq}\\
P^{(0)}_{g q}(z) &=&  C_F  \left( \frac{1 + (1-z)^2}{z}\right)  \label{Pgq} \, , \\ 
P^{(0)}_{q g}(z) &=&  T_R  \left( z^2 + (1-z)^2 \right)  \label{Pqg}  		\, , \\ 
P^{(0)}_{g g}(z) & =& 2  C_A \left(   z \left[\frac{1}{1-z}\right]_+ +    \frac{1-z}{z}+ z (1-z )   \right) + \frac{\beta_0}{2}\delta(1-z) \label{Pgg}\,.
\end{eqnarray}
The color factors in Eqs.~\eqref{Pqq}--\eqref{Pgg} are $C_F = 4/3$, $C_A = 3$, $T_R = 1/2$, while $\beta_0$ is the leading order coefficient of the beta function,
\begin{equation}
\beta_0 = \frac{11}{3} C_A -\frac{4}{3} T_R n_f\,.
\end{equation}
The space-like and time-like splitting functions at $\mathcal O(\alpha_s^2)$ are given in Refs.~\cite{Furmanski:1980cm,Curci:1980uw}, and nicely summarized in Ref.~\cite{Ellis:1991qj}. Space-like splitting functions are known to $\mathcal O(\alpha_s^3)$ \cite{Moch:2004pa,Vogt:2004mw}. 
The non-singlet component of the time-like splitting functions is also known to three-loops \cite{Mitov:2006ic}, while the singlet is, at the moment, unknown. 

The fragmentation functions of light hadrons are non-perturbative matrix elements, which  need to be extracted from data.
In the case of heavy flavored hadrons, the heavy quark mass  
$m_Q$ is large compared to the hadronization scale $\Lambda_{\textrm{QCD}}$. Neglecting corrections of order $\Lambda_{\textrm{QCD}}/m_Q$, one can identify the heavy hadron with a heavy quark or antiquark and  
the fragmentation function  can be computed in perturbation theory at the scale $m_Q$ \cite{Mele:1990cw}.
Expanding in $\alpha_s$,
\begin{equation}\label{alphaexpD}
D_i^j\left(z, \frac{\mu^2}{m_Q^2}\right) = \sum_{n = 0}^{\infty} \left(\frac{\alpha_s}{2\pi}\right)^n D_{i}^{j (n)} \left(z, \frac{\mu^2}{m_Q^2}\right)\,,
\end{equation}
the fragmentation function for a heavy quark into a quark or a gluon, and for a gluon into a heavy quark at $\mathcal O(\alpha_s)$ are
\begin{eqnarray}
D^{Q (0)}_Q \left(z, \frac{\mu^2}{m_Q^2} \right) & = & \delta(1-z)\,, \label{qFF.0} \\
D^{Q (1)}_Q \left(z, \frac{\mu^2}{m_Q^2}\right) &=&  C_F \left\{ \left[\frac{1+z^2}{1-z}\right]_+ \left( -1 +\log\frac{\mu^2}{m_Q^2}\right)   - 2 \left[\frac{1+z^2}{1-z} \log(1-z)\right]_+\right\}\, , \label{qFF.2}
\\
D^{g (1)}_Q \left(z, \frac{\mu^2}{m_Q^2}\right) & = & C_F \frac{(1-z)^2 + 1}{z} \left( - 1 + \log \frac{\mu^2 }{m^2_Q z^2}\right)\, , \label{qFF.3} \\
D^{Q (1)}_g \left(z, \frac{\mu^2}{m_Q^2}\right) & = & T_R (z^2 + (1-z)^2)  \log \frac{\mu^2}{m_Q^2} \, .
\label{gFF.2}
\end{eqnarray}
The  fragmentation functions of a heavy quark, heavy antiquark or light quark into a heavy quark were computed at $\mathcal O(\alpha_s^2)$  in Ref.~\cite{Melnikov:2004bm},  while the fragmentation of a gluon into a heavy quark in Ref.~\cite{Mitov:2004du}.

The fixed order expressions for the heavy quark fragmentation functions are reliable at scales $\mu \sim m_Q$, where logarithms are small.
The fragmentation functions at an arbitrary scale $\mu$  are obtained by taking the fixed order expressions  as initial condition for the DGLAP evolution.
The evolution of the one-loop initial condition \eqref{qFF.0}--\eqref{gFF.2}
with $\mathcal O(\alpha^2_s)$ splitting functions resums all leading and next-to-leading logarithms (NLL), that is all  terms of the form  $\alpha_s^n \log^n (m^2_Q/\mu^2)$ and $\alpha_s^n \log^{n-1} (m_Q^2/\mu^2)$.
The knowledge of the initial condition at $\mathcal O(\alpha_s^2)$, and of the non-singlet splitting functions at $\mathcal O(\alpha_s^3)$, allows to achieve NNLL accuracy for non-singlet combinations of the quark fragmentation function, for example $D_Q^Q - D_Q^{\bar Q}$. The evolution of the gluon distribution $D_g^Q$ and of the singlet  distribution require  the time-like singlet splitting function to $\mathcal O(\alpha_s^3)$, which, at the moment, is not known.

The picture obtained with the partonic initial conditions \eqref{qFF.0}--\eqref{gFF.2}  and the DGLAP evolution is
a  valid description of the fragmentation functions of heavy hadrons, except in the endpoint  region, $1- z\sim \Lambda_{\textrm{QCD}}/m_Q$, corresponding to the peak of the quark distribution. In this region 
soft gluon resummation and non-perturbative  effects become important and a model describing hadronization must be included, and fitted to data \cite{Cacciari:2001cw,Cacciari:2005uk,Neubert:2007je}.

We conclude this section by mentioning two important sum rules obeyed by the fragmentation functions.
The first is the momentum conservation sum rule, 
\begin{equation}\label{sumD}
\sum_H  \int \df z z D_i^H(z ,\mu^2) = 1  \, .
\end{equation}
The sum is extended over a complete set of states.  
Eq.~\eqref{sumD} is the statement that the total energy carried off by all the   fragmentation products  sums to  that of the original parton.  
At the perturbative level,  $H \in \{Q,\bar Q, g, l, \bar l\}$.
Eq.~\eqref{sumD} can be readily verified using the one loop results for the quark distributions in Eqs.~\eqref{qFF.0}--\eqref{qFF.3}.
Using the one loop expression for the splitting functions, and the DGLAP equation \eqref{DGLAP}, one can also  check that the momentum conservation sum rule is not spoiled by renormalization. This is true at all orders \cite{Collins:1981ta}. 

In addition, there are flavor conservation sum rules. 
For heavy quarks,
\begin{eqnarray}\label{sumDflav}
\int \df z  ( D^Q_Q(z,\mu^2)  -  D_{Q}^{\bar Q}(z,\mu^2)  )   = 1\, .
\end{eqnarray}
Eq.~\eqref{sumDflav} is a consequence of the fact that QCD interactions do not change the flavor of the fragmenting quark, 
and therefore the number of quark minus antiquark in the fragmentation products is always equal to one.
At $\mathcal O(\alpha_s)$   $D_{Q}^{\bar Q}$ vanishes and Eqs.~\eqref{qFF.0} and \eqref{qFF.2} explicitly satisfy the flavor sum rule.
The fragmentation functions at  $\mathcal O(\alpha^2_s)$ also satisfy Eq.~\eqref{sumDflav}  \cite{Melnikov:2004bm}. 
DGLAP evolution does not modify the flavor sum rule.

\subsection{Inclusive Jet Functions}\label{JF}

The gauge invariant quark and gluon fields, Eq.~\eqref{chin}, are a natural ingredient for the description of jets in SCET. 
Quark and gluon inclusive jet functions are defined as matrix elements of $\chi_n$ and $\mathcal B_{n \perp}$ \cite{Bauer:2002ie,Bauer:2003pi,Fleming:2003gt}, 
\begin{eqnarray}\label{quarkJ}
 J_{q}(\omega  r^+) &=&    \frac{1}{2 N_c} \int \frac{\df y^-}{4\pi} e^{i r^+ y^-/2}  \sum_X 
\textrm{tr} \left[\frac{\slashchar{\bar n}}{2}  \langle 0 |  \chi_{n}(y^-) |   X \rangle \langle  X | \bar \chi_{n} (0)| 0 \rangle  \right]
\,,  \\
J_{g}(\omega r^+) &=&  -\frac{1}{(d-2) (N^2_c -1) } \int \frac{\df y^-}{4\pi} e^{i r^+ y^-/2}    \sum_X 
 \langle 0 |  \mathcal B^{\mu, a}_{n\perp}(y^-) |  X \rangle \langle  X | \mathcal B^a_{n \perp,\, \mu} (0)| 0 \rangle \, .  \label{gluonJ}
\end{eqnarray}
We work in a frame where the momentum is aligned with the jet direction,  $p^{\mu}_J = (\omega, r^+ ,0)$. 
$\omega$ is the large component of the momentum, of the size of the jet  $p_T$, and the jet invariant mass is $\omega r^+ \ll p_T^2$. To simplify the  notation, in the rest of the paper we drop the superscript  on the plus component of the jet momentum.
The quark and gluon inclusive jet functions are infrared finite quantities, insensitive to the  scale $\Lambda_{\textrm{QCD}}$, and can be computed  in perturbation theory. 
In the case the quark field $\chi_n$ is massive, the quark jet function \eqref{quarkJ} depends on the quark mass, but the dependence is not singular \cite{Fleming:2007qr,Fleming:2007xt}.

Beyond leading order, the quark and gluon jet functions  are UV divergent and require renormalization. The dependence on the renormalization scale $\mu$
is governed by the renormalization group equation (RGE)
\begin{equation}\label{JetEv}
\frac{\df}{\df \log \mu} J_i (\omega r ,\mu^2) = \int \df(\omega s) \gamma_{J_i} (\omega r - \omega s, \mu^2)  J_i (\omega s,\mu^2)\, . 
\end{equation}
The anomalous dimension is 
\begin{eqnarray}\label{gammajet}
\gamma_{J_i} (\omega r, \mu^2) & &=  - 2 \Gamma^i_{\textrm{cusp}}(\alpha_s) \left( \left[\frac{\theta(\omega r)}{\omega r}\right]_+  - \log(\mu^2) \delta(\omega r)\right) + \gamma^i(\alpha_s) \delta(\omega r)\, ,
\end{eqnarray}
where the  plus distribution of the dimensionful variable $\omega r$ is defined as
\begin{equation}
\int \df \, (\omega r)\left[\frac{\theta(\omega r)}{\omega r}\right]_+^{}  \varphi(\omega r) = \int_0^\infty \df (\omega r) \frac{1}{\omega r}\Big( \varphi(\omega r) - \theta(\omega \kappa - \omega r) \varphi (0)\Big) +   \log^{} (\omega \kappa) \varphi(0) \, , \label{dist.1}
\end{equation}
and it is independent of the arbitrary cut-off $\omega \kappa$.

$\Gamma^q_{\textrm{cusp}}$ and $\Gamma^g_{\textrm{cusp}}$  are the quark and gluon cusp anomalous dimensions \cite{Korchemsky:1987wg,Korchemskaya:1992je},
which are known to three loops \cite{Moch:2004pa}. Up to this order, they are related by $\Gamma^g_{\textrm{cusp}}/\Gamma^q_{\textrm{cusp}} = C_A/C_F$.
$\gamma^i$ is the non-cusp component of the anomalous dimension, known to $\mathcal O(\alpha_s^2)$ \cite{Becher:2006qw,Becher:2009th}.

The form of the anomalous dimension \eqref{gammajet}, in particular its dependence on $\log \mu^2$, allows to resum  Sudakov double logarithms.
The RGE  \eqref{JetEv} can be solved  analytically, and, given an initial condition at the scale $\mu_I$, the jet function at the scale $\mu_F$ is 
\begin{equation}\label{JetSol}
J_i (\omega r , \mu^2_F) = \int \df (\omega s)\,  U_J (\omega r - \omega s, \mu^2_I, \mu^2_F) J_i(\omega s, \mu^2_I) \, ,
\end{equation}
where $U_J$ is an evolution function, given, for example in Ref.~\cite{Fleming:2007xt}.

In hadronic collisions, in addition to collinear radiation from final state particles,  one has to account for initial state radiation from the incoming beams. Initial state radiation  is described by beam functions $B_i$  \cite{Stewart:2009yx,Stewart:2010qs}.
The beam functions  depend on the invariant mass $t$ and also on  the momentum fraction $x$ of the incoming parton. 
They satisfy the same RGE as final state jets, Eq.~\eqref{JetEv}.
Large Sudakov logarithms  induced by collinear radiation from the incoming beams are resummed in the same way as logarithms in the jet functions,
\begin{equation}\label{BeamSol}
B_i (t, x , \mu^2_F) = \int \df s\,  U_J (t  - s, \mu^2_I, \mu^2_F) B_i(s, x, \mu^2_I)\, .
\end{equation}
Notice that the beam function evolution does not change the distribution in the momentum fraction $x$.
The beam functions are perturbatively related to the parton distributions \cite{Stewart:2009yx,Stewart:2010qs}, according to 
\begin{equation}\label{BeamToPDF}
B_i(t,x,\mu^2) = \int \frac{\df \xi}{\xi} \mathcal I_{i j} \left(t, \frac{x}{\xi}, \mu^2 \right) f_j(\xi,\mu^2)\, .
\end{equation}
In this case,  the initial condition for the evolution \eqref{BeamSol} cannot be computed purely in perturbation theory, but it is
 obtained  convoluting the perturbative matching coefficients $\mathcal I_{i j}$ with the parton distributions
evaluated at the scale $\mu^2_I$. 

The last  ingredient in factorization theorems for jet cross sections  is a soft function, describing soft interactions between  jets, and between  jets and the beams.
The precise definition of the soft function depends on the observable in consideration, but 
in general its RGE is of similar form as Eq.~\eqref{JetEv}, and resums Sudakov double logarithms.

\subsection{Heavy Quark Fragmenting Jet Functions}\label{HQFJF}

Fragmenting jet functions  were introduced in Refs.~\cite{Procura:2009vm,Jain:2011xz} to 
describe the fragmentation of a hadron inside a quark or gluon jet.  A first application to heavy quarks was discussed in Ref.~\cite{Liu:2010ng}. 
FJFs combine the fragmentation function, given in Eqs.~\eqref{qFF} and \eqref{gFF}, with the inclusive jet function, given in Eqs.~\eqref{quarkJ} and \eqref{gluonJ}.
This can be explicitly seen  in their   definition \cite{Procura:2009vm,Jain:2011xz}:
\begin{eqnarray}\label{quarkFJF}
\mathcal G^H_{q}(\omega  r, z) &=&  \frac{2 (2\pi)^3}{p^-_h}  \frac{1}{2 N_c} \int \frac{\df y^-}{4\pi} e^{i r^+ y^-/2} \int \df^{d-2} p_{\perp h} \sum_X \nonumber\\ 
& &\textrm{tr} \left[\frac{\slashchar{\bar n}}{2}  \langle 0 | \delta\left( \omega-  \bar n \cdot \mathcal P \right) \delta^{(d-2)} \left( \mathcal P_{\perp} \right) \chi_{n}(y^-) | H(p_h) \, X \rangle \langle H(p_h) X | \bar \chi_{n} (0)| 0 \rangle  \right]
\,,  \\
\mathcal G^H_{g}(\omega r, z) &=&  -\frac{2 (2\pi)^3 \omega }{(d-2) (N^2_c -1) p^-_h} \int \frac{\df y^-}{4\pi} e^{i r^+ y^-/2}   \int \df^{d-2} p_{\perp h} \sum_X \nonumber \\
& &
 \langle 0 | \delta\left(\omega - \bar n \cdot \mathcal P\right) \delta^{(d-2)}\left( \mathcal P_{\perp} \right) \mathcal B^{\mu, a}_{n\perp}(y^-) | H(p_h) \, X \rangle \langle H(p_h) X | \mathcal B^a_{n \perp,\, \mu} (0)| 0 \rangle \, . 
\label{gFJF}
\end{eqnarray} 
$\chi_n$ and $\mathcal B_{n \perp}$ are the gauge-invariant  fields defined in Eq.~\eqref{chin}.
Antiquark FJFs are defined in a similar way, by exchanging the fields $\chi_n$  and $\bar \chi_n$ in Eq.~\eqref{quarkFJF}.
As in Eqs.~\eqref{quarkJ} and \eqref{gluonJ}, the large component of the jet momentum is $\omega$, and $\omega r$ is the jet invariant mass.
However, differently from inclusive jets, in the definition of FJF a heavy hadron $H$ in the final state is singled out, and its momentum $p^-_h = \omega z$ is measured.
The FJFs thus depend on the jet invariant mass $\omega r$, on the momentum fraction $z$ and on the heavy quark mass $m_Q$.

The FJFs $\mathcal G_i^{H}$ have several important properties, which were  proven for light partons in Ref.~\cite{Procura:2009vm,Jain:2011xz,Procura:2011aq} and which we now discuss briefly. 

The first relationship states that after integrating over $z$ and summing over all the possible emitted particles, one should recover the inclusive jet function.
This is guaranteed by 
the  momentum \cite{Procura:2009vm,Jain:2011xz} and flavor \cite{Procura:2011aq} sum rules obeyed by the FJFs.
The  momentum conservation sum rule states that  
\begin{equation}\label{sr1}
\frac{1}{2 (2\pi)^3}\sum_H \int_0^1 \df z z \mathcal G_i^{H} (\omega r,  z, m_Q^2,\mu^2)  = J_i (\omega r, m^2_Q, \mu^2)\, , 
\end{equation}
where the sum is over a complete set of states.
The flavor sum rule for a quark is \cite{Procura:2011aq}
\begin{equation}\label{sr2}
 \frac{1}{2 (2\pi)^3} \int_0^1 \df z  \left( \mathcal G_Q^{Q} (\omega r,  z, m_Q^2,\mu^2) - \mathcal G_Q^{\bar Q} (\omega r,  z,m^2_Q,\mu^2) \right)  = J_Q (\omega r, m^2_Q, \mu^2)\, .
\end{equation}
These relations are valid  both in the approximation $\omega r \gg m_Q^2$, and in the regime $\omega r \sim m_Q^2$.
In the former case, $J_i$  are the inclusive quark and gluon jet functions, computed with massless quarks 
\cite{Bauer:2002ie,Fleming:2003gt,Becher:2006qw,Becher:2010pd}. If $\omega r \sim m_Q^2$, $J_Q$ is the massive jet function of   Ref.~\cite{Fleming:2007qr,Fleming:2007xt}.
In both cases, the mass dependence of the jet function on the r.h.s. of Eqs.~\eqref{sr1} and \eqref{sr2} is not singular.

The second property arises again due to the similarity of FJFs with inclusive jet functions. In the UV, the   FJFs $\mathcal G_i^{H}$  look like  inclusive jet functions, initiated by the parton $i$. In particular, the restriction on the final state, requiring the identification of the hadron $H$, does not affect the UV poles of the FJF,  so that  $\mathcal G_i^{H}$  have the same renormalization group equation  as quark or gluon inclusive jet functions \cite{Procura:2009vm,Jain:2011xz} 
\begin{equation}\label{rge}
\frac{\df}{\df \log \mu} \mathcal G_i^H (\omega r,  z, m_Q^2 ,\mu^2) = \int \df (\omega s) \gamma_{J_i} (\omega r - \omega s, \mu^2) \mathcal G_i^H (\omega s,  z , m_Q^2,\mu^2) \, .
\end{equation}
The anomalous dimension $\gamma_{J_i}$ is identical to the inclusive case, given in Eq.~\eqref{gammajet}, and in particular is independent of the momentum fraction $z$ and the mass of the heavy quark $m_Q$. 
The resummation of $\log r_{\tau}$ proceeds as in the inclusive case discussed in Section \ref{JF}, albeit with a different initial condition.

The final relation is due to the fact that the IR sensitivity of the FJFs is completely captured by the unpolarized fragmentation functions.
Therefore, if the jet scale and the heavy quark mass  are well separated, $\omega r \gg m^2_Q$, at leading power in $m_Q^2/\omega r$ one can factorize the dynamics at the two scales by matching the FJFs onto fragmentation functions $D_i^{H}$   
\begin{equation}\label{FJFMatching}
\mathcal G_i^H (\omega r, z, m_Q^2, \mu^2)= \sum_j \int_z^1 \frac{\df \xi}{\xi} \mathcal J_{i j} (\omega r, \xi, \mu^2,\mu^2_F) \, D_j^H \left(\frac{z}{\xi} ,\frac{\mu^2_F}{m_Q^2}\right) 
\,.
\end{equation}
The coefficients $\mathcal J_{i j}$ depend on the jet invariant mass, and on the momentum fraction, but are independent of  the heavy quark mass, up to power corrections
of the size $m_Q^2/\omega r$.   
Eq.~\eqref{FJFMatching} is very similar to the relation between the beam functions and the parton distributions in Eq.~\eqref{BeamToPDF}.

We will further discuss the properties \eqref{sr1}, \eqref{sr2}, \eqref{rge}, and \eqref{FJFMatching}, and illustrate them with 
examples at $\mathcal O(\alpha_s)$ and $\mathcal O(\alpha_s^2)$ in Secs. \ref{OneLoop}, \ref{GluonAlpha2} and \ref{LightQuark}.

\section{Review of how to resum logarithms of $r_Q$ and $r_{\tau}$ }\label{ResumRb}

Consider  single inclusive  production of one (light) hadron $h$ in $p p$ collisions,  $p p  \rightarrow h + X$.
Beyond leading order,  the partonic cross section  contains collinear divergences, when additional emissions become collinear to initial or final state partons. The divergences are physically cut off by non-perturbative  physics, and they need to be  absorbed into non-perturbative matrix elements,  parton distribution functions for  the partons in the initial state, and fragmentation functions for the final state.

In the case of  single inclusive hadroproduction of heavy hadrons, $p p \rightarrow H + X$, 
the final state collinear divergences in the partonic cross section are cut off by the heavy quark mass $m_Q$, a perturbative scale.
Identifying the heavy hadron with a heavy quark, the cross section for the production of a heavy hadron, differential in the  hadron $p_T$ and rapidity $y$, can be expressed as a convolution of the partonic cross section for the production of a heavy quark and two parton distribution functions for the incoming partons \cite{Nason:1989zy}
\begin{equation}\label{ResumRQ.3}
\frac{\df \sigma}{\df p_T^2 \df y} = \int \df x_a \int \df x_b \frac{\df \sigma_{i j} \left( p_T^2,m_Q^2, \mu^2\right) }{ \df p^2_T \df y } f_i (x_a,\mu^2) f_j(x_b,\mu^2) \, .
\end{equation}
Here the functions $f_i (x_a,\mu^2)$ denote the standard parton distributions functions, while $\df \sigma_{i j}$ denotes the partonic cross section for a parton $i$ and parton $j$ to scatter into a heavy quark with transverse momentum $p_T$ and rapidity $y$. We have omitted the dependence of the short-range cross section on the momentum fractions $x_{a,b}$, and on the heavy quark rapidity.

The final state collinear divergences present in the massless case manifest  as logarithms of the heavy quark mass in Eq.~\eqref{ResumRQ.3}.
As the energy increases, logarithms of $r_Q$ in the partonic cross section become large, threatening the validity of the perturbative expansion. In order to resum them, 
one needs to factorize the partonic cross section into two separate pieces, each of which depends on only one of the two scales $m_Q$ and $p_T$. This is achieved by introducing a fragmentation function~\cite{Cacciari:1993mq}
\begin{equation}\label{ResumRQ.4}
\frac{\df \sigma}{\df p_T^2 \df y} = \int \df x_a \int \df x_b \int \frac{\df z}{z^2} \frac{\df \sigma_{i j,\, k} \left( \hat{p}_T^2, \mu^2\right) }{ \df \hat{p}^2_T \df y } f_i (x_a,\mu^2) f_j(x_b,\mu^2) D_k^H  \left(z, \frac{\mu^2}{m_Q^2}\right) .
\end{equation}
In Eq.~\eqref{ResumRQ.4}, the short-range cross section is the cross section for the production of the parton $k$ with transverse momentum $\hat p_T$ and rapidity $y$ in the collision of partons $i$ and $j$, computed with all partons considered massless. The parton $k$ then fragments into a heavy hadron $H$, carrying a transverse momentum $ p_T = z \hat p_T $, and the same rapidity as the original parton $y$. If 
the short-range cross section and the fragmentation function are evaluated at their characteristic scale, respectively $\mu \sim p_T$  and  $\mu \sim m_Q$, no large logarithms arise in the perturbative expressions. Of course, in the end all functions have to be evaluated at a common scale $\mu$, and one therefore has to use the RGE to evolve each function to this scale. 
The RG evolution of the fragmentation function is determined by the DGLAP equation, Eq.~\eqref{DGLAP}.
By evolving the fragmentation function from $\mu \sim m_Q$ to $\mu \sim p_T$ one can sum the logarithms of $m_Q / p_T$. 

As already noted, the short-range cross section in \eqref{ResumRQ.4} is calculated in the limit $m_Q = 0$. Thus, while this approach correctly resums the logarithms of $m_Q / p_T$, it does not contain any dependence on powers of the same ratio. Since the power dependence is correctly reproduced using \eqref{ResumRQ.3}, one can obtain an expression that correctly reproduces both the logarithmic and the power dependence on $m_Q / p_T$ by combining the two ways of calculating. This is the approach taken in FONLL~\cite{Cacciari:1998it}.

Now consider jet cross sections.
As in the previous case, the starting point for a resummation of the large logarithms that arise in cross sections that are differential in a jet resolution parameter $\tau$ is the separation of the dimensionful variables whose ratio gives the value of $\tau$. This is achieved by a factorization of the cross section. There are many jet resolution variables one can choose, and a large body of literature how to obtain the relevant factorization theorems. Since all approaches in the end contain the same physics, and the final factorization theorems look very similar, we simply state the result here for one specific resolution variable, namely $N$-jettiness $\tau_N$ \cite{Stewart:2010tn}. For the purposes of this discussion, the only relevant part of the definition of $N$-jettiness is that $\tau_N$ has dimension one,  $\tau_N \to 0$ as we approach $N$ pencil-like jets, and that $\tau_N$ is linear in the contributions from each jet (both from initial and final state radiation) and soft physics $\tau_N = \tau_N^{(a)} + \tau_N^{(b)} + \tau_N^s + \sum_j \tau_N^{(j)}$. 

The factorization theorem can be written schematically as \cite{Stewart:2010tn}
\begin{eqnarray}\label{cross.0}
\frac{\df\sigma }{\df p_T\df\tau_N  } &=& 
 \int \!\df x_a \! \int \!\df x_b
 \,
 H_{a b,  k_1 \ldots k_{N}  }\left( p_T,\mu^2 \right) 
\int \!\df \tau_N^{(s)} \!\int \! \df \tau_N^{(a)} \ldots \int \! \df  \tau_N^{(N)} 
\nonumber\\
&& \quad
\delta\Big(\tau_N - \tau_N^{(s)} - \tau_N^{(a)} - \tau_N^{(b)} - \sum_j\tau_N^{(j)} \Big )   S_{a b ,k_1\ldots k_N } \left(\tau^{(s)}, \mu \right)  
   \nonumber \\
  & & \times 
 B_{a}(  Q \tau_N^{(a)}, x_a, \mu^2)  B_{b} ( Q \tau_N^{(b)}, x_b,\mu^2)  \prod_{j=1}^N J_{k_j} ( Q \tau_N^{(j)},\mu^2)       
\,,
\end{eqnarray}
and we have only included the dependence on terms that are relevant for our discussion.
$H$ is the hard function for the production of the partons with flavor $k_1, \ldots k_{N}$, and it depends on the $p_T$ of the $N$ signal jets. 
Collinear radiation in the final state is described by the inclusive jet functions $J_{k_i}$, while two beam functions
$B$  describe initial state radiation from the incoming beams, initiated by the partons of flavor $a$ and $b$. The jet and beam functions are function of the jet invariant mass  $ Q \tau_N^{}$, where $Q$ is  of the size of the hard scattering scale. In addition, the beam functions depend on the momentum fraction $x$ of the incoming partons.
The soft function  describes soft interactions between jets, and between jets and  the beams.  It depends on soft momenta $\sim \tau_N^{(s)}$.

After evolving the parton distribution functions to the jet scale, as discussed in Section \ref{SCET}, each function in the factorization theorem \eqref{cross.0} only depends on a single scale, thus one can again calculate each term at its characteristic scale without encountering any large logarithms, and then evolve them to a common scale using the RGEs discussed in Section \ref{JF}.

The factorization formula in \eqref{cross.0} is derived in the limit $\tau_N \to 0$, such that no power corrections of the ratio $\tau_N/p_T$ can be included. In order to derive an expression that is valid in both the limits of small and large $\tau_N / p_T$, one needs to combine the resummed result with the known fixed order expression, which includes this power dependence.

\section{A combined resummation of $r_Q$ and $r_{\tau}$}\label{ResumBoth}

In this section we give the factorization theorems that are required to combine both types of resummation, such that one can study the production of heavy flavor at high energy in the presence of jet vetoes, or perhaps more importantly, such that one can combine calculations which resum the dependence on the heavy quark mass with parton shower algorithms. The later sections in this paper are then devoted to calculating the new ingredients in the resulting factorization theorems perturbatively. 

We will consider two separate cases. The first is the production of identified heavy flavored hadrons in hadronic collisions, with measured momentum of the heavy hadron. Examples are   $ p p \rightarrow H + X$ or the associated production of a heavy flavored hadron and a weak boson,  $p p \rightarrow W + H +X$. In particular, we consider the case in which the heavy hadron is part of an identified jet, and a jet veto limits the total number of jets in the event. The momentum of the heavy hadron is characterized by its fraction $z$ of the total jet momentum it is part of. 
A second interesting application is the production of jets (identified by a regular jet algorithm), which are tagged as $b$ jets, and again extra jet activity is vetoed. Since $b$-tagging algorithms rely on the presence of at least one  weakly decaying $b$-flavored hadron, the situation is related to the previous case. The main difference is that the momentum of the $B$ hadron is not measured in this case, and the momentum fraction $z$ is therefore integrated over.

The factorization theorem is based on the FJFs, defined in Refs.~\cite{Procura:2009vm,Jain:2011xz} for light hadrons, and reviewed in Section \ref{HQFJF}. The main new ingredient in this work is to extend the idea of a FJF to the case of heavy quarks, in which case infrared singularities that were present in the light FJFs manifest themselves as a logarithmic dependence on the heavy quark mass. 

In addition to cases we discuss, logarithmic dependence on  $m_Q$ appears in the flavor excitation channel.
In this channel, one heavy quark is present in the initial state and enters the hard collision. This is similar to the cases discussed above, with the  difference  that the production of the heavy flavor happens in the initial  rather than the final state. 
In this case $\log r_Q$ are resummed by introducing a perturbative $b$-quark parton distribution at the scale $m_Q$, and running it with the DGLAP equation up to the hard scattering scale. Initial state radiation at a scale $t \gg m_Q^2$ can be studied using the same techniques developed in this paper, by introducing a heavy quark beam function.
We leave a detailed discussion for future work, and will not discuss initial state splitting any further.

\subsection{The production of an identified heavy hadron}
We consider first the case of production of an identified heavy flavored hadron in the presence of a veto on extra jet activity. As already discussed, the momentum of the heavy hadron is measured to have a fraction $z$ of the momentum of the jet  it is part of. The extra jet activity is vetoed using a jet resolution variable $\tau_N$, where $\tau_N$ is defined such that it goes to zero when there are at most $N$ pencil-like jets present. Phenomenologically interesting applications are the
 two-jettiness cross section in $p p \rightarrow Q \bar Q + X$, or the one- and two-jettiness cross sections for $ p p \rightarrow W + Q + X$.

In the limit of small $\tau_N$, the factorization theorem for the cross section differential in $\tau_N$ and in the $p_T$ and rapidity of the observed hadron can schematically be written as
\begin{eqnarray}\label{cross.1}
\frac{\df\sigma }{\df p_T^2 \df y \, \df \tau_N  } &=& 
 \int \!\df x_a \! \int \!\df x_b \int \frac{\df z}{z^2}
 \,
 H_{a b,  k_1 \ldots k_{N}  }\left( \frac{p_T}{z},\mu^2 \right) 
\int \!\df \tau_N^{(s)} \!\int \!\df  \tau_N^{(a)} \ldots \int \! \df  \tau_N^{(N)} 
\nonumber\\
&& \times  S_{a b ,k_1\ldots k_N } \left(\tau^{(s)}, \mu \right)  
  B_{a}(  Q \tau_N^{(a)}, x_a, \mu^2)  B_{b} ( Q \tau_N^{(b)}, x_b,\mu^2)  
\prod_{j=1}^{N-1} J_{k_j} ( Q \tau_N^{(j)},\mu^2)
\nonumber\\
&& 
\times
  \mathcal G_{k_N}^H ( Q \tau_N^{(N)} , z,m^2_Q,\mu^2 )
\delta\left(\tau_N - \tau_N^{(a)} - \tau_N^{(b)} - \tau_N^{(s)} - \sum_j\tau_N^{(j)}  \right) \,.
\end{eqnarray}
This factorization theorem is almost identical to the one given in Eq.~\eqref{cross.0}, and, as in that case, it holds up to power corrections in  $\tau_N/p_T$. The only  difference is that a hadron $H$ is observed inside the jet initiated by the parton $k_N$, and its $p_T$ and rapidity are measured. To be able to describe this extra information, the inclusive jet function $J_{k_N}$ needs to be replaced
by  the FJF $\mathcal G_{k_N}^H$. In addition to the argument $\tau_N$, describing the contribution of the inclusive jet to the jet resolution variable, the fragmenting jet function depends on the mass of the heavy quark $m_Q$ as well as the momentum fraction of the heavy hadron $z$. 

As discussed in Section \ref{HQFJF} the RGE of the FJF, Eq.~\eqref{rge}, is identical to that of an inclusive quark or gluon jet function, so that  the resummation of $\log \tau_N$ proceeds as in the inclusive case.

The FJFs are two-scale objects, sensitive to the jet invariant mass and to  the heavy quark mass $m_Q$.
Differently from the light parton FJFs discussed in Refs.~\cite{Procura:2009vm,Jain:2011xz}, the heavy quark FJFs can be computed purely in perturbation theory.
If the jet scale $Q \tau_N $ in Eq.~\eqref{cross.1} is close to $m_Q^2$, the fixed order expression for the FJFs at the scale $\mu_I^2 \sim Q \tau_N$ does not contain large logarithms, and the evolution \eqref{JetSol} resums logarithms of $\tau_N$.
On the other hand, if $Q \tau_N \gg m_Q^2$, there is no choice of initial scale the minimizes the logarithms in the FJFs, and the initial condition for the jet evolution is still plagued by large logarithms. However,
the IR sensitivity of the FJFs is completely captured by the unpolarized fragmentation functions.
Therefore, if the jet scale and the heavy quark mass  are well separated  one can factorize the dynamics at the two scales by matching the FJFs onto heavy quark fragmentation functions $D_i^{H}$, as in Eq.~\eqref{FJFMatching}. 
Since each term on the right-hand side of Eq.~\eqref{FJFMatching} depends only on a single scale, the logarithms of $\log m_Q^2/( Q \tau_N)$ of the FJFs are reproduced through logarithms of $Q\tau_N / \mu^2_F$ and $m^2_Q/\mu^2_F$ on the right hand side, such that they can be resummed through RG evolution.  Evolving the fragmentation function from the mass scale to a scale of order $Q \tau_N$, no large logarithms are left in the initial condition for the FJF evolution. Then, running the hard, beam, soft and jet functions  to a common scale, all large 
logarithms in Eq.~\eqref{cross.1} are correctly resummed.

By matching the FJFs onto fragmentation functions and the beam functions onto parton distributions, we can recast Eq.~\eqref{cross.1} 
in a form that stresses the relation to  single inclusive production discussed in Section \ref{ResumRb}.
\begin{eqnarray}\label{cross.1b}
\frac{\df\sigma }{\df p_T^2 \df y \, \df \tau_N  } &=& 
 \int \! \frac{\df \xi_a}{\xi_a} \int \! \frac{\df\xi_b}{\xi_b} \int \! \frac{\df \zeta}{\zeta} \Bigg (
\int \!\df x_a \! \int \!\df x_b \int \frac{\df z}{z^2}
 \,
 H_{a b,  k_1 \ldots k_{N}  }\left( \frac{p_T}{z},\mu^2 \right) 
\nonumber\\ 
& &  
\int \!\df \tau_N^{(s)} \int \!\df \tau_N^{(a)} 
\ldots \int \!\df\tau_N^{(N)} 
\delta\left(\tau_N - \tau_N^{(a)} - \tau_N^{(b)} - \tau_N^{(s)} - \sum_j\tau_N^{(j)}  \right)
\nonumber \\
& & 
\mathcal{I}_{a  a^{\prime}}( Q \tau_N^{(a)}, x_a/\xi_a,\mu^2)  \,
\mathcal{I}_{b  b^{\prime}}(  Q \tau_N^{(b)}, x_b/\xi_b,\mu^2) \, \mathcal J_{k_N k_N^{\prime}} (Q \tau_N^{(N)} , z/\zeta,\mu^2 )
\nonumber\\
&& \times  S_{a b ,k_1\ldots k_N } \left(\tau^{(s)}, \mu \right)  
\prod_{j=1}^{N-1} J_{j} ( Q \tau_N^{(j)},\mu^2)
 \Bigg) \;  f_{i^{\prime}} (\xi_a,\mu^2) f_{j^{\prime}} (\xi_b, \mu^2) D_{k_n^{\prime}}^H \left(\zeta, \frac{\mu^2}{m_Q^2}\right) \, . \nonumber\\
 \end{eqnarray}
This form is very similar to Eq.~\eqref{ResumRQ.4},  the only difference being that the partonic short range cross section in Eq.~\eqref{ResumRQ.4} has been further  separated into different pieces, each of them dependent on a single scale. 
If $\tau_N \sim Q$,  the hard, jet and soft scales  become equal and the resummation of  $\log \tau_N$  is turned off. The fragmentation function and the parton distributions are evolved up to the hard scale, resumming $\log r_Q$, at the desired logarithmic accuracy. In this situation, Eq.~\eqref{cross.1b} reduces to  the $N$ jet limit of  Eq.~\eqref{ResumRQ.4}.

\subsection{The production of tagged heavy flavor jets}
In this section, we consider the impact of logarithms of the quark mass on observables involving jets containing heavy flavor. The most important application of this is for the description of $b$-tagged jets.
Let us start by giving a closer look to the experimental definition of $b$ jets. In high energy experiments, like ATLAS or CMS, $b$ jets are tagged using a variety of techniques based on the long lifetime of weakly decaying heavy flavored hadrons inside the jets \cite{ATLAS:2011qia,Chatrchyan:2012jua}.
These techniques have in common the requirement of the presence of a weakly decaying $b$-flavored hadron, within a certain distance $\Delta R$ from the jet axis, in rapidity-azimuthal angle space,  and with a minimum $p_T$. Typical choices are $\Delta R < 0.3$, and $p_T > 5$ GeV. $b$ tagging algorithms do not differentiate between jets containing $b$  or $\bar b$ quarks. 

Our goal is to  define a heavy quark tagged ($Q$-tagged) jet function $J_i^{ Q} (\omega r, m_Q^2,\mu^2)$ with these features, such that one can use the factorization formula given in Eq.~\eqref{cross.0} and simply replace the standard jet function by its heavy quark tagged version. A heavy quark tagged jet function is agnostic as to which type of hadron gives rise to the long decay time and therefore the secondary vertex. Furthermore, it is insensitive to the momentum fraction of the heavy hadron, as long as the transverse momentum is above the minimum transverse momentum imposed in the $b$-tagging algorithm. Therefore, the $b$-tagged jet function can be obtained from the heavy flavor FJF by summing over all heavy flavored hadrons as well as integrating over the momentum fraction of the heavy hadron (down to a cutoff $z_0$ related to the minimum $p_T$).

As is the case for a fragmentation function, the hard interaction does not necessarily need to involve the production of a heavy quark $Q$, since this can be produced from the splitting $g \to Q \bar Q$ in the radiation happening within the jet. Thus, heavy quark tagged jets can be initiated by any possible flavor. 
In the case of heavy quark initiated jets we define
\begin{eqnarray}\label{intro.3}
 J_Q^Q (\omega r, m_Q^2, \mu^2) & & = \frac{1}{2 (2\pi)^3} \int_{z_0}^1 \df z  \left(  \mathcal G_Q^{Q} ( \omega r,   z, m_Q^2, \mu^2)  -  \mathcal G_Q^{\bar Q} ( \omega r,  z, m_Q^2, \mu^2) \right)\, ,
\end{eqnarray}
where the subtraction of the antiquark contribution avoids the double counting of configurations in which the heavy quark splits in an additional $Q \bar Q$ pair, $Q \rightarrow Q Q \bar Q$.
When $z_0$ approaches 0, a heavy quark initiated jet should always be tagged. In virtue of the flavor sum rules  obeyed by the quark fragmentation function and FJF,
Eq.~\eqref{intro.3} does indeed guarantee that for $z_0 \rightarrow 0$ one recovers an  inclusive quark jet. In particular, any dependence on the fragmentation function, and thus on logarithms of the mass, disappears. For $J_Q^Q $ there is no need to resum DGLAP logarithms, while Sudakov double logarithms of $ r_{\tau}$ are resummed by the evolution of the  inclusive quark jet function.
Powers of $m_Q^2/(Q \tau_N)$ can be retained by using inclusive massive jet functions \cite{Fleming:2007qr,Fleming:2007xt}.

For gluon and light quark initiated jets, $i \in \{g, l\}$, heavy quarks are always produced in pairs. In this case  we define
\begin{eqnarray}\label{intro.4}
 J_{g, l}^Q (\omega r, m_Q^2, \mu^2) & & =  \frac{1}{2} \int_{z_0}^1 \df z  \frac{1}{2 (2\pi)^3} \left(  \mathcal G_{g,l}^{Q} ( \omega r,   z,m_Q^2, \mu^2)  +  \mathcal G_{g,l}^{\bar Q} ( \omega r,  z, m_Q^2, \mu) \right) \nonumber \\ && = \frac{1}{2 (2\pi)^3} \int_{z_0}^1 \, \df z  \,  \mathcal G_{g,l}^{Q} ( \omega r,  z, m_Q^2, \mu^2) \, ,
\end{eqnarray}
where, in the last step, we used  charge conjugation invariance. 
With this definition, $J_{g, l}^Q$ count the multiplicity of $Q \bar Q$ pairs in a gluon or light quark jet of invariant mass $\omega r$.\footnote{We thank W.~J.~Waalewijn for discussions on this point.}
Since the anomalous dimension of the FJF is $z$-independent,  the RGE of the $Q$-tagged jet is still identical to that of the  inclusive jet function.
For gluon or light quark initiated jets the dependence on the fragmentation function does not drop out.
For small values of $Q \tau_N \sim 4 m_Q^2$, this does not cause problems. The only large logarithms in this case are $\log\tau_N$, which are resummed by using the fixed order expression of $J_{g,l}^Q$ as initial condition for the jet evolution \eqref{JetSol}.
For larger values of $4 m_Q^2 \ll Q \tau_N \ll Q^2$,  resummation of $\log m_Q^2/(Q\tau_N)$ does become necessary and is achieved by running the fragmentation function to the scale $Q \tau_N$.

\section{Heavy Quark Fragmenting Jet Functions at $\mathcal O(\alpha_s)$}\label{OneLoop}

\begin{figure}
\center
\includegraphics[width=14cm]{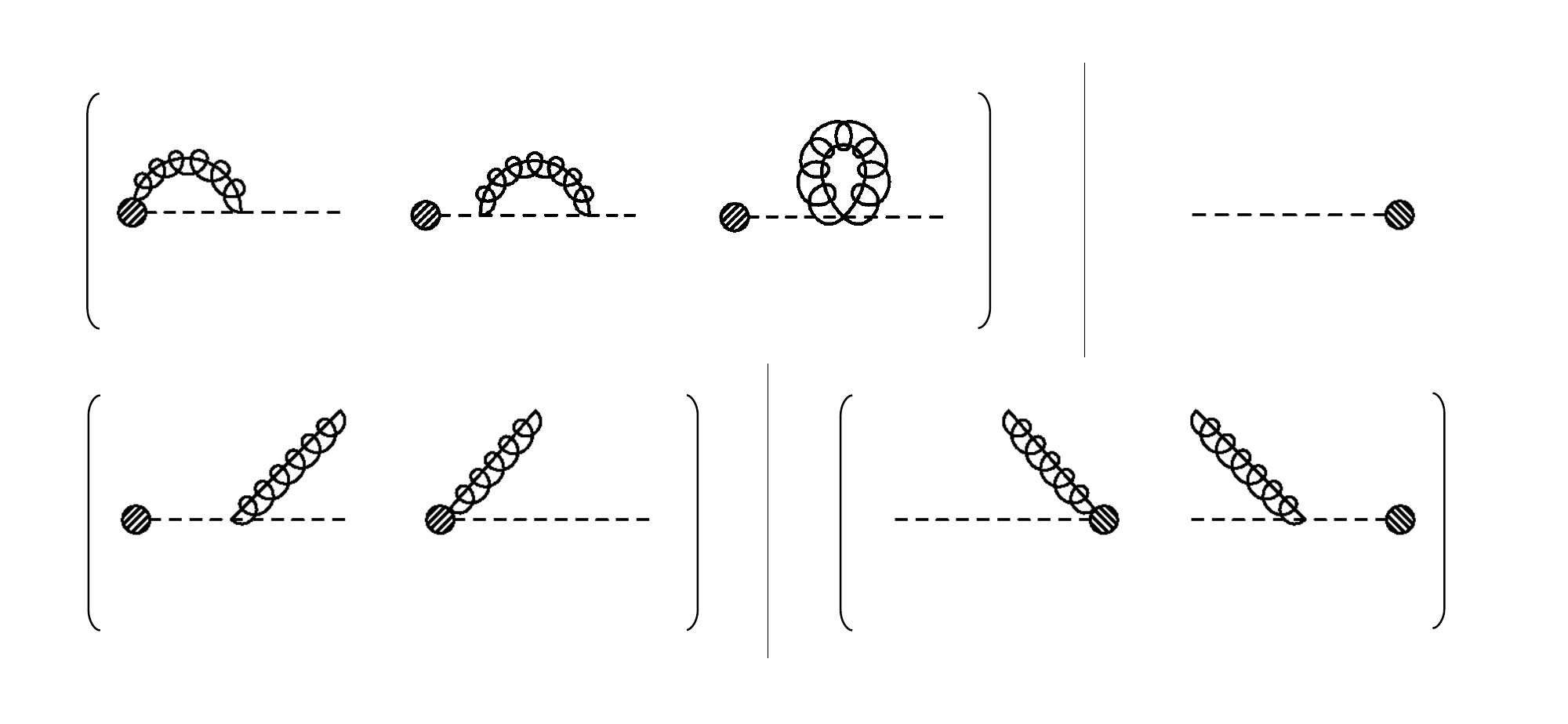}
\caption{$\mathcal O(\alpha_s)$  corrections to the heavy quark FJFs $\mathcal G_Q^Q$ and $\mathcal G_Q^g$.
Dashed lines denote collinear heavy  quarks. Springs denote collinear gluons.
}
\label{alpha}
\end{figure}

\begin{figure}
\center
\includegraphics[width = 16cm]{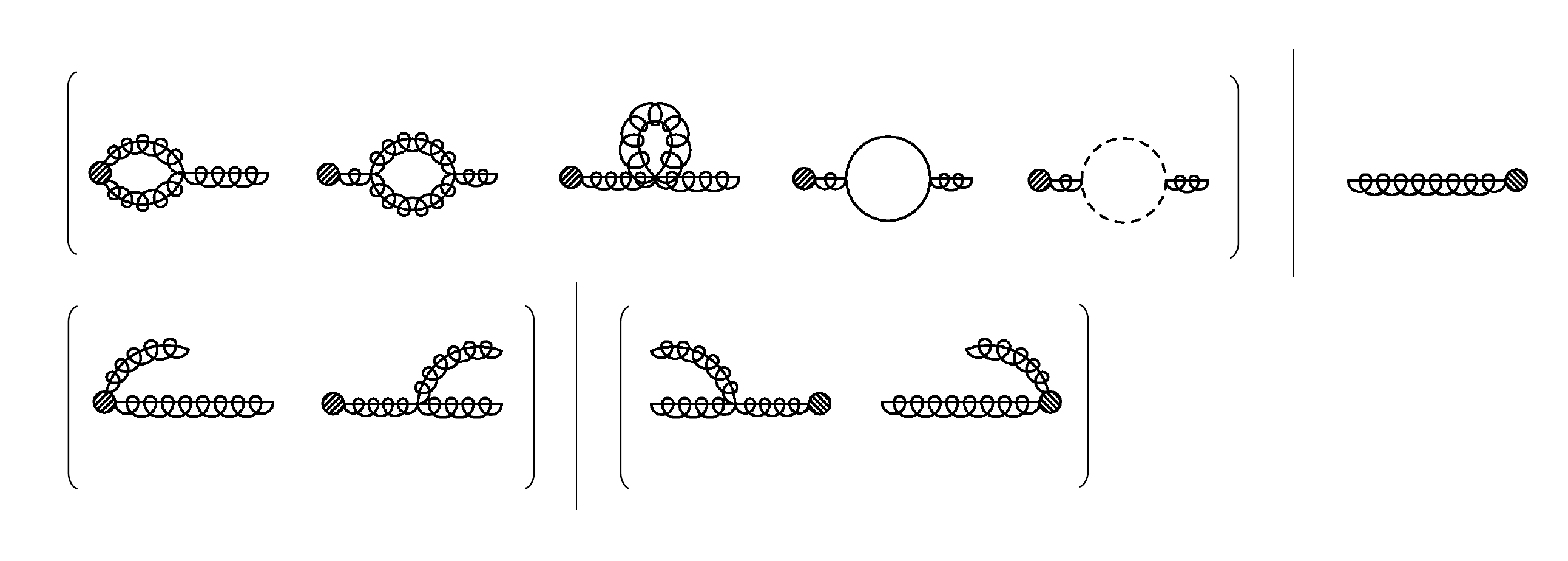}
\caption{ $\mathcal O(\alpha_s)$ corrections to the gluon FJF   $\mathcal G_g^g$. The notation for collinear heavy quarks and gluons is as in Fig.~\ref{alpha}. Collinear light quarks are denoted by a plain line. }\label{ggalpha}
\end{figure}

We now discuss  in more detail the FJFs of massive quarks,  illustrating  the general properties discussed in  Section \ref{HQFJF} with one loop examples.
We work in perturbation theory, identifying the hadron $H$ with one of the parton species, $\{Q, \bar Q, g\}$.
For heavy quark production, the most interesting functions are those with an identified $Q$ or $\bar Q$.
For completeness, and to verify the cancellation of mass dependent terms in the momentum sum rule \eqref{sr1} for the quark and gluon FJFs,  we also consider the effects of the heavy quark mass on the FJF of a heavy quark into a gluon, and of a gluon into a gluon, even though these FJFs are of less practical interest.

We expand the FJFs  $\mathcal G$ and the matching coefficients $\mathcal J$ in powers of $\alpha_s/(2\pi)$ , 
\begin{equation}\label{alphaexpJ}
\mathcal G_i^j = \sum_{n  = 0}^{\infty}  \left( \frac{\alpha_s}{2\pi}\right)^n \mathcal G_i^{j (n)}, \qquad  
\mathcal J_{i j}
= \sum_{n = 0}^{\infty} \left(\frac{\alpha_s}{2\pi} \right)^n \mathcal J^{(n)}_{i j}
\, .
\end{equation}
At tree level the heavy quark FJF, $\mathcal G_{Q}^{Q}$, and the gluon FJF, $\mathcal G_g^g$, are the product of a delta function on the jet invariant mass, and a delta function on the observed momentum fraction,
\begin{equation}
\frac{1}{2 (2\pi)^3} \mathcal G^{Q (0)}_Q (\omega r, z, m_Q^2, \mu^2) = \delta(1-z) \delta(\omega r), \quad
\frac{1}{2 (2\pi)^3} \mathcal G^{g (0)}_g (\omega r, z, m_Q^2, \mu^2) = \delta(1-z) \delta(\omega r),
\end{equation}
with the factor of $2 (2\pi)^3$ due to the choice of normalization of Refs.~\cite{Procura:2009vm,Jain:2011xz}. 
All other FJFs vanish at tree level. 

\begin{figure}
\center
\includegraphics[width = 6cm]{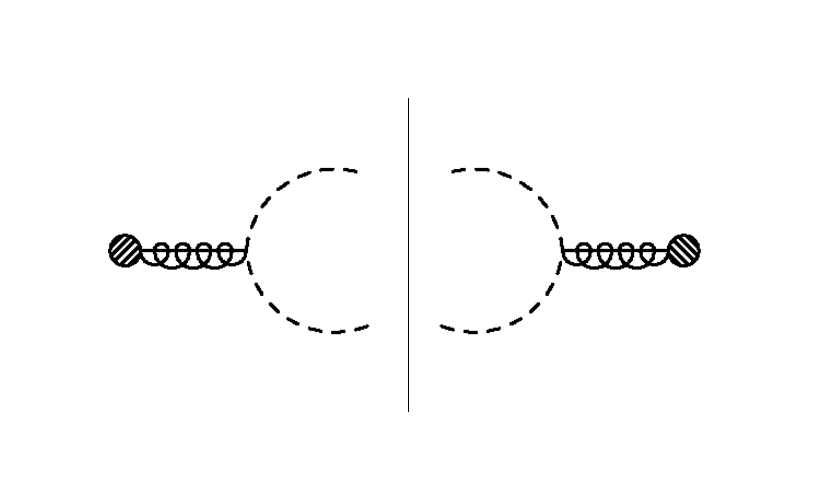}
\caption{$\mathcal O(\alpha_s)$ contribution to $\mathcal G_g^Q$.}\label{gQQalpha}
\end{figure}

At order $\mathcal O(\alpha_s)$, $\mathcal G_Q^{Q}$ and $\mathcal G_g^g$ receive corrections from virtual one loop diagrams, and real diagrams, with the emission of an additional parton. We show the diagrams contributing to $\mathcal G_Q^Q$  in Fig.~\ref{alpha}, and to $\mathcal G_g^g$ in Fig.~\ref{ggalpha}.
At this order, one finds the first contributions to $\mathcal G_Q^g$ and $\mathcal G_g^Q$. They originate purely from real emissions, the real diagrams in Fig.~\ref{alpha} for $\mathcal G_Q^g$, and the diagram in Fig.~\ref{gQQalpha} for $\mathcal G_g^Q$.
We compute the diagrams in Figs.~\ref{alpha}, \ref{ggalpha} and \ref{gQQalpha} with a finite quark mass $m_Q$, and take the limit $m_Q^2 \ll \omega r$ at the end of the calculation. We present here the results in this limit, which we refer to as ``massless limit'', and relegate the one loop expressions for finite $m_Q$ to Appendix \ref{AppC}.

The individual diagrams  contributing to  $\mathcal G_Q^Q$  contain UV and IR divergences. We regulate  UV divergences in dimensional regularization, and IR divergences by introducing a $\Delta$
regulator, as defined in Ref.~\cite{Chiu:2009yx}. Double counting  of the collinear and usoft regions is eliminated via zero-bin subtractions \cite{Manohar:2006nz}.
We choose different regulators in the IR and UV to explicitly check that, after zero-bin subtraction, all infrared divergences cancel between real and virtual emission diagrams, and the remaining  $1/\varepsilon$ poles are UV in nature.

The UV divergences of the diagrams in Fig.~\ref{alpha} are canceled by introducing the counter-term $Z_{J_q}$ relating the renormalized and unrenormalized FJF
\begin{equation}\label{JetRenQ}
( \mathcal G_Q^Q (\omega r, z,  m_Q^2, \mu^2) )_{\textrm{ren}} = \int \df (\omega s) Z_{J_q} (\omega r - \omega s,\mu^2) \mathcal G_Q^{Q}(\omega s, z, m_Q^2, \mu^2) ,
\end{equation}
with, at one loop,
\begin{equation}\label{ZJ}
Z_{J_q}(\omega r, \mu^2) =\delta(\omega r)  - \frac{\alpha_s }{4\pi} (4 C_F )\left\{  \delta(\omega r) \left( \frac{1}{\varepsilon^2} + \frac{1}{ \varepsilon }\right)
-  \frac{1}{\varepsilon} \left(  \left[\frac{\theta(\omega r)}{\omega r}\right]_{+} -  \log (\mu^2) \delta(\omega r) \right)   \right\}.
\end{equation}
As expected $Z_{J_q}$ does  not depend on $m_Q$, which is an IR scale in SCET.  $Z_{J_q}$ is also independent of  the momentum fraction $z$, implying  that the evolution of the quark FJF from the jet scale to the hard scale  does not change the shape of the momentum fraction  distribution.

From Eq.~\eqref{ZJ}, one can derive the RGE of $\mathcal G_Q^Q$. It is of the form \eqref{JetEv},
with anomalous dimension
\begin{eqnarray}\label{gammajet2}
\gamma_{J_q} (\omega r ,\mu^2) & &= \int \df (\omega s) Z_2 Z_{J_q}^{-1}\left( \omega r - \omega s,\mu^2 \right) \frac{d}{d\log \mu} \left( Z_2^{-1} Z_{J_q}\left( \omega s, \mu^2\right) \right)  \nonumber \\
& & =\frac{\alpha_s}{4\pi} \left( - 2 (4 C_F) \left( \left[\frac{\theta(\omega r)}{\omega r}\right]_+  - \log(\mu^2) \delta(\omega r)\right) +6 C_F \delta(\omega r)
\right).
\end{eqnarray}
 $Z_2$ is the quark field renormalization. At one loop, in the $\overline{\textrm{MS}}$ scheme  and in Feynman gauge, 
\begin{equation}
 Z_2 = 1 - \frac{\alpha_s C_F}{4\pi \varepsilon}.
\end{equation}
Eq.~\eqref{gammajet2} has the same form as Eq.~\eqref{gammajet}. The coefficient of the plus distribution is the one-loop value of the quark cusp anomalous dimension $\Gamma^q =  \alpha_s/ (4 \pi) (4 C_F)$ \cite{Korchemsky:1987wg,Korchemskaya:1992je}. The coefficient of the delta function reproduces the one-loop value of the
non-cusp component of the anomalous dimension of the inclusive quark jet function, $\gamma^{q}  =  \alpha_s /(4\pi) (6 C_F)$.
The anomalous dimension $\gamma_{J_q}$ is thus identical to the anomalous dimension that governs the evolution of inclusive quark jets.

In the massless limit, the quark FJF depends on the jet invariant mass $\omega r$ through plus distributions of the form
\begin{eqnarray}
\int \df (\omega r) \left[\frac{\theta(\omega r)\log^n (\omega r)}{\omega r}\right]_+^{}  \varphi(\omega r)&= & \int_0^\infty \df (\omega r) \frac{\log^n (\omega r)}{\omega r}\Big( \varphi(\omega r) - \theta(\omega \kappa - \omega r) \varphi (0)\Big) \nonumber \\ & &+   \frac{1}{n+1}\log^{n+1} (\omega \kappa) \varphi(0)\,. \label{dist.2} 
\end{eqnarray}
Since the integration range extends to infinity, one has to introduce an arbitrary cut-off $\omega \kappa$ in the subtraction term, $\varphi(0)$. The cut-off dependence  is canceled by the second line of Eq. \eqref{dist.2}, and the distributions   do not depend on $\omega \kappa$. 
It is also convenient to introduce the notation
\begin{eqnarray}
 \left[\frac{\theta(\omega r)\log^n (\omega r/\mu^2)}{\omega r}\right]_+^{(\mu^2)}  \equiv   \left[\frac{\theta(\omega r)\log^n (\omega r/\mu^2)}{\omega r}\right]_+^{} + (-1)^{n+1} \frac{1}{n+1}\log(\mu^2) \delta(\omega r)  .\label{dist.3} 
\end{eqnarray}
In terms of the distribution in Eq.~\eqref{dist.3},
the renormalized  heavy quark FJF at one loop  is  
\begin{eqnarray}
\frac{\mathcal G^{Q (1)}_Q( \omega r, z, m_Q^2, \mu^2)}{2 (2\pi)^3} &=& 
C_F \left\{ 
     \delta(1-z)  \left( 2 \left[\frac{\theta(\omega r) \log (\omega r/\mu^2)}{\omega r}\right]_+^{(\mu^2)}   - \frac{3}{2}   \left[\frac{\theta(\omega r)}{\omega r}\right]_{+}^{(\mu^2)} \right)  \right. \nonumber \\  & &  \left.
  + \left[\frac{1+z^2}{1-z}\right]_+ \left(  \left[\frac{\theta(\omega r)}{\omega r}\right]_+^{(\mu^2)} + \delta(\omega r) \left(
  \log \left( \frac{ \mu^2 z}{m^2_Q} \right) - 1
   \right) \right)
   \right. \nonumber \\ & & \left.
  + \delta(\omega r) \left( \delta(1- z) \left(   -\frac{\pi^2}{6} + \frac{7}{4}\right) 
  -     \left[\frac{1+z^2}{1-z} \log(1-z) \right]_+  + (1-z) \right)
 \right\}. \nonumber\\\label{qFJF.2} 
\end{eqnarray}

The gluon FJF $\mathcal G_g^g$ is affected by the quark mass only via quark loop corrections to the gluon propagator, the last virtual diagram in Fig.~\ref{ggalpha}. 
The remaining diagrams in Fig.~\ref{ggalpha} are unchanged with respect to the massless case discussed in Ref.~\cite{Jain:2011xz}, and we did not evaluate them.
The correction to $\mathcal G_g^g$ is obtained by multiplying the tree level result by the contribution of massive quarks to the residue of the gluon propagator, and we obtain 
\begin{eqnarray}
\frac{\mathcal G^{g (1)}_g(\omega r, z, m_Q^2, \mu^2)}{2 (2\pi)^3} & =& \left. \frac{\mathcal G^{g (1)}_{g}(\omega r, z,\mu^2)}{2 (2\pi)^3} \right|_{\textrm{light}}- \delta(1-z)\delta(\omega  r)  T_R \frac{2}{3} \log \frac{\mu^2}{m_Q^2}, \label{gFJF.3}
\end{eqnarray}
where $\left.\mathcal G^{g}_{g}(\omega r, z,\mu^2) \right|_{\textrm{light}}$ is the gluon FJF computed with $n_l$ massless quarks. 
The same correction  to the gluon propagator affects the fragmentation function for a gluon into a gluon, $D_g^g$, so that it cancels in the matching and the matching coefficients are mass independent.
The heavy quark mass does not affect the anomalous dimension of $\mathcal G_g^g$, which, as showed in
 Ref.~\cite{Jain:2011xz}, is the same as that of an inclusive gluon jet.

At order $\mathcal O(\alpha_s)$, the first contributions to heavy quark fragmentation into a gluon, $\mathcal G_{Q}^g$, and gluon fragmentation into heavy quark, $\mathcal G_{g}^Q$, arise. 
$\mathcal G_{Q}^g$ receives contributions from the real emission diagrams in Fig.~\ref{alpha}, when one integrates over the heavy quark phase space and fixes  the gluon momentum fraction to $z$. The lowest order diagram contributing to $\mathcal G_{g}^{Q}$ is showed in Fig.~\ref{gQQalpha}.
The diagrams are UV and IR finite, and, in the massless limit, they give
\begin{eqnarray}
\frac{\mathcal G^{g (1)}_Q(\omega r,z,m_Q^2, \mu^2)}{2 (2\pi)^3} &=& 
C_F \Bigg\{ \frac{(1-z)^2 + 1}{z} \left( \left[\frac{\theta(\omega r)}{\omega r}\right]_+^{(\mu^2)} + \delta(\omega r)   \log \left( \frac{\mu^2 }{m^2_Q }\right) \right)  \nonumber \\  & &   + \delta(\omega r) \left( - \frac{(1-z)^2 + 1}{z}\left(  \log(z) -\log(1-z)  + 1 \right)   + z  \right)  \Bigg\} \label{qFJF.3}\\
\frac{\mathcal G^{Q (1)}_g(\omega r,z,m_Q^2, \mu^2)}{2 (2\pi)^3} &=& 
T_R \Bigg\{ \left((1-z)^2 + z^2 \right) \left(\left[\frac{\theta(\omega r)}{\omega r}\right]_+^{(\mu^2)} + \delta(\omega r) \log \left(\frac{\mu^2}{m^2_Q}\right) \right)  \nonumber \\ & &  + \delta(\omega r) \left(  (z^2 + (1-z)^2 ) \log (z(1-z)) +  2 z (1-z) \right) \Bigg\}.   \label{gFJF.2}  
\end{eqnarray}
Notice that  Eqs.~\eqref{qFJF.3} and \eqref{gFJF.2} are independent of $\mu$.
The evaluation of the anomalous dimension of $\mathcal G_Q^g$ and $\mathcal G_g^Q$ requires the calculation of UV poles at $\mathcal O(\alpha_s^2)$. We explicitly verify in Section \ref{GluonAlpha2} that $\mathcal G_g^Q$ has, as expected, the same RGE as an inclusive gluon jet.
FJFs for heavy antiquarks, $\mathcal G_{\bar Q}^{\bar Q}$, $\mathcal G_{\bar Q}^g$ and $\mathcal G_{g}^{\bar Q}$ have the same expressions as the quark FJFs in Eqs.~\eqref{qFJF.2}, \eqref{qFJF.3} and \eqref{gFJF.2}.
FJFs of a heavy antiquark $\bar Q$, or of a light quark (or antiquark) $l$ into a heavy quark $Q$ vanish at $\mathcal O(\alpha_s)$.

The only dependence on the quark mass in Eqs.~\eqref{qFJF.2}, \eqref{qFJF.3} and \eqref{gFJF.2} comes in front of one loop  splitting functions, and it is matched exactly by the quark and gluon fragmentation functions in Eqs.~\eqref{qFF.2}--\eqref{gFF.2}. 
Below the jet scale, we can therefore match the FJFs onto heavy quark fragmentation functions. 
At tree level, Eq.~\eqref{FJFMatching} implies $\mathcal J^{(0)}_{Q Q} = \mathcal G_Q^{Q (0)}$ and  $\mathcal J^{(0)}_{g g} = \mathcal G_g^{g (0)}$,
while, taking $H$ to be  either a heavy quark or a gluon, and expanding in $\alpha_s$ as in Eqs.~\eqref{alphaexpD} and \eqref{alphaexpJ}, the one loop matching condition reads
\begin{eqnarray}
\mathcal J^{(1)}_{i j} \left(\omega r, z,\mu^2\right) &=& \mathcal G_i^{j (1)} (\omega r, z, m_Q^2, \mu^2) - \delta(\omega r) D_i^{j (1)}\left(z, \frac{\mu^2}{m_Q^2} \right)\, .
\end{eqnarray}
At one loop, the quark matching coefficients $\mathcal J_{Q Q}$ are 
\begin{eqnarray}
\frac{ \mathcal J^{(0)}_{Q Q}(\omega r ,z,\mu^2 ) }{2 (2\pi)^3 } &=& \delta(\omega r)\delta(1-z), \nonumber \\
\frac{ \mathcal J^{(1)}_{Q Q}(\omega r,z,\mu^2 )  }{2 (2\pi)^3 } &=& C_F  
\left\{ \delta(1-z)  \left(   2    \left[\frac{\theta(\omega r) \log (\omega r/\mu^2)}{\omega r}\right]_+^{(\mu^2)}  - \frac{3}{2}  \left[\frac{\theta(\omega r)}{\omega r}\right]_{+}^{(\mu^2)} \right)
 \right. \nonumber  \\ & & \left.  
 + \left[\frac{1+z^2}{1-z}\right]_+  \left[\frac{\theta(\omega r)}{\omega r}\right]_+^{(\mu^2)} 
 +  \delta(\omega r) \left(  \delta(1-z)\left(   -\frac{\pi^2}{6} + \frac{7}{4}\right )  \right.  \right. \nonumber
 \\ & & \left. \left.
+ \left[\frac{1+z^2}{1-z}\right]_+  \log ( z) 
 +  \left[\frac{1+z^2}{1-z} \log(1-z) \right]_+ 
 + (1-z) \right) 
 \right\} \label{match.2}.
\end{eqnarray}
Eq.~\eqref{match.2} reproduces the results in Ref.~\cite{Liu:2010ng,Jain:2011xz}, as one expects, since all the IR dependence should cancel in the matching.
Similarly, we obtain 
\begin{eqnarray}
\frac{\mathcal J^{(1)}_{Q g}(\omega r,z,\mu^2)}{2 (2\pi)^3 } &=& C_F \left\{ \frac{(1-z)^2 + 1}{z} \left(\left[\frac{\theta(\omega r)}{\omega r}\right]_+^{(\mu^2)} + \delta(\omega r) \log \left(  z( 1-z)  \right) \right) + \delta(\omega r) z \right\}\, ,\nonumber\\  \label{match.3}\\
\frac{ \mathcal J^{(1)}_{g Q}(\omega r,z,\mu^2)}{2 (2\pi)^3 } &=& T_R  \Bigg\{ \left((1-z)^2 + z^2 \right) \left(\left[\frac{\theta(\omega r)}{\omega r}\right]_+^{(\mu^2)} + \delta(\omega r) \log \left( z(1-z)  \right)\right)  \nonumber \\ & & + \delta(\omega r) 2 z (1-z) \Bigg\}\, , \label{match.4}
\end{eqnarray}
which  agree with Ref.~\cite{Jain:2011xz}.

The gluon matching coefficient $\mathcal J_{gg}$ is also unaffected by $m_Q$, since the correction to gluon propagator cancels between the gluon FJF and fragmentation function. For completeness, and since it is needed in the $\mathcal O(\alpha_s^2)$ calculation, we report the result of Ref.~\cite{Jain:2011xz}.
\begin{eqnarray}
\frac{ \mathcal J^{(0)}_{gg}(\omega r, z,\mu^2)}{2 (2\pi)^3 } & = & \delta(\omega r) \delta(1-z) \nonumber\\
\frac{ \mathcal J^{(1)}_{gg}(\omega r, z,\mu^2)}{2 (2\pi)^3 } & = &  C_A \left\{  2 \delta(1-z)  \left[\frac{ \theta(\omega r)\log(\omega r/\mu^2)}{\omega r}\right]_+^{(\mu^2)} + p_{gg}(z)  \left[\frac{\theta(\omega r)}{\omega r}\right]_+^{(\mu^2)}   \right. \nonumber \\  
& & \left. + \delta(\omega r) \left(  \left[ \frac{\log(1-z)}{1-z} \right]_+ \frac{2 (1-z + z^2)^2}{z} + p_{gg}(z) \log z - \delta(1-z) \frac{\pi^2}{6} \right) \right\},\nonumber \\ 
\label{match.g2}
\end{eqnarray}
with $p_{gg}(z)$ given by
\begin{equation}
p_{gg}(z) = 2 z \left[\frac{1}{1-z}\right]_+ + 2 (1-z) \frac{1+z^2}{z}.
\end{equation}

The explicit expressions for the FJFs,  Eqs.~\eqref{qFJF.2}--\eqref{gFJF.2},
allow to check the sum rules \eqref{sr1} and \eqref{sr2}.
At the perturbative level, 
the sum over a complete set of states in Eq.~\eqref{sr1} is  a sum over partons, $H \in \{Q, \bar Q, g, l, \bar l\}$. Integrating the fixed order results  \eqref{qFJF.2}, \eqref{qFJF.3} over $z$ one finds that 
\begin{eqnarray}
 J_Q (\omega r,\mu^2)& =  &   \frac{1}{2 (2\pi)^3}\int_0^1 \df z z  \left( \mathcal G^Q_Q (\omega r, z,m_Q^2,\mu^2)   +  
\mathcal G^g_Q (\omega r, z, m_Q^2, \mu^2)   \right) \nonumber \\
& = &    \frac{1}{2 (2\pi)^3} \int_0^1 \df z  \left( \mathcal G^Q_Q (\omega r, z, m_Q^2, \mu^2)  - \mathcal G^{\bar Q}_Q (\omega r, z,m_Q^2,\mu^2) \right) 
    \nonumber \\  & =  &   
 \delta(\omega r)   + \frac{\alpha_s C_F}{2\pi} \left\{ 
2   \left[\frac{\theta(\omega r) \log(\omega r/\mu^2)}{\omega r} \right]_+^{(\mu^2)} 
-  \frac{3}{2}   \left[\frac{\theta(\omega r)}{\omega r}\right]_{+}^{(\mu^2)}  
  +
 \delta(\omega r) \left(  \frac{7}{2} -\frac{\pi^2}{2}  \right)  
 \right\} , \nonumber\\
\end{eqnarray}
that agrees with the massless quark inclusive jet function at one loop, given in Ref.~\cite{Bauer:2003pi}.
In Appendix \ref{AppC} we prove the analogous relations in the regime $\omega r \sim m_Q^2$.

The momentum and flavor conservation sum rules are not affected by the evolution of the  fragmentation functions. Using the momentum and flavor sum rules for the fragmentation function, Eqs.~\eqref{sumD} and \eqref{sumDflav},  Eqs.~\eqref{sr1} and \eqref{sr2} can be translated into relations for the perturbative coefficients $\mathcal J_{ij}$. Specifying again to the case of heavy quark FJF,
\begin{eqnarray}
J_Q (\omega r, \mu^2 ) &  = & \sum_{j \in \{ Q, \bar Q, g, l, \bar l \}} \frac{1}{2 (2\pi)^3}   \int_0^1 \df z\,  z   \mathcal J_{Q j}(\omega r, z,\mu^2) 
, \nonumber\\
&  = & \frac{1}{2 (2\pi)^3} \int_0^1 \df z  \left( \mathcal J_{Q Q} (\omega r, z,\mu^2)  - \mathcal J_{Q \bar Q} (\omega r, z,\mu^2) \right). \label{sr3}
\end{eqnarray}
Eq.~\eqref{sr3} can be explicitly checked at one loop, using Eqs.~\eqref{match.2}--\eqref{match.4}.
In particular, after integrating over the full range of $z$, the dependence on the fragmentation function, and thus the logarithmic dependence on the quark mass, drops out. 
The definition  
 \eqref{intro.3} and the property \eqref{sr3} imply that 
the heavy quark initiated $Q$-tagged jet function  does not depend logarithmically on the quark mass, up to terms proportional to the minimum $B$ meson momentum fraction $z_0$. 
If details of the $B$ meson inside the $Q$-tagged jet are not observed, then, factorization theorems can be  expressed in terms of inclusive jet functions, 
massless if $Q \tau_N \gg m^2_Q$, or massive in the case $Q \tau_N \sim m^2_Q$.

The dependence on the heavy quark mass cancels in the inclusive gluon jet function.  
The combination
\begin{equation}
\int \df z z \left(\mathcal G_g^{Q} (\omega r,z,m_Q^2,\mu^2) + \mathcal G_{g}^{\bar Q}(\omega r, z,m_Q^2,\mu^2) + \mathcal G_g^{g} (\omega r, z,m_Q^2,\mu^2)
+ 2 n_l \mathcal G_g^{l} (\omega r, z, \mu^2)
 \right)
\end{equation}
is indeed equal to  the inclusive gluon jet function, and mass independent, up to power corrections.
However, if one insists on tagging the heavy quark, she is left with some mass dependence. 
Retaining terms of $\mathcal O(\alpha_s)$ in the matching coefficients $\mathcal J_{i j}$, the $Q$-tagged jet function is
\begin{equation}\label{resummed}
  J_g^Q   (\omega r,m_Q^2,\mu^2)    =  
   \int_{z_0}^1 \df z  \int_z^1 \frac{\df \xi}{\xi} \left( 
\mathcal J_{g g}(\omega r, \xi,\mu^2) D_{g}^Q \left(\frac{z}{\xi},\frac{ \mu^2}{m_Q^2}\right)+
\mathcal J_{g Q}(\omega r, \xi,\mu^2) D_{Q}^Q \left(\frac{z}{\xi},\frac{ \mu^2}{m_Q^2}\right) \right).
\end{equation}
Logarithms of the ratio $m_Q^2/(Q\tau_N)$ are resummed at  NLL accuracy
by solving the DGLAP equation, with two-loop splitting functions and one loop initial conditions for $D_g^Q$ and $D_{Q}^Q$ at $\mu_0 \sim m_Q$, given in Eqs.~\eqref{gFF.2} and \eqref{qFF.2}.

Setting $z_0 = 0$ in Eq.~\eqref{resummed} would 
express $J_g^Q$ in terms of the first Mellin moment of the quark and gluon fragmentation function. However, the DGLAP equation for the first moment of $D_g^Q$ is not well defined, since the Mellin transform of $P_{gg}$ has a pole for $N = 1$. In common approaches for the study of heavy quark multiplicity in gluon jets  \cite{Mueller:1985zz,Mangano:1992qq}, the DGLAP equation is modified at small $z$ to regulate  the singularity of $P_{gg}$, by including coherence effects. In this work, we will assume that $z_0$ is large enough that small $z$ effects can be neglected.

\section{Gluon and light quark fragmentation into heavy quarks at $\mathcal O(\alpha_s^2)$ }\label{Alpha2}

We have seen in Section \ref{OneLoop} that  gluon initiated $Q$-tagged jets 
are sensitive to the  scale of the quark mass, and that large logs of the ratio $m^2_Q/(Q\tau_N)$ can be resummed by solving the DGLAP equation for the fragmentation function.
In this section we calculate the gluon and light quark FJFs into a heavy quark at $\mathcal O(\alpha_s^2)$, both of which involve gluons splitting into $Q\bar Q$ pairs. 
In the case of the gluon FJF $\mathcal G_g^Q$, the leading order is $\mathcal O(\alpha_s)$, and the calculation of this section  amounts to the NLO contribution. The light quark FJF starts at  $\mathcal O(\alpha_s^2)$, and we give here the leading order term.

There are several reason to go beyond the lowest order in processes involving gluon splitting into heavy quark pairs. 
First of all, one can study the renormalization group properties of $\mathcal G_g^Q$. We  explicitly show that the RGE for 
$\mathcal G_g^Q$ is the same as for the gluon  inclusive jet. Furthermore, 
knowing  the fixed order expression of $\mathcal G_g^Q$   at NLO,
together with two loop cusp anomalous dimension and one loop non-cusp anomalous dimension,
allows to resum Sudakov double logarithms of $r_{\tau}$ at NNLL accuracy. Finally is interesting to explicitly check that at $\mathcal O(\alpha_s^2)$ the infrared sensitivity 
is exactly reproduced by the heavy quark fragmentation functions, computed at $\mathcal O(\alpha_s^2)$ in Ref.~\cite{
Mitov:2004du}.

But perhaps the most important reason for this calculation is that for many  interesting SM processes involving $b$ jets fixed order calculations are available at NLO accuracy \cite{Campbell:2008hh,Caola:2011pz}. In order to match the resummed result
to these fixed order results, the knowledge of the gluon  and  the light quark FJFs into heavy quarks at $\mathcal O(\alpha_s^2)$ is required.

\subsection{Gluon fragmentation into  heavy quarks at $\mathcal O(\alpha_s^2)$ }\label{GluonAlpha2}

\begin{figure}
\center
\includegraphics[width=16cm]{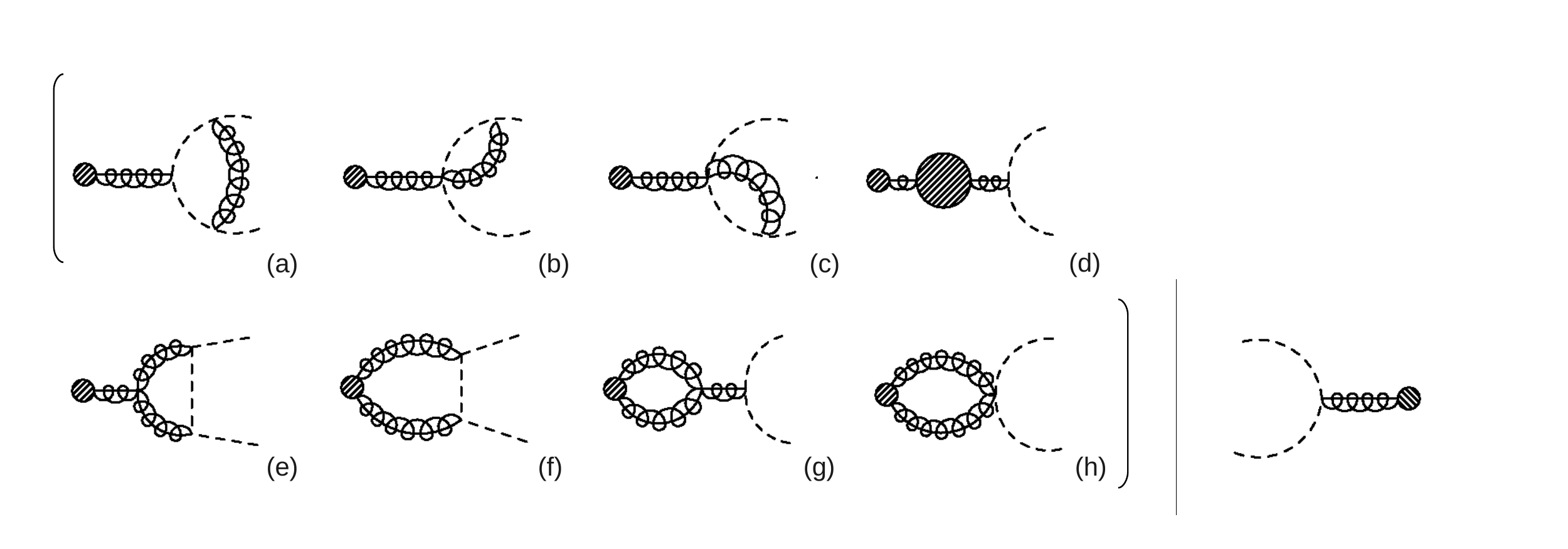}
\caption{Virtual diagrams contributing to $\mathcal G_g^Q$ at $\mathcal O(\alpha_s^2)$. The shaded circle in diagram (d) denotes one loop corrections to the gluon propagator, shown in Fig.~\ref{ggalpha}.}\label{FigV}
\end{figure}

\begin{figure}
\center
\includegraphics[width=16cm]{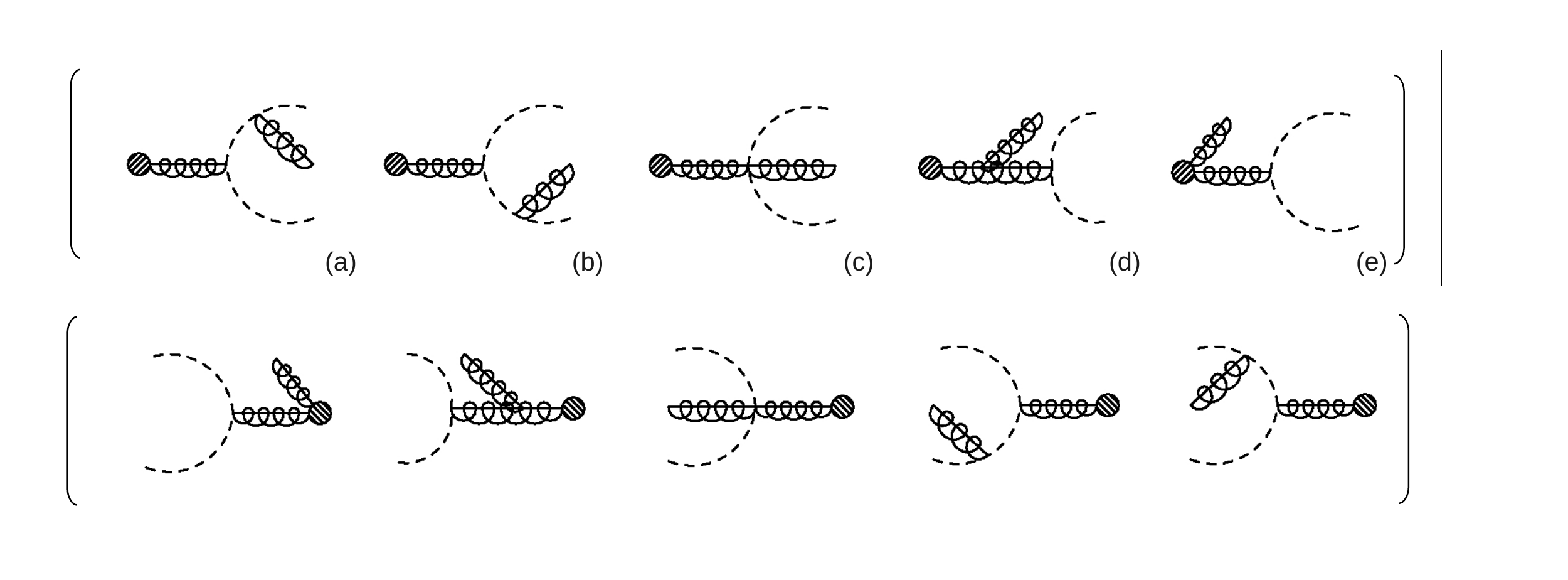}
\caption{Real emission diagrams contributing to $\mathcal G_g^Q$ at $\mathcal O(\alpha_s^2)$.}\label{FigR}
\end{figure}

The virtual and real diagrams contributing to $\mathcal G_g^Q$ at $\mathcal O(\alpha_s^2)$ are shown, respectively, in Figs.~\ref{FigV} and \ref{FigR}.
We decompose $\mathcal G_g^Q$ in terms of color factors
\begin{equation}
\mathcal G_g^{Q (2)} =  C_F T_R \, \mathcal G_g^{C_F T_R} + C_A T_R\,  \mathcal G_{g}^{C_A T_R} + T_R^2\,  \mathcal G_{g}^{T^2_R} + T_R^2 n_l\, \mathcal G_{g}^{T^2_R n_l} , 
\end{equation}
where $n_l = n_f - 1$ is the number of light quarks. 
At $\mathcal O(\alpha_s^2)$, the matching condition Eq.~\eqref{FJFMatching} reads  
\begin{eqnarray}\label{rematch}
\mathcal G_g^{Q (2)}(\omega r, z, m_Q^2, \mu^2) &=&  
  \int \frac{\df \xi }{\xi}  \left(  \mathcal J^{(1)}_{g  Q} (\omega r, \xi, \mu^2) D^{Q (1)}_{Q} \left(\frac{z}{\xi}, \frac{\mu^2}{m_Q^2}\right) + \mathcal J^{(1)}_{g  g}(\omega r,\xi, \mu^2)  \right. \nonumber \\ & & \left. 
 \times   D^{Q (1)}_{g} \left(\frac{z}{\xi}, \frac{\mu^2}{m_Q^2}\right) 
 \right) + \mathcal J^{(2)}_{g  Q} (\omega r, z, \mu^2) + \delta (\omega r)\, D^{Q (2)}_{g} \left(z, \frac{\mu^2}{m_Q^2}\right) \,,  \nonumber\\ 
\end{eqnarray}
where we used that at tree level $D_Q^{Q (0)}$ and $\mathcal J_{g g}^{(0)}$ are delta functions.
The matching coefficients $\mathcal J^{(2)}_{g Q }$ have an identical color decomposition as $\mathcal G_g^{Q (2)}$.
The one loop fragmentation functions and matching coefficients are given in Eqs.~\eqref{qFF.2}--\eqref{gFF.2} and Eqs.~\eqref{match.4} and \eqref{match.g2}, respectively.
The gluon fragmentation function  at  $\mathcal O(\alpha_s^2)$, $D_g^{Q (2)}$, has been computed in Ref.~\cite{Mitov:2004du}.
Therefore the calculation of $\mathcal G_g^Q$ allows for the extraction of the  matching coefficient $\mathcal J_{g Q}$ at $\mathcal O(\alpha_s^2)$. 
We stress that the matching  is meaningful if the coefficients are  independent of $m_Q$, which is an infrared scale in the problem, and, thus,  all the singular mass dependence of $\mathcal G_g^Q$ needs to be reproduced by  the fragmentation functions. We will see that this is indeed the case.

We use dimensional regularization to regulate UV and IR divergences and work in Feynman gauge.
We compute the diagrams in Figs.~\ref{FigV} and \ref{FigR} analytically, for finite value of $m_Q$, and take the massless limit, $m_Q^2 \ll \omega r$, at the end of the calculation.
This limit has to be taken carefully, and we refer the reader to Appendix \ref{AppA} for details. 
We find that  $\mathcal G_{g}^Q$ 
depends on $\omega r$ through plus distributions
$\left[\log^n(\omega r)/(\omega r) \right]_+$, defined  in Eq.~\eqref{dist.2}, with $n\leq2$. 
As discussed in more detail in Appendix \ref{AppA}, the coefficients of the plus distributions and the logarithms of the mass in the $\delta(\omega r)$ piece are determined by the ``naive''
massless limit of $\mathcal G_{g}^Q$. The mass independent component of $\delta(\omega r)$, on the other hand, requires to integrate the result for $\mathcal G_g^{Q}$, obtained at fixed $m_Q$, 
from the minimum invariant mass required for the production of a $Q \bar Q$ pair,  $\omega r = m_Q^2/(z(1-z))$, to $\infty$. 
Most of the integrals can be performed analytically, and expressed in terms of polylogarithms up to rank three. A few integrals from the real emission diagrams 
with color structure $C_A T_R$ had to be solved numerically.

$\mathcal G_{g}^{C_F T_R}$ receives contributions from the virtual diagrams $(a)$, $(b)$ and $(c)$ in Fig.~\ref{FigV}, and from the square of the real diagrams $(a)$, $(b)$ and $(c)$ in Fig.~\ref{FigR}. The ultraviolet divergences in these diagrams are canceled by charge and mass renormalization, while infrared divergences cancel between the virtual and real emission diagrams. The color structures $T_R^2$ and $T^2_R n_l$ receive contributions from heavy and light quark loop corrections to the gluon propagator,  diagram $(d)$ in Fig.~\ref{FigV}. The diagrams are IR finite, and the UV divergences are renormalized by charge renormalization.

The situation is more interesting for $\mathcal G_g^{C_A T_R}$. Even after charge renormalization, the result is still divergent and are rendered finite only by an operator renormalization.
Defining the jet renormalization $Z_{J_g}$, in the same way as in Eq.~\eqref{JetRenQ}, one finds
\begin{eqnarray} \label{ZJg}
Z_{J_g}(\omega r, \mu^2) &=& \delta(\omega r ) +   \frac{\alpha_s }{4\pi}  (4 C_A )\left\{  \delta(\omega r ) \left(-  \frac{1}{\varepsilon^2} - \frac{1}{\varepsilon}  \left( \log \mu^2  + \frac{1}{2} \right)\right) 
+ \frac{1}{\varepsilon} \left[\frac{\theta(\omega r)}{\omega r  }\right]_+^{}
\right\}.  \nonumber\\
\end{eqnarray}
This leads to the RGE for $\mathcal G_g^Q$
\begin{equation}
\frac{\df}{\df \log \mu} \mathcal G_g^{Q}(\omega r, z,m_Q^2, \mu^2) =  \int \df\, (\omega s ) \, \gamma_{J_g} \left(\omega r- \omega s,\mu^2\right) \mathcal G_g^{Q}(\omega s, z,m_Q^2, \mu^2),
\end{equation}
with anomalous dimension
\begin{eqnarray}\label{gammajetg}
\gamma_{J_g} (\omega r,\mu^2) & &= \int \df (\omega s) Z_3 Z_{J_g}^{-1}\left( \omega r - \omega s,\mu^2 \right) \frac{d}{d\log \mu} Z_3^{-1} Z_{J_g}\left( \omega s,\mu^2\right)  \nonumber \\
& & = \frac{\alpha_s}{4\pi} \left\{ - 2 (4 C_A) \left( \left[\frac{\theta(\omega r)}{\omega r}\right]_+  - \log(\mu^2) \delta(\omega r)\right) + 2 \beta_0 \delta(\omega r) \right\}.
\end{eqnarray}
 $Z_3$ is the gluon field strength renormalization in the $\overline{\textrm{MS}}$ scheme, which at one loop and in  Feynman gauge is given by
 \begin{equation} 
 Z_3 = 1 - \frac{\alpha_s}{4\pi \varepsilon} \left(2 C_A - \beta_0 \right).
 \end{equation}
 Eq.~\eqref{gammajetg} is of the form \eqref{gammajet}, and it is the same anomalous dimension that governs the running of an inclusive gluon jet function.
As in the case of the quark FJF, the anomalous dimension of $\mathcal G_g^Q$ is mass and momentum fraction independent.

We now give the  expression of the renormalized gluon FJF $\mathcal G_g^{Q (2)}$, in the massless limit.
We work in the pole mass scheme, and use as mass counterterm
\begin{equation}
\delta m_Q = - m_Q \frac{\alpha_s C_F}{4\pi} \left( \frac{3}{\varepsilon} + 3 \log \frac{\mu^2}{m_Q^2} + 4\right),
\end{equation}
which subtracts 
the entire one loop correction to the quark mass.

The color structure $\mathcal G_g^{C_F T_R }$ is given by
\begin{eqnarray}\label{CFTR}
& &\frac{\mathcal G_g^{C_F T_R}(\omega r, z,m_Q^2, \mu^2)}{2 (2\pi)^3} = 
 \nonumber \\ & &  \left( \left[\frac{\theta(\omega r)\log(\omega r /\mu^2)  }{\omega r }\right]_+^{(\mu^2)}
+ \left[\frac{\theta(\omega r)  }{\omega r }\right]_+^{(\mu^2)} \log\left(\frac{\mu^2 }{m_Q^2}\right)
  + \frac{1}{2} \log^2\left(\frac{\mu^2}{m_Q^2}\right) \delta (\omega r) \right ) \nonumber  \\ & & \times \left\{  2 (z^2 + (1-z)^2) \log (1-z) 
- ( 1-2 z + 4 z^2 )  \log(z) -\frac{1 - 4 z}{2} \right\} \nonumber \\ 
& & - \left( \left[\frac{\theta(\omega r)}{\omega r} \right]_+^{(\mu^2)}  + \log\left(\frac{\mu^2}{m_Q^2}\right) \delta(\omega r)\right) \bigg\{ 
 \frac{4 + z}{2} - \frac{\pi^2}{6} (1 -2 z + 4 z^2) + 
(1-4z^2) \log(z)   \nonumber\\ & &   + \frac{3-4 z + 8 z^2}{2} \log(1-z)+  (1-2z + 4 z^2) \log^2(z) 
 + (z^2 + (1-z)^2) \log^2(1-z )  \nonumber\\ & &  
 - 4 \left(z^2 + (1-z)^2\right) \log(1-z)\, \log(z) - (3 - 6 z + 4 z^2) \Li_2(z)  \bigg\} + \delta (\omega r)  g^{C_F T_R}(z).
 \label{GCFmassless}
\end{eqnarray}

\begin{figure}
\includegraphics[width= 8cm]{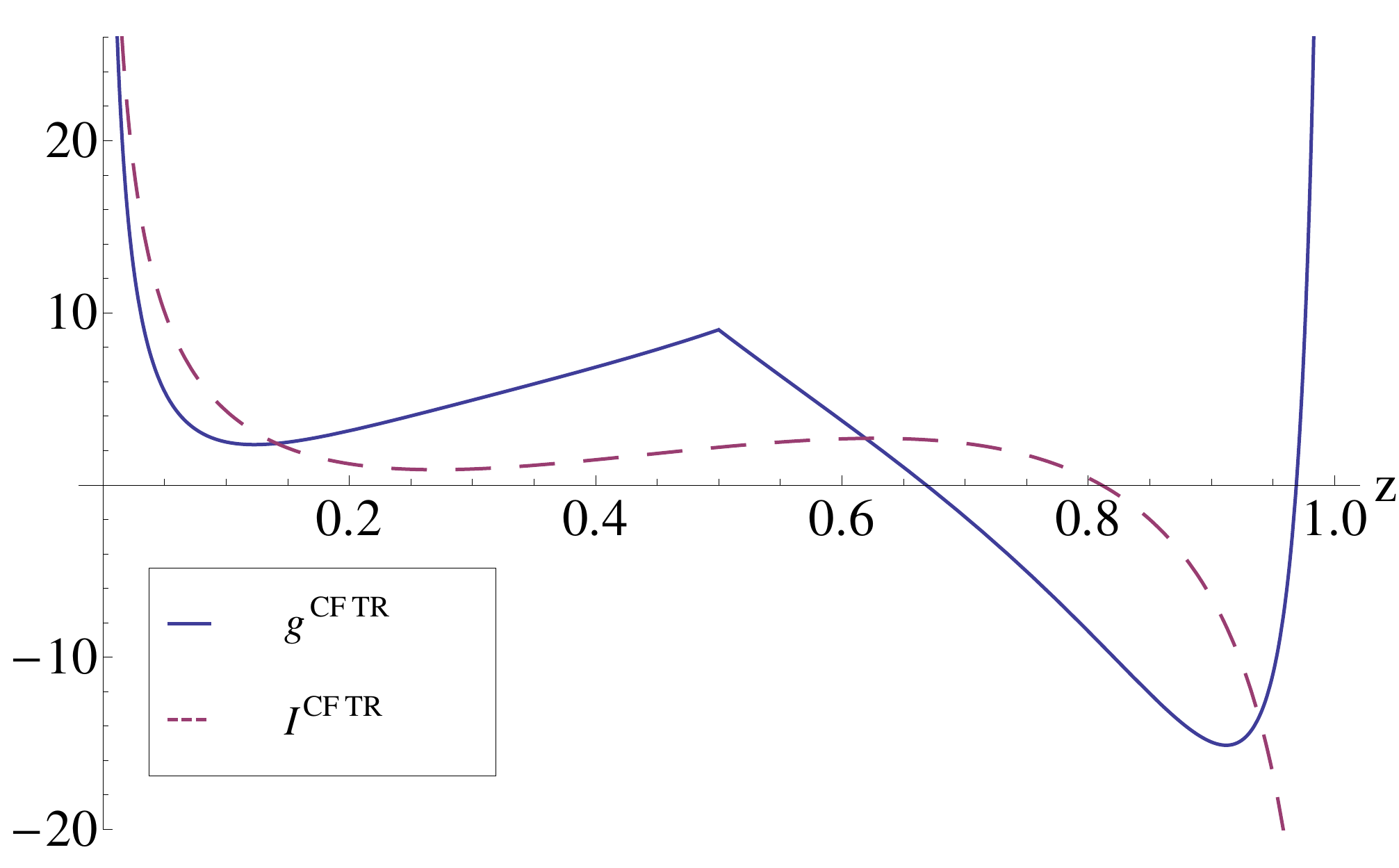}
\includegraphics[width= 8cm]{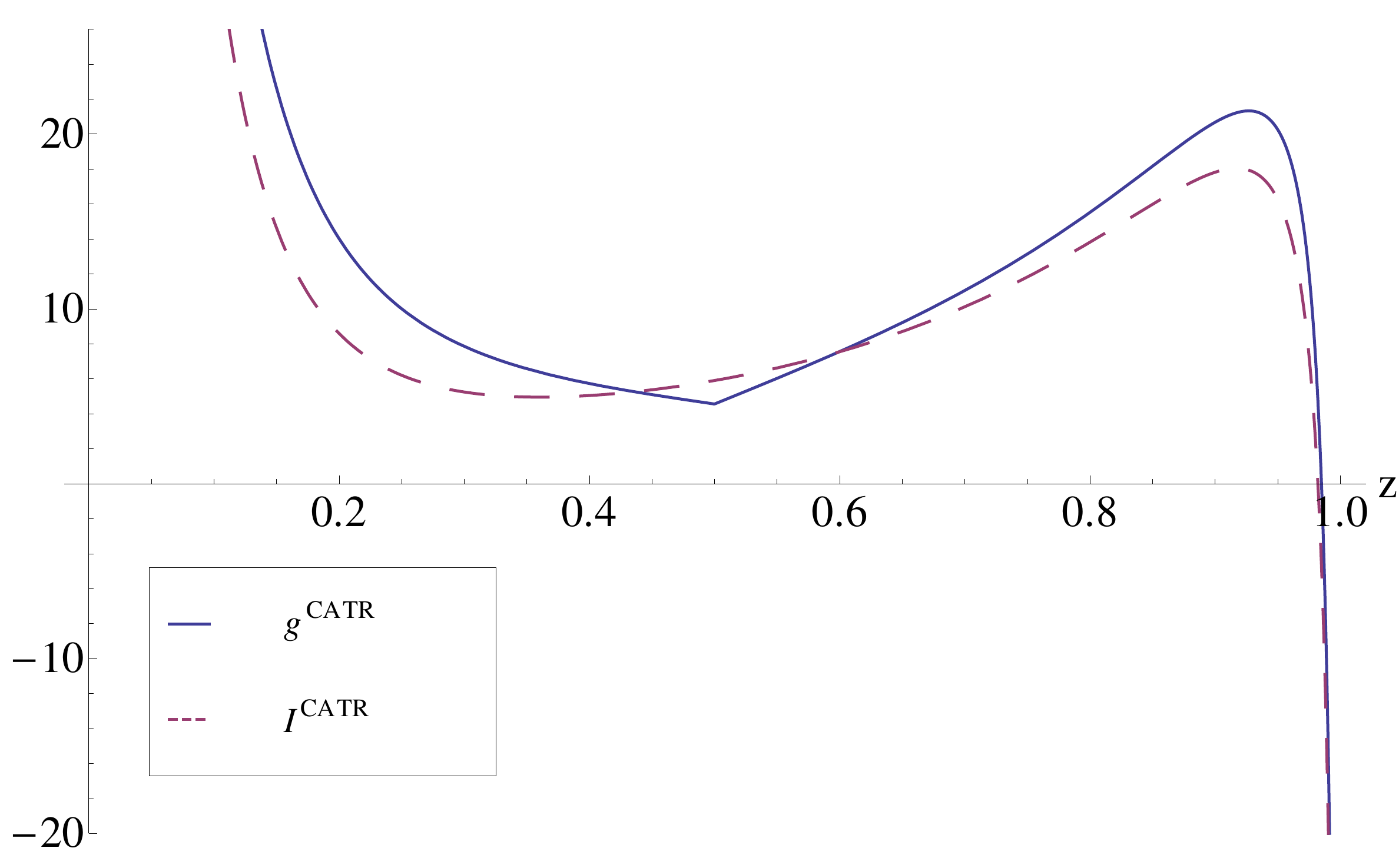}
\caption{\textit{Left Panel}: functions    $g^{C_F T_R}(z)$ (solid blue) and $\mathcal I^{C_F T_R}(z)$ (dashed magenta), in Eqs.~\eqref{gCFTR} and \eqref{ICFTR1}. 
\textit{Right Panel}: functions $g^{C_A T_R}(z)$ (solid blue) and $\mathcal I^{C_A T_R}(z)$ (dashed magenta), entering the FJF $\mathcal G_g^{C_A T_R}$
and the matching coefficient $\mathcal J_{g Q}^{C_A T_R}$, in Eqs.~\eqref{CATR} and \eqref{ICATR}
  } \label{MatchingCF}
\end{figure}

The massless limit of the  color structure $C_A T_R$ is given by
\begin{eqnarray}\label{CATR}
& & \frac{ \mathcal G_g^{C_A T_R}(\omega r,z, m_Q^2, \mu^2) }{2 (2\pi)^3} = 3 \left(z^2 + (1-z)^2 \right)  \left[ \frac{\theta(\omega r) \log^2 (\omega r/\mu^2)}{\omega r} \right]_+^{(\mu^2)}   \nonumber \\ & & +
   \left[  \frac{\theta(\omega r) \log (\omega r /\mu^2)}{\omega r} \right]_+^{(\mu^2)}  
\left\{ 2 \left( z^2 + (1-z)^2 \right)  \log \frac{\mu^2}{m_Q^2} +   4 (1 + z + z^2 ) \log (z)  \right. \nonumber \\ && \left. 
+ 4 ( z^2 + (1-z)^2 ) \log (1-z)  + \frac{4}{3 z} - \frac{ 8 - 58 z + 65 z^2 }{3}  \right\}  +  \left[ \frac{\theta(\omega r)}{\omega r}  \right]_+^{(\mu^2)}  \nonumber \\
& &  \times 
\bigg\{ \log \frac{\mu^2}{m_Q^2} 
 \left(  2 (1+4 z ) \log (z) + 2 (z^2 + (1-z)^2 ) \log (1-z) + \frac{4}{3 z} + 1 + 8 z -\frac{31}{3} z^2  \right)  
   \nonumber \\ && 
+ \frac{\pi^2}{6} (-3 + 14 z - 10 z^2) - \frac{7}{9 z} +  \frac{16 - 173 z +240z^2}{9} \nonumber  \\ && 
+ \left( \frac{4}{3 z} - \frac{3 - 12 z + 13 z^2}{3} \right) \log ( z )
 +  \left( \frac{4}{3 z} + \frac{ 3 + 36 z - 43  z^2}{3}  \right) \log (1-z ) 
\nonumber  \\ & &  + 2 ( z^2 + (1-z)^2 ) \log^2 (1-z) +  2 (1 + 5 z) \log^2 z  - 2 (z^2  + (1-z)^2 ) \log (1-z)\log(z)   \nonumber \\ & &  - 2 (1 + 2 z + 2 z^2 ) \log (z) \log (1+z)
 -  2 ( 1 + 2 z + 2 z^2 ) \textrm{Li}_2 (-z) -  4 (2 -  z + 3 z^2 ) \textrm{Li}_2 (z) 
   \bigg\}\nonumber \\
& &+ \delta(\omega r) \bigg\{    \left( \frac{4}{3 z} + \frac{14 + 2 z  - 9 z^2}{3}  +  2 (z^2 + (1-z)^2 ) \log ( 1-z) + 2 (1+ 4 z) \log (z) \right) \frac{1}{2} \log^2 \frac{\mu^2}{m_Q^2}  \nonumber \\ && 
+ \bigg[ - \frac{\pi^2}{6} (1 - 10 z + 6 z^2 ) - \frac{7}{9 z} + \frac{16 - 107 z + 174 z^2}{9}
+ \left( \frac{4}{3z }  + \frac{ 8 - 10 z + 9 z^2}{3}  \right)\log (z)   \nonumber \\ &&   
 + \left( \frac{4}{3 z} + \frac{7}{3} (2 + 2 z -3 z^2) \right) \log(1- z) 
 +  2 ( z^2 + (1-z)^2  ) \log^2 (1-z) + 2 (1 + 5 z) \log^2 z
  \nonumber \\ && 
-  2 \left(1 + 2 z + 2 z^2 \right) \log(z) \, \log (1+z)
-  2 (z^2  + (1-z)^2) \log(z) \, \log(1-z ) 
  \nonumber \\ &&    -  2(1 + 2 z + 2 z^2) \Li_2(-z)  + 4 ( - 2 + z - 3 z^2 )\Li_2 (z)
 \bigg] \log \frac{\mu^2}{m_Q^2}   \bigg\} + \delta(\omega r) g^{C_A T_R} (z). \label{GCAmassless}
\end{eqnarray}

The functions $g^{C_F T_R}(z)$ and
$g^{C_A T_R} (z)$ are functions of $z$ only,  independent on the mass.  
We plot them in Fig.~\ref{MatchingCF}. 
We give the analytic expression of $g^{C_F T_R}$ in Eq.~\eqref{gCFTR}.  
A few contributions to $g^{C_A T_R}$ from the interference of the real diagrams $(d)$ and $(e)$ with $(a)$, $(b),$ and $(c)$
had to be computed numerically, so that we do not have the full analytic expression of this function.
$g^{C_F T_R}(z)$ and
$g^{C_A T_R} (z)$ are not smooth at $z=1/2$, and they exhibit the same behavior as the gluon fragmentation function, $D_g^{Q (2)}$, described in Ref.~\cite{Mitov:2004du}. The non-smooth terms can be traced back to the virtual diagram \ref{FigV}(a), with color structure $C_F - C_A/2$ and are related to the production of the heavy quark pair at threshold.
In matching the FJFs onto fragmentation functions, the non-smooth terms cancel and the matching coefficients are  smooth functions of $z$.

The color structures $T_R^2 n_l$ and $T_R^2$ arise from 
light and heavy quark loop corrections to the gluon propagator, respectively.
In the massless limit, the plus distribution and the mass dependence of   $\mathcal G_g^{T_R^2 n_l}$ and $\mathcal G_g^{T_R^2}$ are the same, the only difference between the two functions is in the mass-independent part of the  coefficient of the delta function.
Thus, we can write
\begin{eqnarray}\label{TRa}
\frac{\mathcal G_g^{T_R^2 n_l}(\omega r, z,m_Q^2,\mu^2)}{2 (2\pi)^3} &=& f(\omega r,z,m_Q^2,\mu^2) + \delta(\omega r)\, g ^{T_R^2 n_l}(z),
\nonumber\\
\frac{\mathcal G_g^{T_R^2}(\omega r, z,m_Q^2,\mu^2)}{2 (2\pi)^3} &=& f(\omega r,z,m_Q^2,\mu^2) + \delta(\omega r)\, g ^{T_R^2 n_l}(z),
\end{eqnarray}
with the common function $f$ given by
\begin{eqnarray}\label{fdef}
 f(\omega r,z,m^2_Q,\mu^2) & = & - \frac{4}{9} (1 - 2 z + 2 z^2) \left\{ - 3  \left[ \frac{\theta(\omega r)\log(\omega r/\mu^2)}{\omega r}\right]_+^{(\mu^2)}
+ 5  \left[ \frac{\theta(\omega r)}{\omega r}\right]_+^{(\mu^2)}
\right\} 
\nonumber \\ & &
+ \delta(\omega r) \left\{ - \frac{2}{3} \left(z^2 + (1-z)^2\right) \log^2 \frac{\mu^2}{m_Q^2}
\right. \nonumber \\ & &+  \left.
\left(  
\frac{4}{9} \left( - 5 + 4 z (1-z) \right) - \frac{4}{3} (z^2 + (1-z)^2) \log (z (1-z))
\right)\log \frac{\mu^2}{m_Q^2} \right\}
\,. \nonumber \\
\end{eqnarray}
The mass independent functions $g(z)$ are given by
\begin{eqnarray}\label{TR2}
g^{T_R^2 n_l}(z) &=& - \frac{16}{9} z (1-z) - \frac{4}{9} (5 - 4 z (1-z) ) \log (z (1-z)) -\frac{2}{3} (z^2 + (1-z)^2) \log^2 (z (1-z))
\nonumber\\
g^{T_R^2}(z) &=& 
\frac{32}{45} \left( -9 z + 13 z^2 - 8 z^3 + 4 z^4\right) - \frac{4}{45} \left( 25 - 50 z + 20 z^2 + 40 z^3 - 80 z^4 + 32 z^5\right) \log(z) 
\nonumber \\ & & + \frac{4}{45}\left( 13 - 50 z+ 20 z^2 + 40 z^3 - 80 z^4 + 32 z^5\right) \log (1-z) 
\nonumber \\ & &
 - \frac{2}{3 } \left(  z^2 + (1-z)^2 \right) \left( \log^2(z) + \log^2 (1-z) -2 \log(z)\log(1-z) \right)
\,.
\end{eqnarray}

\begin{figure}
\includegraphics[width=8.1cm]{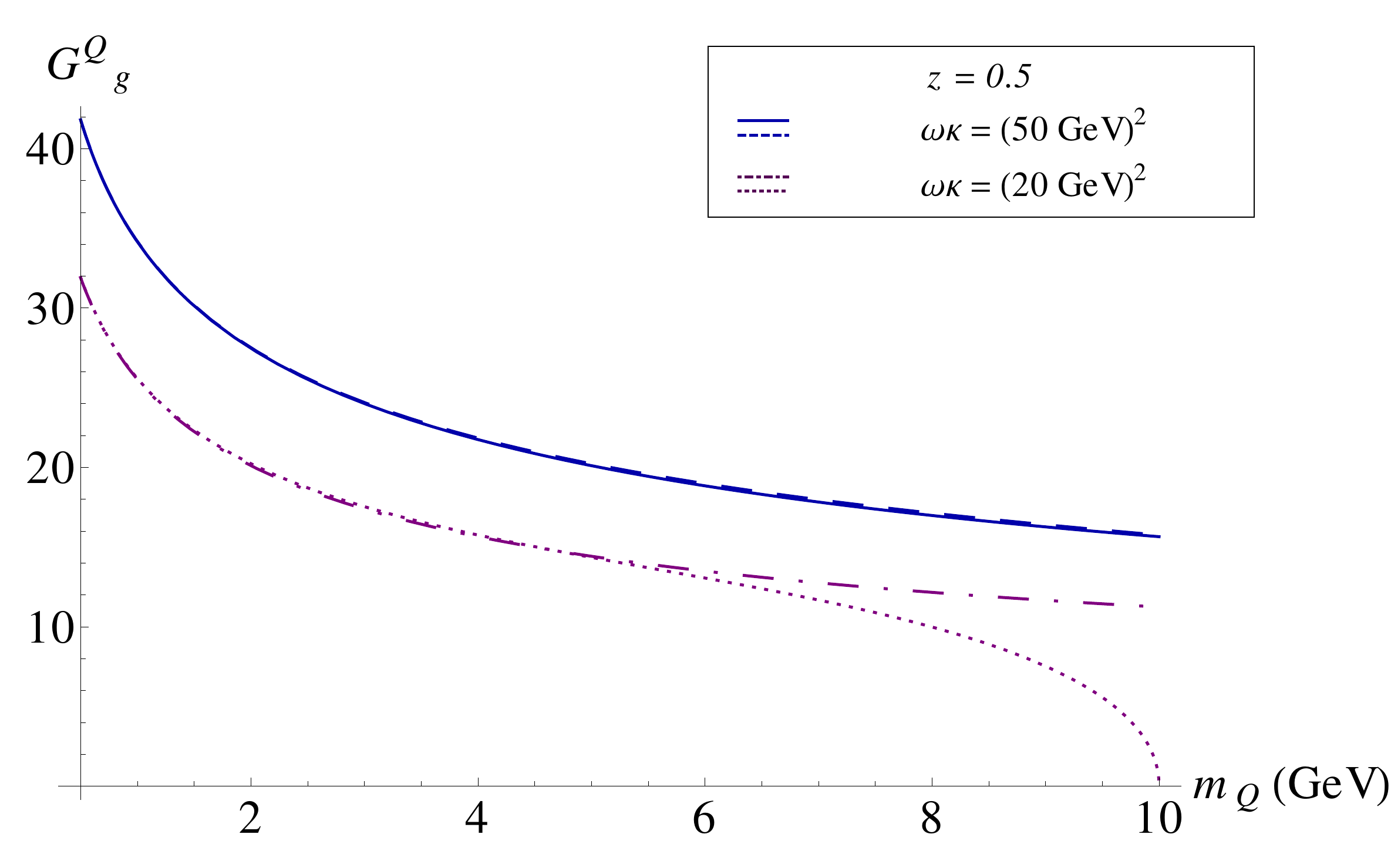}
\includegraphics[width=8.1cm]{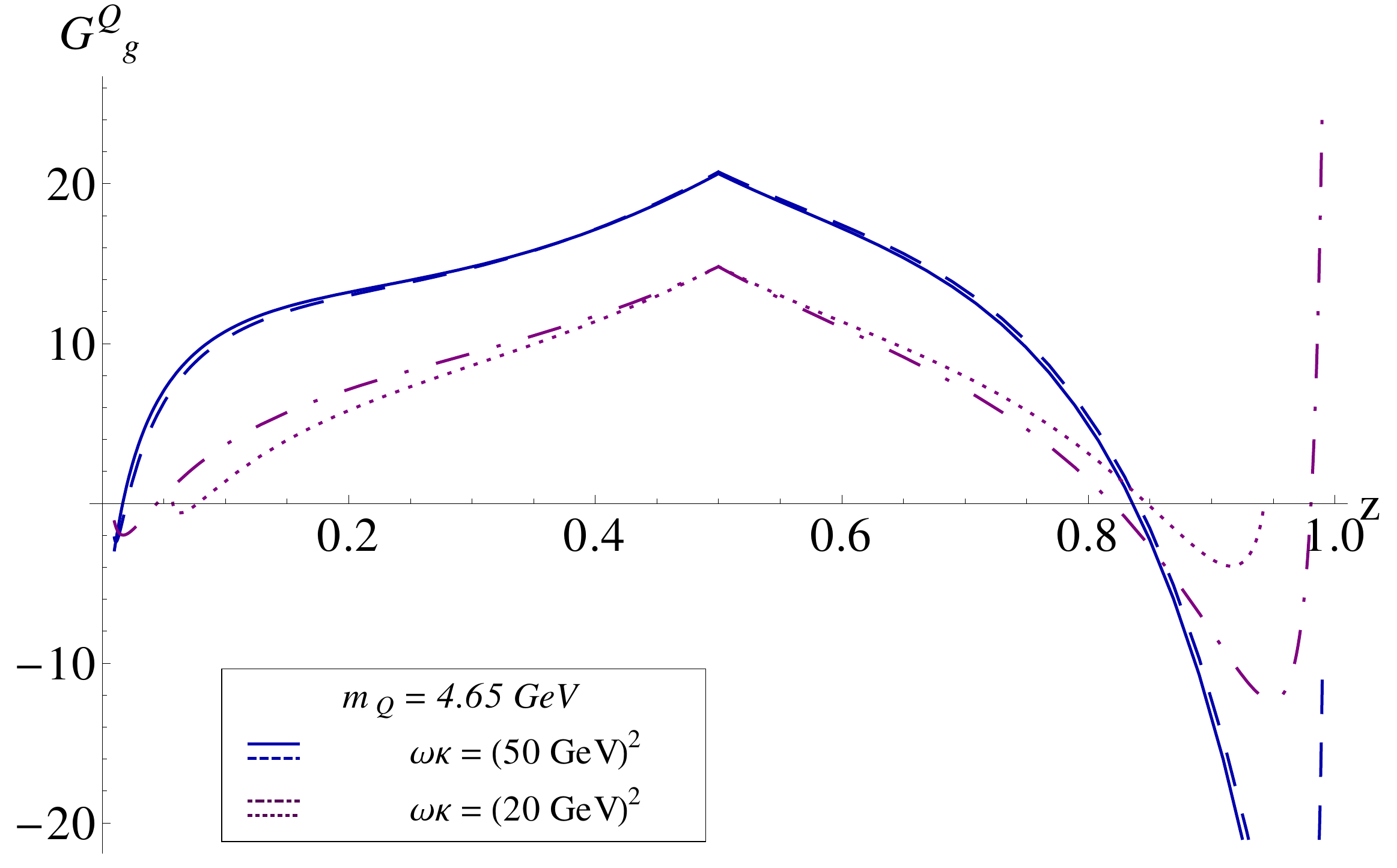}
\caption{$\mathcal G_g^{C_F T_R}$ with full mass dependence (dashed blue line and dotted magenta line) and in the massless limit (solid blue line and dot-dashed magenta line), for two values of the jet invariant mass, $\omega \kappa = ( 50 \;  \textrm{GeV})^2$  and $\omega \kappa = (20 \; \textrm{GeV})^2$ .
}\label{CompareToMassCF}
\end{figure}

\begin{figure}
\includegraphics[width=8.1cm]{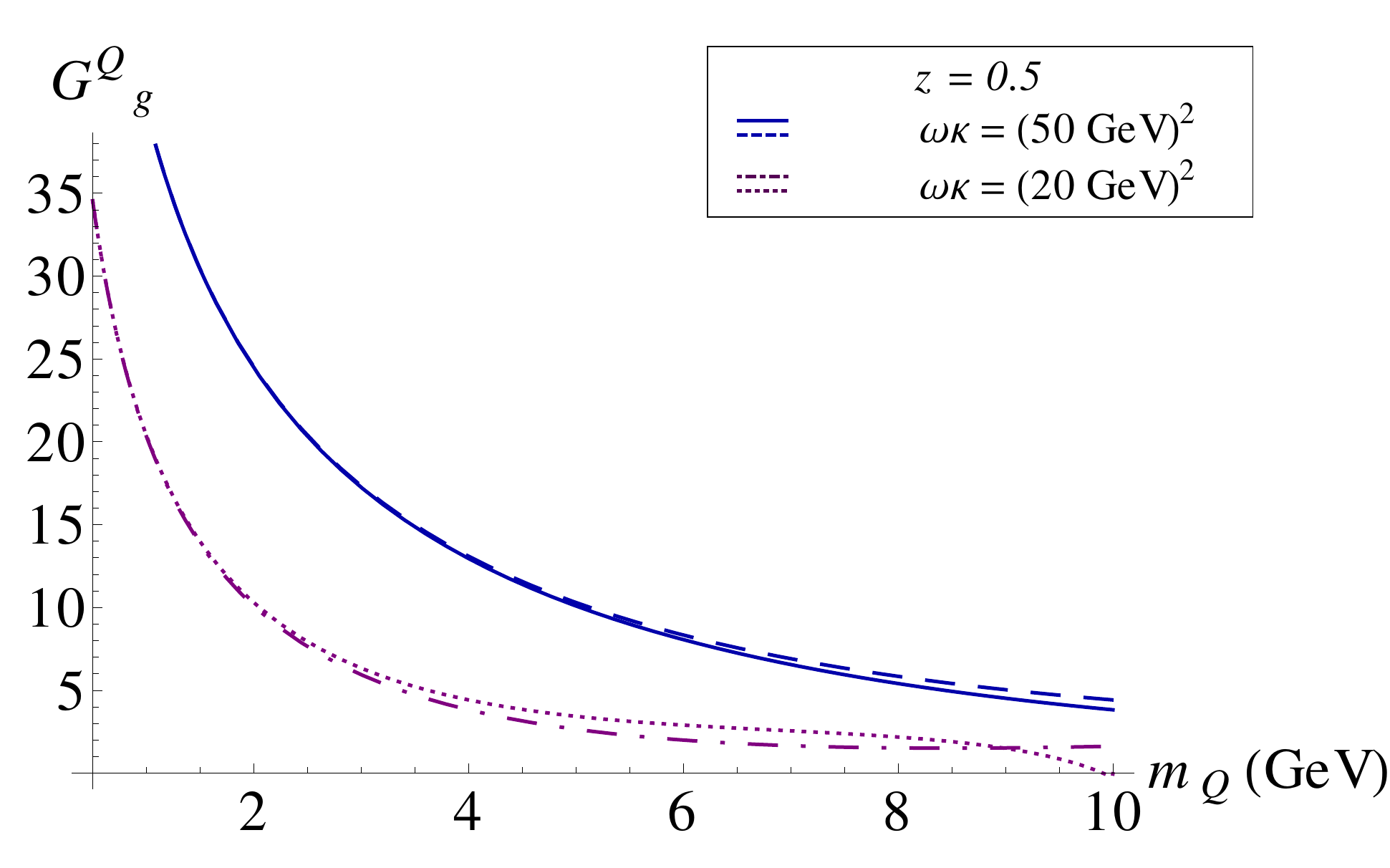}
\includegraphics[width=8.1cm]{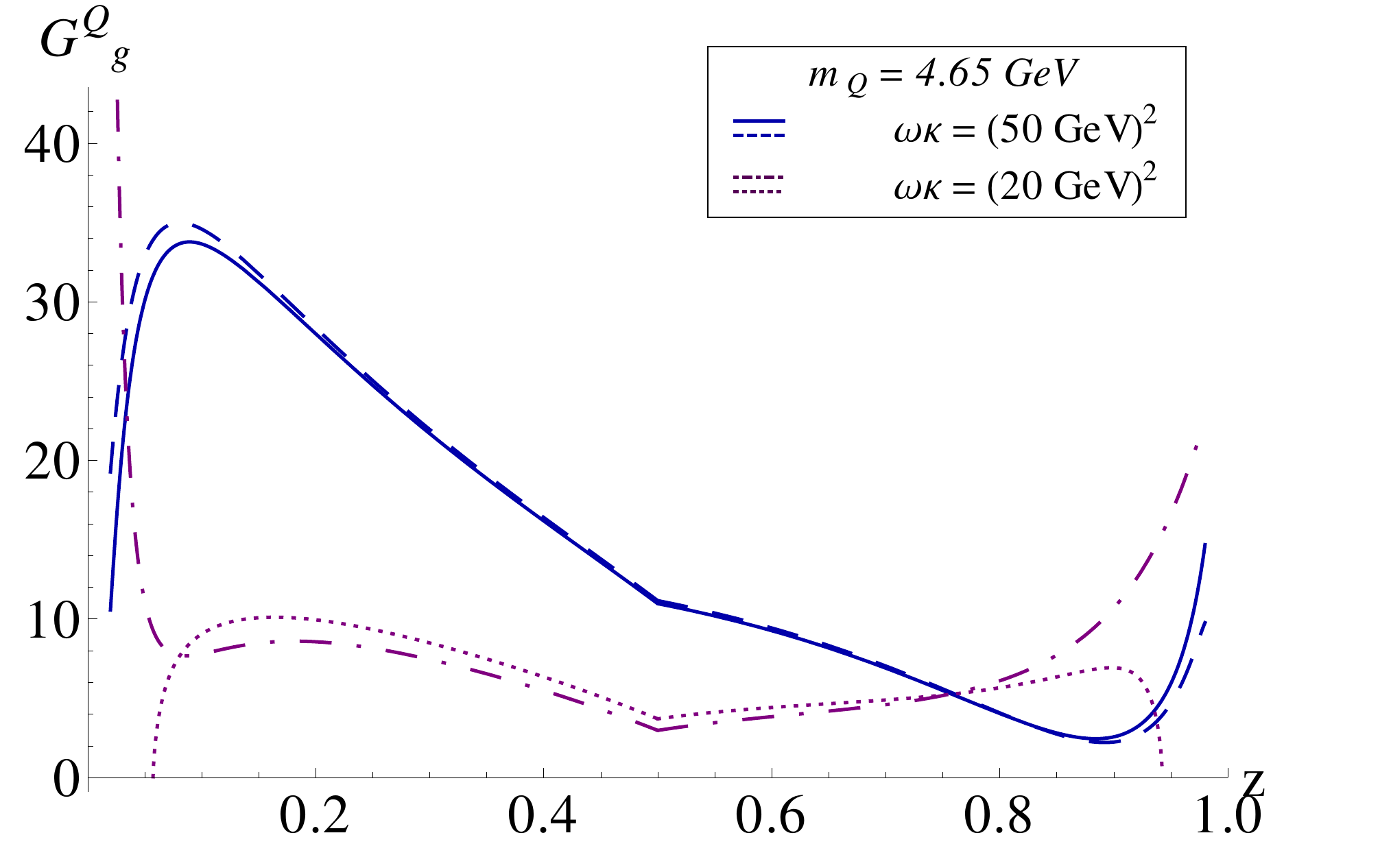}
\caption{$\mathcal G_g^{C_A T_R}$ with full mass dependence (dashed blue line and dotted magenta line) and in the massless limit (solid blue line and dot-dashed magenta line), for two values of the jet invariant mass, $\omega \kappa = ( 50 \;  \textrm{GeV})^2$  and $\omega \kappa = (20 \; \textrm{GeV})^2$ .}\label{CompareToMassCA}
\end{figure}

\begin{figure}
\includegraphics[width=8.1cm]{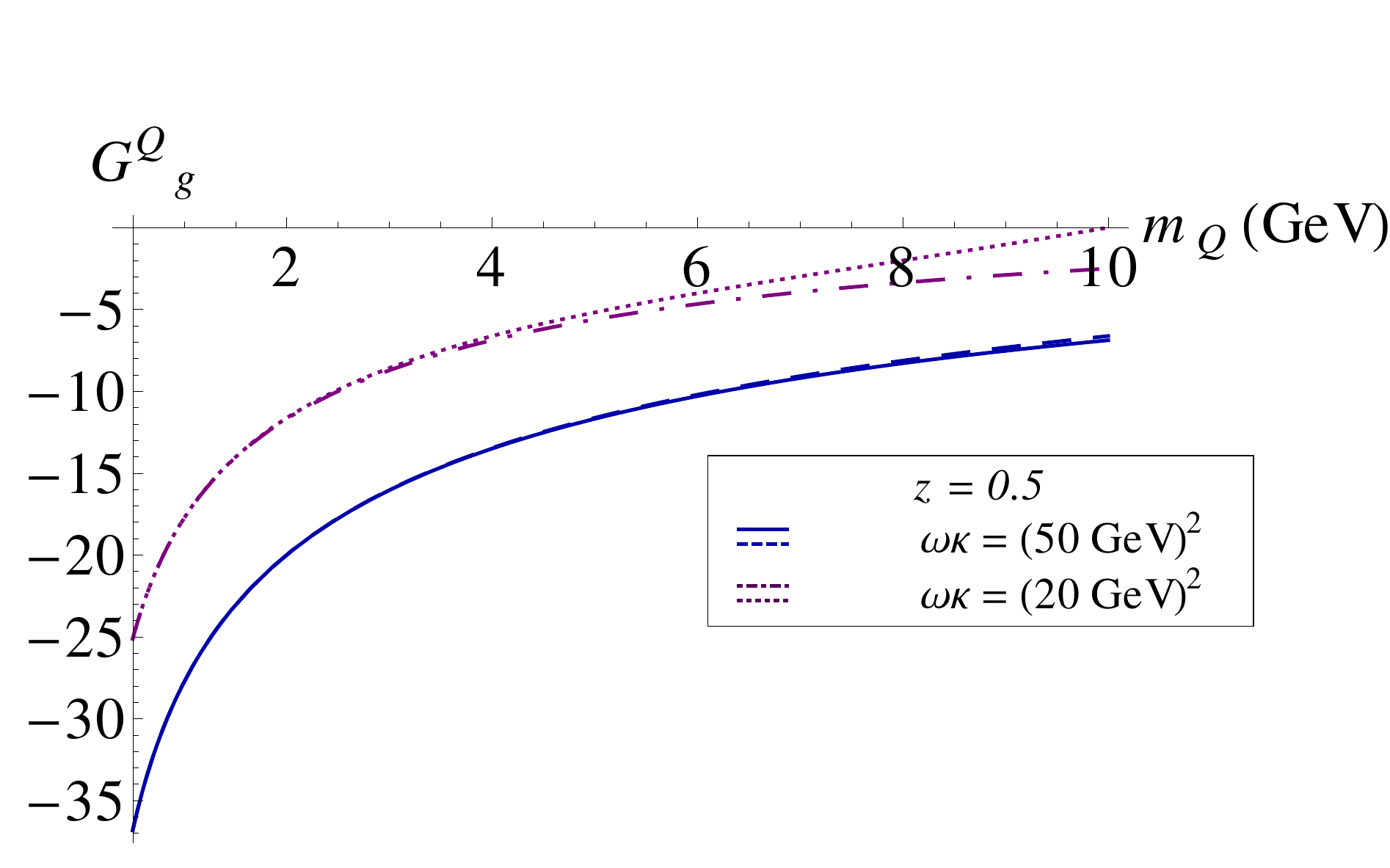}
\includegraphics[width=8.1cm]{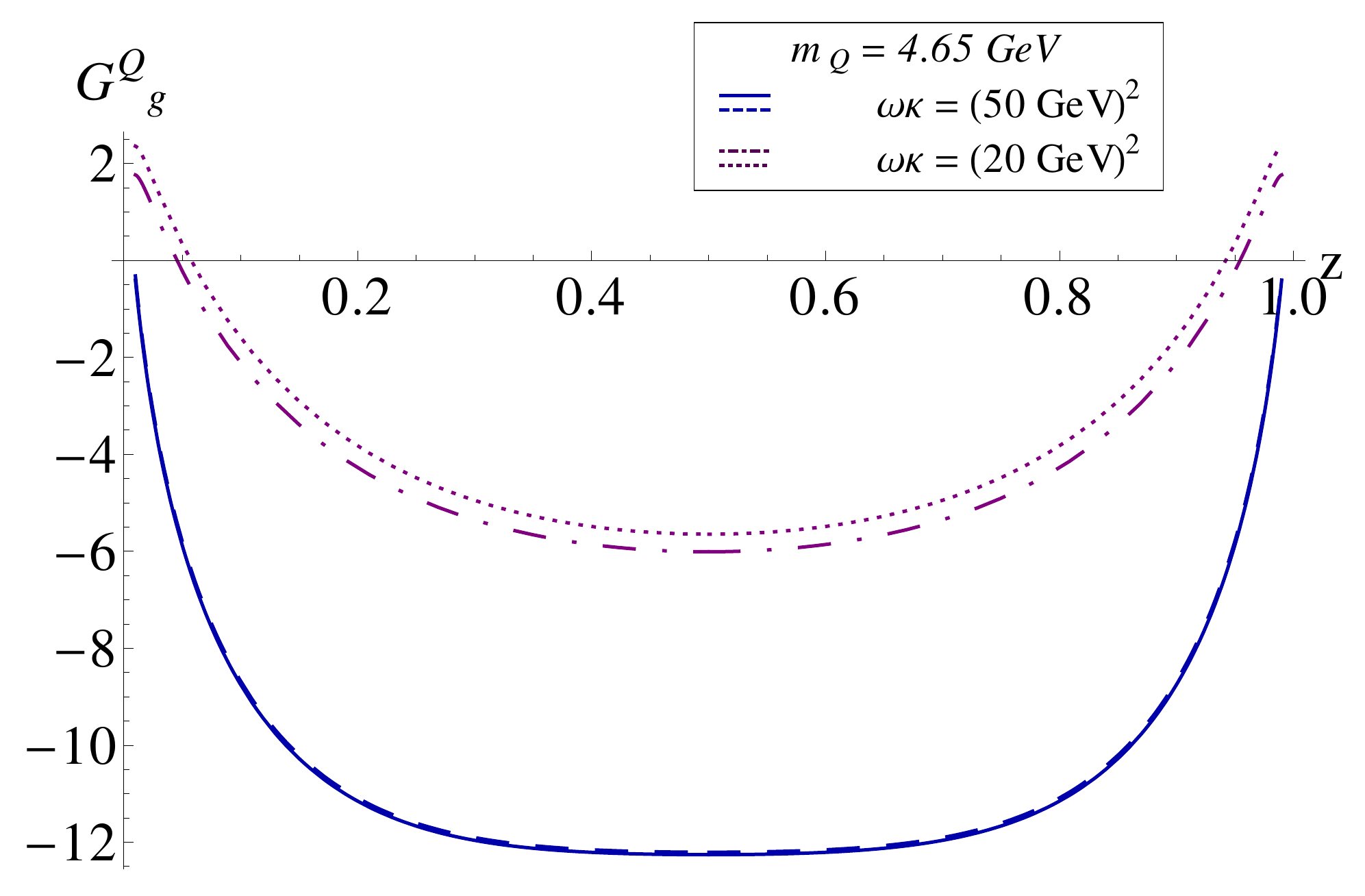}
\caption{$\mathcal G_g^{T^2_R}$ with full mass dependence (dashed blue line and dotted magenta line) and in the massless limit (solid blue line and dot-dashed magenta line), for two values of the jet invariant mass, $\omega \kappa = ( 50 \;  \textrm{GeV})^2$  and $\omega \kappa = (20 \; \textrm{GeV})^2$.}\label{CompareToMassTR}
\end{figure}

In Figs.~\ref{CompareToMassCF}, \ref{CompareToMassCA}, and  \ref{CompareToMassTR} we compare the massless limits of the functions $\mathcal G_g^{C_F T_R}$, 
$\mathcal G_g^{C_A T_R}$ and $\mathcal G_g^{T^2_R}$ with the results with full mass dependence. We do not show results for $\mathcal G_g^{T_R^2 n_l}$, which are qualitatively similar to $\mathcal G_g^{T^2_R}$.
To capture the contribution of terms proportional to $\delta(\omega r)$, we integrate  massive and massless FJFs  against a constant  test function $\varphi(\omega r)$, with cut-off  $\omega \kappa$
\begin{equation}\label{testf}
\varphi(\omega r) = \theta(\omega \kappa  - \omega r).
\end{equation} 
$\omega \kappa$ is representative of the jet scale, and we set the renormalization scale $\mu^2 = \omega \kappa$. The 
integration of  FJFs in the massless limit can be carried out very easily, while we integrate the massive FJFs numerically. The massive jet has an additional theta function, that constrains the jet invariant mass to be larger than $m_Q^2/(z (1-z))$, the minimum invariant mass to produce a $Q\bar Q$ pair at fixed $z$. 
We show results for two choices of jet invariant masses, $\omega \kappa = ( 50\, \textrm{GeV})^2$  (solid and dashed blue lines), and $\omega \kappa = ( 20\, \textrm{GeV})^2$ (dot-dashed and dotted magenta lines). The solid blue and  dot-dashed magenta lines denote the massless limits of $\mathcal G_g^{C_F T_R}$, 
$\mathcal G_g^{C_A T_R}$ and $\mathcal G_g^{T^2_R}$, Eqs.~\eqref{CFTR}, \eqref{CATR} and \eqref{TRa}, while the blue dashed and magenta dotted lines the results with full mass dependence.
In the left panel of Figs.~\ref{CompareToMassCF}, \ref{CompareToMassCA}, and  \ref{CompareToMassTR}, we show plots for fixed $z = 0.5$, and vary the heavy quark mass between $0.5$ and 10 GeV.
For all color structures and both choices of $\omega \kappa$, the agreement between massive and massless result is very good. For values of $z$ closer to 0 and 1, the agreement at the $b$ mass, $m_Q \sim 5$ GeV, is worse, but we checked that massive and massless result agree very well for  $m_Q \rightarrow 0$. 

On the right panel, we show results at the value of the bottom mass in the $1S$ scheme, $m_Q = 4.65$ GeV. As expected, power corrections are more important for smaller values of $\omega \kappa$. 
The importance of power corrections grows at small and large $z$, due to the fact  that the expansion parameter is really $m_Q^2/(\omega r z (1-z))$, rather than $4 m_Q^2/\omega r$.
Figs.~\ref{CompareToMassCF}, \ref{CompareToMassCA}, and  \ref{CompareToMassTR} are only indicative of the importance of power corrections. Having both the massive and massless expressions of $\mathcal G_g^Q$, it will be possible to repeat the analysis for phenomenologically interesting observables.

By comparing our results with the expressions for the fragmentation functions given in Ref.~\cite{Mitov:2004du}, one finds that the mass dependence of $\mathcal G_{g}^{Q}$ is exactly reproduced by the one and two loop fragmentation functions, as it should. It therefore cancels in the matching,  leaving  mass independent matching coefficients. 
For the color structure $C_F T_R$, the matching coefficient  $\mathcal J_{g Q}^{C_F T_R}$  at $\mathcal O(\alpha_s^2)$ is given by
\begin{eqnarray}\label{ICFTR}
& & \frac{\mathcal J_{g Q}^{C_F T_R}(\omega r, z,\mu^2) }{2 (2\pi)^3}  = \nonumber \\ & &  \left[\frac{\theta(\omega r) \log (\omega r/\mu^2)}{\omega r} \right]_+^{(\mu^2)}  
  \left\{   2(z^2 + (1-z)^2 )\log(1-z)  - (1-2z + 4z^2) \log(z)  - \frac{1 - 4 z}{2} \right\}  \nonumber \\
 & & -  \left[\frac{\theta(\omega r)}{\omega r} \right]_+^{(\mu^2)} \Bigg\{  \frac{4 + z}{2} - \frac{\pi^2}{6} (1 -2 z + 4 z^2) +  (1-4z^2)\log(z) + \frac{3-4 z + 8 z^2}{2} \log(1-z)  \nonumber\\ & &  
+  (1-2z + 4 z^2) \log^2(z)  + \left(z^2 + (1-z)^2\right) \log^2(1-z )  - 4 \left(z^2 + (1-z)^2\right) \log(1-z)\, \log(z) \nonumber\\ & & 
 - (3 - 6 z + 4 z^2) \Li_2(z)  \Bigg\} + \delta(\omega r)  \mathcal{I}^{ C_F T_R}(z)  .
\end{eqnarray}
with
\begin{eqnarray}\label{ICFTR1}
& & \mathcal I^{C_F T_R} (z)  =  
 \frac{1}{4} \left( 45 - 83 z + 56 z^2\right) - \frac{\pi^2}{6} (3 + 3 z - 5 z^2) \nonumber \\ & & 
 + \left( \frac{31z - 56 z^2 }{4 }  - \frac{\pi^2}{3} (1- 2z + 3z^2) \right) \log(z) 
 - \left( \frac{9 z- 8 z^2}{2 }  + \frac{\pi^2}{3} (2 - 4 z + 5 z^2) \right) \log(1-z)
 \nonumber  \\ & &  
 -\frac{  9  + 28 z - 44 z^2}{8} \log^2 (z) - \frac{5 - 9 z + 7z^2 }{2}  \log^2 (1-z)  - \frac{3(1 - 6 z + 6 z^2)}{2} \log (1-z)\log(z)   \nonumber \\ & &   
+  ( 2+10 z - 14 z^2) \Li_2(z) -\frac{5}{12} (1 -2 z + 4 z^2) \log^3 z + \frac{1}{6} \left( z^2 + (1-z)^2 \right) \log^3 (1-z)  \nonumber \\ &&  +  (4 - 8 z + 9 z^2) \log^2 (1-z) \log z
 + \frac{3}{2} (z^2 + (1-z)^2) \log (1-z) \log^2 (z)
 \nonumber \\ & & 
  +  ( 5 - 10 z+ 12 z^2) \left( \Li_3(1-z) + \log(1-z) \Li_2(z) \right) + 2 (3 - 6 z + 8 z^2) (\Li_3(z)   -  \zeta(3) ).  
  \end{eqnarray}

For the color structure $C_A T_R$, we find
\begin{eqnarray}\label{ICATR}
& & \frac{\mathcal J_{g Q}^{C_A T_R}(\omega r, z,\mu^2)}{2 (2\pi)^3}  =  3 \left(z^2 + (1-z)^2 \right)  \left[\frac{\theta(\omega r)\log^2 (\omega r/\mu^2)}{\omega r} \right]_+^{(\mu^2)} + \left[\frac{\theta(\omega r) \log (\omega r/\mu^2)}{\omega r} \right]_+^{(\mu^2)}
\nonumber  \\ & &  \times
 \left\{  4 (1 + z + z^2 ) \log (z) + 4 ( z^2 + (1-z)^2 ) \log (1-z) 
+ \frac{4}{3 z} - \frac{ 8 - 58 z + 65 z^2 }{3}
  \right\} \nonumber  \\ 
& & +  \left[\frac{\theta(\omega r)}{\omega r} \right]_+^{(\mu^2)}  \bigg\{ \frac{\pi^2}{6} (-3 + 14 z - 10 z^2) - \frac{7}{9 z} +  \frac{16 - 173 z +240z^2}{9} \nonumber \\ && 
+ \left( \frac{4}{3 z} - \frac{3 - 12 z + 13 z^2}{3} \right) \log ( z )
 +  \left( \frac{4}{3 z} + \frac{ 3 + 36 z - 43  z^2}{3}  \right) \log (1-z ) 
\nonumber  \\ & &  + 2 ( z^2 + (1-z)^2 ) \log^2 (1-z) + 2 (1 + 5 z) \log^2 z  - 2 (z^2  + (1-z)^2 ) \log (1-z)\log(z)   \nonumber \\ & &  -  2 (1 + 2 z + 2 z^2 ) \log (z) \log (1+z)
 -   2 ( 1 + 2 z + 2 z^2 ) \textrm{Li}_2 (-z) - 4 (2 -  z + 3 z^2 ) \textrm{Li}_2 (z) 
\bigg\} \nonumber \\
& &+ \delta(\omega r)  
 \, \mathcal{I}^{C_A T_R } (z).
\end{eqnarray}
The matching condition \eqref{rematch} implies that the function $\mathcal I^{C_A T_R}(z)$ is obtained subtracting
from $g^{C_A T_R}$ 
the $C_A T_R$ components of the fragmentation function $D_g^{Q (2)}$, evaluated at $\mu = m_Q$. In the notation of Ref.~\cite{Mitov:2004du}
\begin{eqnarray}\label{relation}
 \mathcal I^{C_A T_R}(z) &=& \left. g^{C_A  T_R}(z) - F_g^{(C_A  T_R)} (z)\right|_{\mu_0 = m_Q}
\,,
\end{eqnarray}
where $F_g^{(C_A T_R)}$ is given in Eq. (21) of Ref.~\cite{Mitov:2004du}. Setting $\mu_0 = m_Q$ eliminates the logarithmic terms in the fragmentation function, leaving only the finite pieces. 
The functions $\mathcal I^{C_F T_R}$ and $\mathcal I^{C_A T_R}$ are plotted in Figs.~\ref{MatchingCF}, and are smooth functions of $z$.

The difference between $\mathcal G_g^{T_R^2}$ and $\mathcal G_g^{T_R^2 n_l}$ is accounted for by the difference between the contributions of light and heavy quark loops to $D_{g}^{Q (2)}$, and thus it cancels in the matching.
The matching coefficient is the same for heavy and light flavors
\begin{eqnarray}
\mathcal J_{g Q}^{T_R^2 n_l} =  \mathcal J_{g Q}^{T_R^2},
\end{eqnarray}
and we find
\begin{eqnarray}\label{ITR}
& &\frac{\mathcal J_{g Q}^{T_R^2} (\omega r, z,\mu^2)}{2 (2\pi)^3} = - \frac{4}{9} (1 - 2 z + 2 z^2) \left\{ - 3  \left[ \frac{\theta(\omega r)\log(\omega r/\mu^2)}{\omega r}\right]_+^{(\mu^2)}   
+ 5
 \left[ \frac{\theta(\omega r)}{\omega r}\right]_+^{(\mu^2)}
\right\} \nonumber \\
& & + \delta(\omega r) \left\{  -\frac{1}{3} \left(z^2 + (1-z)^2 \right) \left(\log^2 (z (1-z)) + \pi^2\right)
- \frac{2}{9} \left( 5 - 4 z (1-z)\right) \log (z (1-z)) \right. \nonumber \\ && \left. + \frac{8}{27} \left(7 - 17 z (1-z )\right) 
\right\}.
\end{eqnarray}

\begin{figure}
\center
\includegraphics[width=11cm]{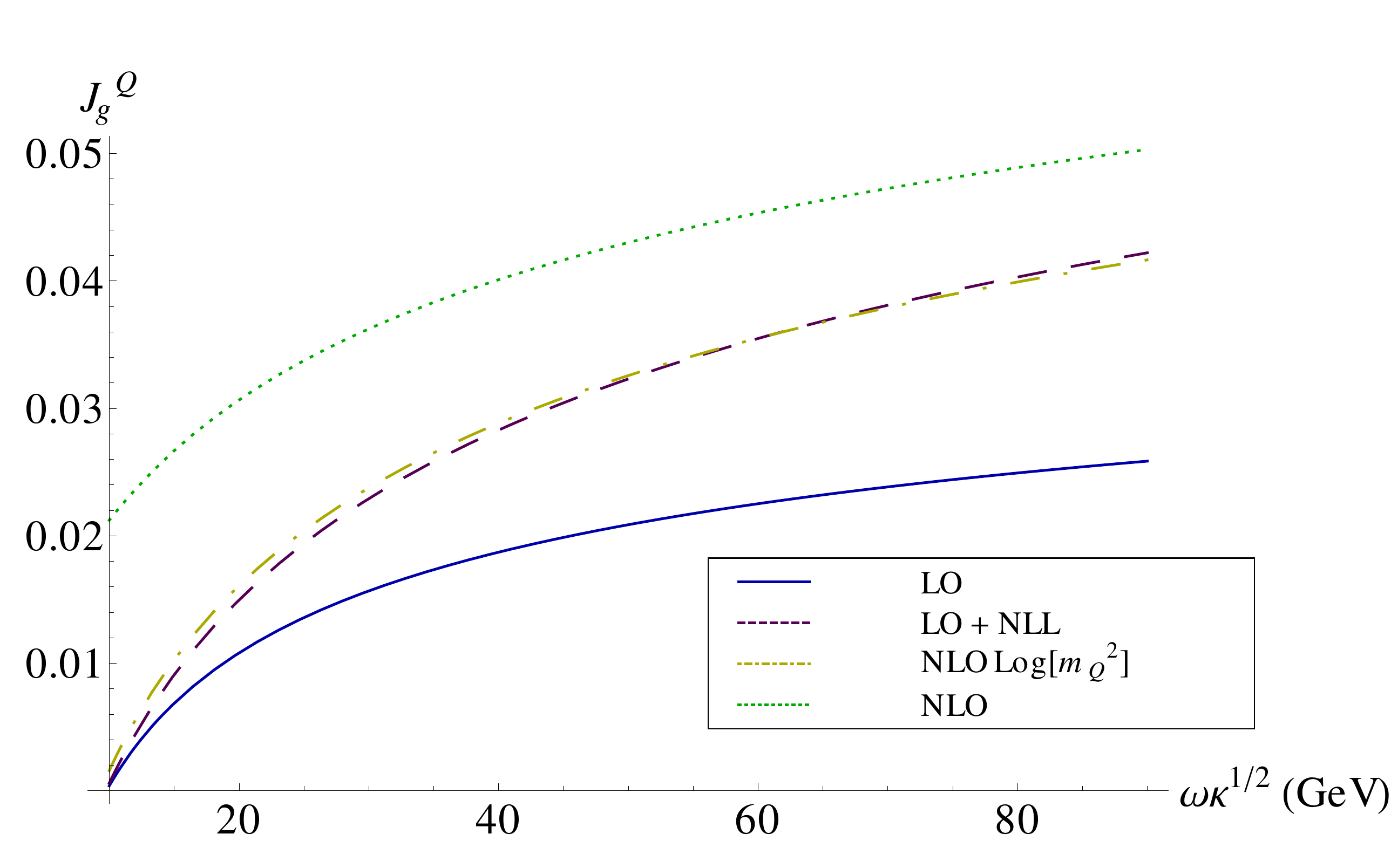}
\caption{$Q$-tagged jet function $J_g^Q$, with $z_0 = 0.05$. Different curves are explained in the text.}\label{DGLAPrun}
\end{figure}

In Fig.~\ref{DGLAPrun}, we show the effects of the DGLAP evolution on the $Q$-tagged jet function $J_g^Q$.
We fix $\omega = 100$ GeV, and integrate the FJF $\mathcal G_g^Q$ from a minimum $z_0 = 0.05$ to 1.
To show the effects of terms proportional to $\delta(\omega r)$, we integrate the jet function with a constant test function \eqref{testf}. We set the renormalization  scale $\mu^2 = \omega \kappa$.

The solid blue line denotes the fixed order result, obtained integrating Eq.~\eqref{gFJF.2}  between $z_0$ and $1$.
The dashed-magenta line uses Eq.~\eqref{resummed}, with the fragmentation functions $D_g^Q$ and $D_Q^Q$ evolved  from the scale $\mu^2_0 \sim m_Q^2$, to $\mu^2 = \omega \kappa$. We work at NLL accuracy, using two-loop time-like splitting functions, in the conventions of Ref.~\cite{Ellis:1991qj}. \footnote{The code for the DGLAP evolution of the fragmentation functions was developed in collaboration with M. Fickinger. It is discussed  in more detail in Ref.~\cite{Chul}.}
The dot-dashed yellow and dotted green lines show the fixed-order $\mathcal O(\alpha_s^2)$ expression, obtained integrating Eq.~\eqref{gFJF.2}, and Eqs.~\eqref{CFTR},
\eqref{CATR},  
 \eqref{TRa}, from $z_0$ to 1. The dotted green line is the complete $\mathcal O(\alpha_s^2)$ result, while the yellow dot-dashed line includes only the logarithmic enhanced terms.
From Fig.~\ref{DGLAPrun} we see that the logarithmically enhanced terms in the NLO result are, for the moderate value of the jet invariant mass showed here, reproduced by the DGLAP evolution of the quark and gluon fragmentation functions with $\mathcal O(\alpha_s)$ initial condition. 
The non-logarithmic terms in $\mathcal G_g^Q$ also provide an important correction to the $Q$-tagged jet function.

\subsection{Light quark fragmentation into heavy quarks at $\mathcal O(\alpha_s^2)$}\label{LightQuark}
We compute in this section the FJF for  light quark fragmenting into a heavy quark at $\mathcal O(\alpha_s^2)$. 
In the case of light quark fragmentation, the only color structure at this order is $C_F T_R$, and we write
\begin{equation}
\mathcal G_l^{Q (2)} =   C_F T_R \mathcal G_{l}^{C_F T_R}.
\end{equation}
The matching condition \eqref{FJFMatching}, specified to $H = Q$, reads
\begin{eqnarray}
\mathcal G^{Q (2)}_{l}(\omega r,z,m_Q^2,\mu^2) &=&  \int \frac{\df \xi}{\xi}     \mathcal J^{(1)}_{l g}(\omega r,\xi,\mu^2) D^{Q (1)}_g \left(\frac{z}{\xi},\frac{\mu^2}{m_Q^2}\right)     + \delta (\omega r) D^{Q (2)}_l \left( z ,\frac{\mu^2}{m_Q^2}\right) 
\nonumber \\  & &
+  \mathcal J^{(2)}_{l Q}(\omega r,z,\mu^2) .
\end{eqnarray}
where we used that
$D^{Q (0)}_Q$ and $\mathcal J^{(0)}_{l l}$ are delta functions.
The one loop result for the matching coefficient  $\mathcal J_{l g}$ is the same as for heavy quark, and it is given in Eq.~\eqref{match.3}, while the $\mathcal O(\alpha_s)$ fragmentation function $D_l^{Q (2)}$
is given in Ref.~\cite{Melnikov:2004bm}. Our calculation, then, allows to extract $\mathcal J_{l Q}$
at $\mathcal O(\alpha_s^2)$.

\begin{figure}
\center
\includegraphics[width=15cm]{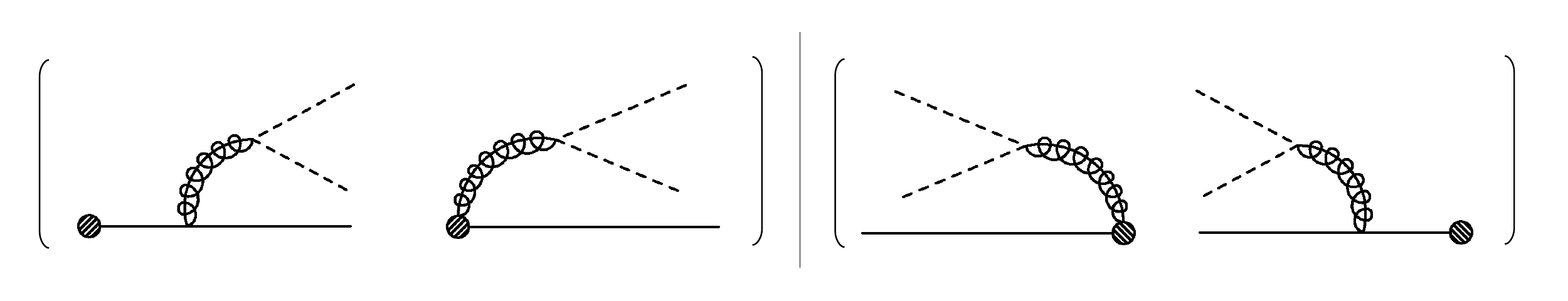}
\caption{Diagrams contributing to the light quark FJF $\mathcal G_l^Q$. The plain line denotes the light quark that initiates the jet, while the dashed lines heavy collinear $Q$ and $\bar Q$. }\label{FigL}
\end{figure}

The diagrams in Fig.~\ref{FigL} are UV and IR finite, and we find 
\begin{eqnarray}\label{lightFJF}
& &\frac{\mathcal G_l^{C_F T_R}(\omega r, z,m^2_Q,\mu^2)}{2 (2\pi)^3}  =   \nonumber \\ & &
 \left( \left[\frac{\theta(\omega r) \log \left(\omega r /\mu^2\right)}{\omega r}\right]_+^{(\mu^2)}  
+ \left[\frac{\theta(\omega r) }{\omega r}\right]_+^{(\mu^2)} \log \left(\frac{\mu^2 }{m_Q^2}\right)
+ \frac{1}{2} \log^2\left( \frac{\mu^2}{m_Q^2}\right)  \delta (\omega r)\right)\nonumber \\ & &
 \times \left( \frac{4}{3z} + \frac{3 (1-z) - 4 z^2}{3} + 2 (1+z)\log(z)\right)   \nonumber \\ & & + \left( \left[ \frac{\theta(\omega r)}{\omega r}\right]_+^{(\mu^2)}  + \log\left(\frac{\mu^2}{m_Q^2}\right) \delta(\omega r) \right)\Bigg\{ 
 -\frac{7}{9 z } + \frac{-60 +42  z + 25 z^2}{9} + \frac{\pi^2}{3} (1+z)  \nonumber \\ & & 
\left(\frac{4}{3 z}+ \frac{3(1-z) - 4 z^2}{3} \right) \log (1-z) +\left( \frac{4}{3z} - \frac{3 + 12 z  + 4 z^2}{3}\right) \log (z) + 2 (1+z) \log^2(z)  \nonumber \\ && 
 - 2 (1+z) \Li_2(z)   \Bigg\} 
 + \delta(\omega r) \Bigg\{ 
-\frac{\pi^2}{6} (2 + 3 z) + \frac{5}{54 z} + \frac{687 - 363 z - 329 z^2 }{54}  \nonumber \\ & &  
+ \left( \frac{\pi^2}{3} (1+z) - \frac{7}{9 z} + \frac{1}{9} \left( - 42 + 87 z + 25 z^2 \right) \right)\log(z) 
 \nonumber \\ & & 
+  \left( \frac{\pi^2}{3} (1+z) - \frac{7}{9 z} + \frac{1}{9} \left( - 60 + 42 z + 25 z^2 \right) \right)\log(1-z) 
\nonumber \\ & & 
+\left( \frac{2}{3z } - \frac{9 + 21 z + 4 z^2 }{6} \right) \log^2(z)
+\left( \frac{2}{3z } + \frac{3 - 3 z - 4 z^2 }{6} \right) \log^2(1-z) \nonumber \\ & & 
+\left( \frac{4}{3z } + \frac{3 - 3 z - 4 z^2 }{3} \right) \log(z) \log(1-z)
+\left(2 +  3 z - 2 (1+z) \log ( z (1-z) ) \right) \Li_2(z)
 \nonumber \\ & & 
+ (1+z)\log^3(z) - (1+z) \log^2(1-z) \log(z) -2 (1+z)\Li_3(1 - z)
 \Bigg\}.
\end{eqnarray}
The logarithmic dependence on $m_Q$ is canceled by the fragmentation function, and the matching coefficient is
\begin{eqnarray}\label{Ilight}
& &\frac{\mathcal J_{l Q}(\omega r, z,\mu^2)}{2 (2\pi)^3} =   \left[ \frac{\theta(\omega r) \log \left(\omega r /\mu^2\right)}{\omega r}\right]_+^{(\mu^2)} \left( \frac{4}{3z} + \frac{3 (1-z) - 4 z^2}{3} + 2 (1+z)\log(z)\right)   \nonumber \\ & & + \left[ \frac{\theta(\omega r)}{\omega r}\right]_+^{(\mu^2)}  \left\{ 
 -\frac{7}{9 z } - \frac{60 - 42  z - 25 z^2}{9} + \frac{\pi^2}{3} (1+z) + \left(\frac{4}{3 z}+ \frac{3(1-z) - 4 z^2}{3} \right) \log (1-z)  \right. \nonumber \\ & & \left. 
+\left( \frac{4}{3z} - \frac{3 + 12 z  + 4 z^2}{3}\right) \log (z) 
+ 2 (1+z) \log^2(z)
 - 2 (1+z) \Li_2(z)   \right\} \nonumber \\
& &+ \delta(\omega r) \left\{ 
-\frac{\pi^2}{6} (2 + 3 z) - \frac{107}{54 z} - \frac{132 + 48 z - 287 z^2 }{54} \right. \nonumber \\ & & \left. 
+ \left( \frac{\pi^2}{3} (1+z) - \frac{7}{9 z} - \frac{1}{9} \left(  90 + 81 z + 31 z^2 \right) \right)\log(z) 
\right. \nonumber \\ & & \left. 
+  \left( \frac{\pi^2}{3} (1+z) - \frac{7}{9 z} + \frac{1}{9} \left( - 60 + 42 z + 25 z^2 \right) \right)\log(1-z) +\left( \frac{2}{3z } - \frac{ 1 + 5 z }{4} \right) \log^2(z)
\right. \nonumber \\ & & \left. 
+\left( \frac{2}{3z } + \frac{ 3 - 3 z - 4 z^2 }{6} \right) \log^2(1-z) +\left( -\frac{4}{3 z} + \frac{3  + 12 z + 4 z^2}{3} \right) \Li_2(z)   -2 (1+z) \right. \nonumber \\ & & \left.
\times  \left( \frac{1}{2} \log^2(1-z) \log(z) - \frac{5}{12} \log^3(z) 
  + \log(1-z ) \Li_2 (z)   +  \Li_3(1 - z) + 2 \Li_3(z)- 2 \zeta(3) \right)
 \right\} \,.\nonumber\\
\end{eqnarray}

\section{Conclusion}\label{Conclusion}

In this paper we outlined a framework for the simultaneous resummation of logarithms of the heavy quark mass $m_Q$ and of a jet resolution variable $\tau_N$. These logarithms arise in heavy quark production, when, for  experimental  or theoretical reasons, the final state is not fully inclusive.
Examples are hadroproduction of heavy flavored hadrons, when one  hadron is detected and its momentum measured, and the final state is restricted to contain a maximum number of jets, or  $b$ jets cross sections, when the tagged $b$ or $\bar b$ is found in a jet initiated by a light parton.
Furthermore, a resolution variable is always needed when combining NLO calculations with parton shower  MonteCarlo. The possibility to carry out the two resummations simultaneously is an important step towards the combination of FONLL-like calculations, which resum the logarithmic dependence on the heavy quark mass, with parton shower algorithms, which require a resummation on a jet resolution variable.

Simultaneously resumming the logarithms of $m_Q$ and $\tau_N$ requires a factorization theorem that separates these two scales. We  discussed the generic structure of factorization theorems, using  the inclusive event shape $N$-jettiness as the jet resolution variable. The crucial ingredient in this factorization theorem is the fragmenting jet function  for a heavy quark, which describes a jet of particles with fixed invariant mass $\omega r$ containing a heavy-flavored hadron of given momentum fraction $z$. The heavy quark FJFs are generalizations of the massless parton FJFs introduced in Ref.~\cite{Procura:2009vm,Jain:2011xz}. Up to power corrections of size $m_Q^2/(Q\tau_N)$, the heavy quark FJFs can then be further factorized into a heavy quark fragmentation function that contains only the dependence on $m_Q$ and a matching coefficient that contains only the dependence on the jet resolution scale $\tau_N$. 

We computed the heavy quark  FJFs at $\mathcal O(\alpha_s)$, giving their expressions in the massless limit $m^2_Q \ll Q \tau_N$ in Eqs.~\eqref{qFJF.2}, \eqref{qFJF.3}, and 
\eqref{gFJF.2}, and their full mass dependence in Eqs.~\eqref{MassiveJet1}, \eqref{MassiveJet2} and \eqref{MassiveJet3}.
Our calculation  explicitly verified that to the calculated order the heavy quark fragmentation functions reproduce the entire dependence on the heavy quark mass, such that the matching coefficients are independent of $m_Q$ and reproduce the results of Ref.~\cite{Jain:2011xz,Liu:2010ng}. We have also shown that the anomalous dimension of the FJFs are independent of the mass $m_Q$, and are in fact equal to the anomalous dimension of the inclusive jet functions. Thus, the resummation of the jet resolution variables is identical to the more familiar case of massless jet functions.

Final state splitting of gluons into heavy quarks is a particularly important source of uncertainty in heavy quark production. They are included at NLL accuracy in  FONLL \cite{Cacciari:1998it}, which, however, can be applied only to fully inclusive final states. 
For more exclusive observables, one might rely on  NLO plus  parton shower Monte Carlo programs.
While the shower evolution resums all the leading  logarithms of $m_Q$ originating by emission of light partons from massive legs, MC@NLO and POWHEG  only include gluon splitting at fixed order \cite{Frixione:2003ei,Frixione:2007nw}.  Fixed-order NLO calculations of $b$-jet cross sections are also affected by $\log m_Q$ originating in final state splittings of  gluons, or  light quarks, into  heavy quarks \cite{Frixione:1996nh,Campbell:2008hh}. 
While the fixed-order approach is sufficient for $b$ jets with moderate $p_T$, at high $p_T$ we expect the resummation to become important.

To compare with  fixed order NLO calculations, and to improve on Monte Carlo parton shower, we computed the  gluon and light quark FJFs into heavy quarks to $\mathcal O(\alpha_s^2)$. We give their expressions in the massless limit in Eqs.~\eqref{CFTR}, \eqref{CATR}, \eqref{TRa} and \eqref{lightFJF}. 
The calculation of the gluon FJF at $\mathcal O(\alpha_s^2)$, that is at NLO, is particularly interesting  because it allows to explicitly check that the RGE of $\mathcal G_g^Q$ is that of an inclusive gluon jet. 
We verified that the singular mass dependence of $\mathcal G_g^Q$ and $\mathcal G_l^Q$ is reproduced by the heavy quark fragmentation functions at 
$\mathcal O(\alpha_s^2)$, and that the matching coefficients $\mathcal J_{g Q}^{(2)}$ and $\mathcal J_{l Q}^{(2)}$ are independent on the quark mass, up to power corrections.
The FJFs $\mathcal G_g^{Q (2)}$ in Eqs.~\eqref{CFTR}, \eqref{CATR} and \eqref{TRa}, and $\mathcal G_l^{Q (2)}$ in Eq.~\eqref{lightFJF}, and the matching coefficients
in Eqs.~\eqref{ICFTR}, \eqref{ICATR}, \eqref{ITR} and \eqref{Ilight} are the main results of this work.

Our calculation of the heavy quark FJFs provides a key ingredient for using the factorization formulae \eqref{cross.1} and \eqref{cross.1b} to describe phenomenologically interesting heavy quark production cross sections.  In most cases, the remaining functions in Eqs.~\eqref{cross.1} and \eqref{cross.1b} can be computed with massless quarks without encountering divergences, and exist in the literature. If some of the inclusive quark jets in Eq.~\eqref{cross.1} and \eqref{cross.1b} are massive, one should use  massive quark jet functions \cite{Fleming:2007qr,Fleming:2007xt}.  
The effects of the heavy quark mass on inclusive light quark jet functions, and on the thrust hemisphere  soft function, which start at $\mathcal O(\alpha_s^2)$ and are relevant in the case the jet or soft scale are close to $m_Q$, have been considered in  Refs.~\cite{Gritschacher:2013pha,Gritschacher:2013tza}

The exception is the flavor excitation channel. In this case logarithms of $m_Q$ originate in initial state splittings of gluons and light quarks into $Q \bar Q$ pairs, and one of the two heavy quarks enters the hard collision.
These logarithms are resummed by introducing a perturbative $b$-quark parton distribution, and evolving it to the hard scale.
In the presence of a resolution variable $m^2_Q \ll Q \tau_N \ll p^2_T$, one needs to introduce a heavy quark beam function, using techniques very similar to those discussed in this paper.

While we have given all perturbative ingredients required to perform the simultaneous resummation of logarithms of $m_Q$ and $\tau_N$, we have left this resummation for future work. 

\section*{Acknowledgement}
We thank S.~Alioli, Z.~Ligeti, J.~Walsh and S.~Zuberi, for many helpful discussions in all the stages of the project.
We thank S.~Fleming, C.~Kim, M.~Procura, I.~Stewart, M.~Ubiali, and W.~J.~Waalewijn for comments on the manuscript. 
EM acknowledges M.~Fickinger for useful discussions and, in particular, for the collaboration on the DGLAP evolution code used in Fig. \ref{DGLAPrun}. 
This work was supported by the Department of Energy Early Career Award with Funding
Opportunity No. DE-PS02-09ER09-26 (CWB), and the Director, Office of Science, Office of High Energy Physics of
the U.S. Department of Energy under the Contract No. DE-AC02-05CH11231 (CWB, EM).

\appendix

\section{Details on the $m_Q \rightarrow 0$ limit}\label{AppA}

In this appendix, we describe in more detail  some technical aspects of the limit $m_Q^2 \ll \omega r$.
The expression of the quark  FJFs at $m_Q \neq 0$ contains  terms proportional to $(\omega r - m_Q^2)^{-1}$, coming from the quark propagator in Fig.~\ref{alpha}, and a theta function, which sets $\omega r$ to be larger than the minimum invariant mass needed for the radiation of a gluon out of a heavy quark.  
Introducing the variable $s = \omega r - m_Q^2$,   $\mathcal G_Q^Q$ is subjected to the constraint 
$s - m_Q^2 (1-z)/z > 0$, while  $\mathcal G_Q^g$, for which   $z$  is the gluon momentum fraction, to  
$s - m_Q^2 z/(1-z) >0$.
Similarly, the gluon FJF $\mathcal G_g^Q$ contains inverse powers of $\omega r$, coming from the gluon propagator, but, for $m_Q \neq 0$, the invariant mass  
$\omega r$ is constrained to be at least $m_Q^2/(z (1-z))$.
The presence of the quark mass, then, forbids the jet invariant mass to reach the values for which the quark or gluon propagators have a pole. 
When taking the limit $m_Q^2 \ll \omega r$ one has to be careful not introduce new singularities in the results. The limit $m_Q \rightarrow 0$ has to be taken in the sense of generalized functions, and it will generate logarithms of the mass $m_Q$.

We describe in some detail an example of term which can be found in the gluon FJF $\mathcal G_g^Q$. 
With similar techniques we derived the expressions of the quark and gluon FJFs in Eqs.~\eqref{qFJF.2}--\eqref{gFJF.3}, 
the gluon FJF in Eqs.~\eqref{CFTR}, \eqref{CATR}, and \eqref{TRa},
and  the light quark FJF, in Eq.~\eqref{lightFJF}.

Consider an expression of the form
\begin{equation}\label{ex.1}
\mathcal F(\omega r, z ,m_Q^2) =  \frac{\theta\left( \omega r z (1-z) - m_Q^2\right)}{\omega r} f\left(\frac{m_Q^2}{\omega r},z\right),
\end{equation}
with $f$ a regular function of $\omega r$, with expansion
\begin{equation}\label{ex.2}
f\left( \frac{m_Q^2}{\omega r} , z\right) = \sum_{n = 0}^{\infty} f_n(z) \left( \frac{m_Q^2}{\omega r} \right)^n.
\end{equation}
The massless limit $\mathcal F_0(\omega r, z)$  is that distribution that applied to a test function $\varphi$ satisfies
\begin{equation}\label{limit1}
\lim_{m_Q \rightarrow 0 }\int \df(\omega r)\frac{\theta\left( \omega r z (1-z) - m_Q^2\right)}{\omega r} f\left(\frac{m_Q^2}{\omega r}, z\right)  \varphi(\omega r) = 
\int \df (\omega r) \mathcal F_0(\omega r, z) \varphi(\omega r).
\end{equation}
On the l.h.s of Eq.~\eqref{limit1}, we regulate the  $\omega r \sim 0$ region by subtracting the test function  evaluated at  0. More precisely, we write
\begin{eqnarray}\label{ex.3}
& & \lim_{m_Q \rightarrow 0} \left(\int^{+ \infty}_{m_Q^2/(z (1-z))}  \frac{\df \omega r}{\omega r}  \sum_{n = 0}^{\infty} f_n(z) \left( \frac{m_Q^2}{\omega r} \right)^n  \left\{  \varphi(\omega r) 
- \theta\left(\omega \kappa - \omega r \right) \left(\sum_{k = 0}^n \frac{\varphi^{(k)} (0)}{k!} (\omega r)^k \right) \right\} \right. \nonumber \\
& &+ \left. \sum_{n = 0}^{\infty} (m_Q^2)^n f_n(z) \left( \frac{\log (\omega \kappa)}{n!}  \varphi^{(n)}(0) - \sum_{k = 0}^{n-1}  \frac{1}{n-k} \frac{1}{(\omega \kappa)^{n-k}}\frac{\varphi^{(k)}(0)}{k!} \right) \right) \nonumber \\
& & + \lim_{m_Q \rightarrow 0} \left( \int^{\omega \kappa}_{m^2_Q/(z(1-z))} \df \omega r
\sum_{n = 0}^{\infty} f_n(z)  \sum_{k = 0}^n   \frac{(m_Q^2)^n}{ ( \omega r)^{n- k +1} } \frac{\varphi^{(k)} (0)}{k!}   \right. \nonumber \\ & & \left. -  \sum_{n = 0}^{\infty} (m_Q^2)^n f_n(z) \left( \frac{\log (\omega \kappa)}{n!}  \varphi^{(n)}(0) - \sum_{k = 0}^{n-1}  \frac{1}{n-k} \frac{1}{(\omega \kappa)^{n-k}}\frac{\varphi^{(k)}(0)}{k!} \right) \right). 
\end{eqnarray}
In the curly brackets in the first line of  Eq.~\eqref{ex.3} we subtracted to the test function enough powers of its series expansion to make the integral convergent at $\omega r \sim 0$. We added  back the same terms in the third line. Since $\omega r$ takes value in $(0,\infty)$, the subtraction terms contain an arbitrary cut-off $\omega \kappa$, but the combined expressions are cut-off independent. The terms in the second and fourth lines, which are identical, are needed to make
the integral, in the first line of Eq.~\eqref{ex.3},  and the contribution to the delta function, in the third line, separately cut-off independent. 
Now one  can take the massless limit of the first two lines of Eq.~\eqref{ex.3} without generating additional divergences.
\begin{eqnarray}
& &\lim_{m_Q \rightarrow 0} \left(\int^{+ \infty}_{m_Q^2/(z (1-z))}  \frac{\df \omega r}{\omega r}  \sum_{n = 0}^{\infty} f_n(z) \left( \frac{m_Q^2}{\omega r} \right)^n  \left\{  \varphi(\omega r) 
- \theta\left(\omega \kappa - \omega r \right) \left(\sum_{k = 0}^n \frac{\varphi^{(k)} (0)}{k!} (\omega r)^k \right) \right\} \right. \nonumber \\
& &+ \left. \sum_{n = 0}^{\infty} (m_Q^2)^n f_n(z) \left( \frac{\log (\omega \kappa)}{n!}  \varphi^{(n)}(0) - \sum_{k = 0}^{n-1}  \frac{1}{n-k} \frac{1}{(\omega \kappa)^{n-k}}\frac{\varphi^{(k)}(0)}{k!} \right) \right) \nonumber \\
&& =  \sum_{n = 0}^{\infty} f_n(z) \left[\frac{(m_Q^2)^n}{(\omega r)^{n+1}}\right]_+ = f_0(z) \left[\frac{1}{\omega r}\right]_+ + \ldots.
\label{ex.4}
\end{eqnarray}
All the powers of $1/\omega r^n$ with $n>2$ are power corrections in the massless limit, and can be discarded. The plus distribution was defined in Eq.~\eqref{dist.2}, following \cite{Russian}.

The third and fourth line of Eq.~\eqref{ex.3} contain terms proportional to $\delta(\omega r)$, and derivatives of the delta function. Derivatives of  $\delta(\omega r)$ are power corrections, one can discard them and retain only terms with $k =0$. One is thus left with 
\begin{eqnarray}
& & \lim_{m_Q \rightarrow 0} \left( \int^{\omega \kappa}_{m^2_Q/(z(1-z))} \df (\omega r)
\sum_{n = 0}^{\infty} f_n(z)  \sum_{k = 0}^n   \frac{(m_Q^2)^n}{ ( \omega r)^{n - k +1} } \frac{\varphi^{(k)} (0)}{k!}   \right. \nonumber \\ & & \left. -  \sum_{n = 0}^{\infty} (m_Q^2)^n f_n(z) \left( \frac{\log (\omega \kappa)}{n!}  \varphi^{(n)}(0) - \sum_{k = 0}^{n-1}  \frac{1}{n-k} \frac{1}{(\omega \kappa)^{n-k}}\frac{\varphi^{(k)}(0)}{k!} \right) \right)  \nonumber \\
& & =
\delta(\omega r) \left\{ \sum_{n = 1}^{\infty}  (m_Q^2)^{n} f_n(z)  \left( \int_{m_Q^2/(z(1-z))}^{\omega\kappa} \df \omega r   \frac{1}{(\omega r)^{n+1}} + \frac{1}{n} \frac{1}{\omega \kappa^n}\right) \right. \nonumber \\
& & \left. + f_0(z) \left( \int_{m_Q^2/(z(1-z))}^{\omega \kappa}\df (\omega r)  \frac{1}{\omega r} - \log (\omega \kappa)\right)   \right\} \nonumber \\.
& & = 
\delta(\omega r) \left\{ \int_{m_Q^2/(z(1-z))}^{\infty}  \frac{\df(\omega r) }{\omega r}  \left(f\left (\frac{m_Q^2}{\omega r}, z \right)  -f_0(z) \right) - f_0(z) \log \left(\frac{m_Q^2}{z(1-z)}  \right) \right\}. \label{ex.5}
\end{eqnarray}
In the last step we used the fact that the result is cut-off independent to set $\omega \kappa$ to  $\infty$, except in the logarithmic term.
Combining Eqs.~\eqref{ex.5} and \eqref{ex.4}, we obtain
\begin{eqnarray}\label{ex.6}
\lim_{m_Q \rightarrow 0} \frac{\theta\left( \omega r z (1-z) - m_Q^2\right)}{\omega r} f\left(\frac{m_Q^2}{\omega r}, z\right) & = &   f_0(z) \left( \left[\frac{\theta(\omega r)}{\omega r} \right]_+ - \delta(\omega r) \log \left(\frac{m_Q^2}{z(1-z)}  \right) \right) \nonumber \\ 
& & + \delta(\omega r)
  \int_{m_Q^2/(z(1-z))}^{\infty}  \frac{\df (\omega r) }{\omega r}  \left(f\left (\frac{m_Q^2}{\omega r}, z \right)  -f_0(z) \right),\nonumber\\
\end{eqnarray}
where $f_0$ is the first term is the series expansion of $f$.
In the example discussed here, the plus distribution and the $\log m_Q^2$ are determined by the value of the function $f$ at $m_Q = 0$, while the coefficient of the delta function requires the complete knowledge of the function $f$. We find the same behavior in more general cases.

\section{Fragmenting jet functions for $\omega r \sim m_Q^2$}\label{AppC}
In Secs. \ref{OneLoop}, \ref{GluonAlpha2} and \ref{LightQuark}, we considered quark and gluon FJFs in the limit of jet invariant mass much larger than $m_Q^2$.
In many cases it may be important to provide a description of heavy flavor production encompassing a wide range of the values of the resolution variable $\tau_N$.
As already discussed, for $Q \tau_N \gg m^2_Q$, it is important to resum logarithms of $m_Q^2/(Q\tau_N)$. The resummation is achieved by matching the FJFs onto fragmentation functions, systematically neglecting powers of $m_Q^2/(\omega r)$.
For smaller value of $\tau_N$, logarithms of $m_Q^2/(Q \tau_N)$ are not large and need not to be resummed. In this regime, it might be more important to retain the full dependence on $m_Q$. In this section we provide the expressions of the quark and gluon FJF  at one loop, keeping all powers of  $m^2_Q/(\omega r)$.

A first important observation is that the renormalization of the FJFs is not affected by assumptions on the relative size of the jet invariant mass and $m_Q$. 
The UV divergences of diagrams Fig.~\ref{alpha}, as well as those of the $\mathcal O(\alpha_s^2)$ diagrams, Figs.~\ref{FigV} and \ref{FigR}, are not changed.
Consequently, even in the limit $\omega r \sim m_Q^2$, the RGEs of the quark and gluon FJFs are given by Eq.~\eqref{rge}. 

We now give the expression of the FJF at one loop. For the quark FJFs, it is convenient to introduce the variable $s  = \omega r - m_Q^2$, while we keep expressing $\mathcal G_g^Q$ as a function of $\omega r$. 
Renormalizing the quark FJF $\mathcal G_Q^Q$ as in Eq.~\eqref{JetRenQ}, we find
\begin{eqnarray}
\frac{\mathcal G^{Q (0)}_{Q} (s, z)}{2 (2\pi)^3} & = & \delta(s)\, \delta(1-z) \nonumber \\
\frac{\mathcal G^{Q (1)}_Q(s,z ) }{2 (2\pi)^3}&=& 
C_F \Bigg\{ 
\delta(1-z) \left(  \delta \left( s \right) \left(
 \log^2 (m_Q^2) + \log^2 (\mu^2) + \frac{3}{2} \log \frac{\mu^2}{m_Q^2}  + \frac{11}{4} +\frac{\pi^2}{6}   \right)    \right. \nonumber \\ && \left.   
 + 4 \left[ \frac{\theta(s) \log s }{s}\right]_+^{} 
- 2 \log \left( \mu^2 \left(s+ m_Q^2 \right) \right) \, \left[\frac{\theta(s)}{s }\right]_+^{}
  - \frac{\theta\left( s\right) (3 s + 4 m_Q^2)}{ 2 \left(s + m_Q^2 \right)^2} \right ) \nonumber \\
& &
+    \left[\frac{1+z^2}{1-z}\right]_+ \left[\frac{\theta\left( s -  m_Q^2\frac{1-z}{z}\right)}{s}\right]_+^{(\kappa)}
- 2 m_Q^2 \left[\frac{\theta\left( s  -  m_Q^2 \frac{1-z}{z} \right)}{ s^2 }\right]_+^{(\kappa)} \nonumber \\
& &   - \delta \left( s \right) \left( \left[ \frac{1+z^2}{1-z} \log \left( m_Q^2  (1-z)\right) \right]_+  
-  \frac{1+z^2}{1-z} \log z \right)
 \nonumber \\ & &  
+ \theta\left( \kappa -  m_Q^2 \frac{1-z}{z} \right)  \left( \delta(s) \left[\frac{2 z}{1-z}\right]_+  - 2 m_Q^2  \log\left( m_Q^2 \frac{1-z}{z} \right) \delta^{\prime} \left(s \right)  \right)
 \nonumber \\ && 
+ \left[ \theta \left( m_Q^2 \frac{1-z}{z} -\omega \kappa \right) \frac{1+z^2}{1-z} \log \left( m_Q^2 \frac{1-z}{z}\right) \right]_+
\Bigg\} , \label{MassiveJet1} \\
\frac{\mathcal G_{Q}^g (s, z)}{2 (2\pi)^3}&=& C_F \; \theta\left(s -m_Q^2 \frac{z}{1-z}\right) \frac{1}{s} \left\{ \frac{1+ (1-z)^2}{z} - \frac{2 m_Q^2}{s}\right\} , \label{MassiveJet2}\\
\frac{ \mathcal G_{g}^Q (\omega r,z)}{2 (2\pi)^3} &=& T_R \; \theta\left (\omega r - \frac{m_Q^2}{z (1- z)} \right)
\frac{1}{\omega r} \left\{  z^2 + (1-z)^2 + \frac{2 m_Q^2}{\omega r} \right\}. \label{MassiveJet3} 
 \end{eqnarray}
It is easy to check that the massless limit of $\mathcal G_Q^g$ and $\mathcal G_g^Q$ in Eqs.~\eqref{MassiveJet2} and \eqref{MassiveJet3} is given by the expressions \eqref{qFJF.3} and \eqref{gFJF.2} in section \ref{OneLoop}.

The expression \eqref{MassiveJet1} is more convoluted than Eq.~\eqref{qFJF.2}, and the distributions in $z$ and $s$ do not factor as nicely. 
Comparing Eq.~\eqref{MassiveJet1}  to Eq.~\eqref{qFJF.2} at a first sight it seems that the logarithmic mass dependence appears  not only in front of the DGLAP splitting function, and thus will not be completely canceled by the fragmentation function. However this is just an artifact of the form of Eq.~\eqref{MassiveJet1}, that still contain expressions, as $1/(s+m_Q^2)^2$, which need to be regulated in the $m_Q \rightarrow 0$ limit.

We now define the  distributions in Eq.~\eqref{MassiveJet1}, applied to a test function factorized as $\psi(s) \varphi(z)$
\begin{eqnarray}
& &\int \df z \int \df s \left[\frac{1+z^2}{1-z}\right]_+ \left[\frac{\theta\left( s - m_Q^2 \frac{1-z}{z}\right)}{s}\right]_+^{(\kappa)}  \varphi(z) \psi(s)=  \int_0^1 \df z  \frac{1+z^2}{1-z}
  (\varphi(z) - \varphi(1)) \label{distM.1} \nonumber\\
& & 
\left(   \int \df s  \frac{\theta\left( s -  m_Q^2 \frac{1-z}{z}\right)}{s}   \Big( \psi(s) - \theta(\kappa- s) \psi\left(0\right)\Big)  + \theta\left( \kappa - m^2_Q \frac{1-z}{z}\right)\log (  \kappa) \psi(0)  \right) . 
\end{eqnarray}
and
\begin{eqnarray}
\int \df s \left[\frac{\theta\left( s -  m_Q^2 \frac{1-z}{z}\right)}{s^2}\right]_+^{(\kappa)} \psi(s) & = & 
\int \df s  \frac{\theta\left( s -  m_Q^2 \frac{1-z}{z}\right)}{s}  \Big( \psi(s) - \theta(\kappa- s) \left( \psi\left(0\right) + s \psi^{\prime}(0) \right) \Big)
\nonumber \label{distM.2} \\
& & 
    + \theta\left(\kappa - m^2_Q \frac{1-z}{z}\right)\left( -\frac{1}{\kappa} \psi(0) + \log\kappa \, \psi^{\prime}(0)   \right). 
\end{eqnarray}
Differently from Eq.~\eqref{dist.2}, the distributions \eqref{distM.1} and \eqref{distM.2} have some dependence on the cut off $\kappa$, which is canceled by the remaining $\kappa$-dependent terms in Eq.~\eqref{MassiveJet1}.  Other  distributions in Eq.~\eqref{MassiveJet1} are defined as in  Eq.  \eqref{dist.2}.

Eqs.~\eqref{distM.1} and \eqref{distM.2} have a simple limit for $m_Q^2\rightarrow 0$,
\begin{eqnarray}
\lim_{m_Q \rightarrow 0} \left[\frac{1+z^2}{1-z}\right]_+ \left[\frac{\theta\left( s -  m_Q^2  \frac{1-z}{z}\right)}{s}\right]_+^{(\kappa)}
&=& \left[\frac{1+z^2}{1-z}\right]_+ \left[\frac{\theta\left( \omega r \right)}{\omega r}\right]_+^{} ,  \label{lim.1}\\
\lim_{m_Q \rightarrow 0} \left[\frac{\theta\left( s -  m_Q^2 \frac{1-z}{z}\right)}{s^2}\right]_+^{} & =  &
\left[\frac{\theta(\omega r)}{(\omega r)^2}\right]_+. \label{lim.2}
\end{eqnarray}
Of the remaining terms in Eq.~\eqref{MassiveJet1}, only two have a non-trivial massless limit 
\begin{eqnarray}
\lim_{m_Q \rightarrow 0} \log\left(s + m_Q^2 \right) \left[\frac{\theta(s)}{s}\right]^{}_+
&=& \left[\theta(\omega r)\frac{\log (\omega r)}{\omega r}\right]^{}_+  + \delta(\omega r) \left( \frac{\pi^2}{6} + \frac{1}{2} \log^2 m_Q^2 \right)  \label{distM.3}\\
\lim_{m_Q \rightarrow 0 } \frac{\theta\left( s\right)(3 s + 4 m_Q^2) }{2  \left(s + m_Q^2  \right)^2}  & = & \frac{3}{2} \left[\frac{\theta(\omega r)}{\omega r}\right]_+^{}  + \delta(\omega r) \left( \frac{1}{2} - \frac{3}{2}\log m_Q^2 \right).
\label{distM.4}
\end{eqnarray}
The factors of $\log^2 m_Q^2$ and $\log m_Q^2$ are exactly what needed to cancel the mass dependence of the $\delta(s)\delta(1-z)$ term in Eq.~\eqref{MassiveJet1}.
Using Eqs.~\eqref{lim.1}, \eqref{lim.2}, \eqref{distM.3} and \eqref{distM.4} one can easily prove that the massless limit of Eq.~\eqref{MassiveJet1} is \eqref{qFJF.2}.

Finally, integrating Eq.~\eqref{MassiveJet1}, we can verify the flavor sum rule
\begin{equation}
\frac{1}{2 (2\pi)^3} \int_0^1 \df z \, \mathcal G_{Q} (s, z) = J_Q ( s, m_Q^2). 
\end{equation}
We find
\begin{eqnarray}
& &J_Q ( s,m_Q^2) = \delta(s) + \frac{\alpha_s C_F}{2\pi} \left\{ 
 \delta(s) \left( 
  \log^2 (m_Q^2) + \log^2( \mu^2) + \frac{1}{2} \log m_Q^2 + \frac{3}{2} \log (\mu^2) + 4 - \frac{\pi^2}{6} \right) \right. \nonumber \\ &&  \left.  
 - 2 \left[\frac{\theta(s)}{s}\right]_+^{} \left(
  \log (\mu^2)+ 1\right)
+ 4 \left[\frac{\theta(s) \log s}{s}\right]_+^{}  - 2 \log \left(s + m_Q^2 \right) \left[ \frac{\theta(s)}{s} \right]_+^{} + \frac{1}{2} \frac{s}{\left(s + m_Q^2 \right)^2} 
\right\}. \nonumber \\\label{MassiveJet4}
\end{eqnarray}
Eq.~\eqref{MassiveJet4} reproduces the result for an inclusive quark massive jet  \cite{Fleming:2007qr,Fleming:2007xt}. Also in this case, one can take the limit $m_Q \rightarrow 0$ without encountering divergences.

We also computed the expressions of $\mathcal G_g^Q$ and $\mathcal G_l^Q$ at $\mathcal O(\alpha_s^2)$, at fixed $m_Q$. 
Their dependence on the jet invariant mass is not as simple as in Eqs.~\eqref{qFJF.3}, \eqref{CFTR}, \eqref{CATR}, \eqref{TRa}, and  \eqref{lightFJF}. They are expressed analytically in terms of logarithms and polylogarithms up to rank two. Their expression is too lengthy to be reproduced here.

\section{Analytic expression of $g^{C_F T_R}$}\label{AppE}
We give here the analytic expression of the function $g^{C_F T_R}$, which enters $\mathcal G_g^{C_F T_R}$ 
in Eq.~\eqref{CFTR}.
\begin{eqnarray}
& &  g^{C_F T_R} (z)  =  
 \frac{1}{6} \left( -3 + 22 z - 16 z^2\right) + \frac{\pi^2}{36} (7 + 36 z - 72 z^2 +32 z^3) \nonumber \\ & & 
 - \left( \frac{9-42 z + 40z^2 -24 z^3 }{3 }  + \frac{\pi^2}{6} (1-2z) \right) \log(z)  \nonumber \\ & & 
 + \left( \frac{40 -119 z + 112 z^2 -48z^3}{6 }  + \frac{\pi^2}{6} (7 - 14 z + 20 z^2) \right) \log(1-z)
 \nonumber  \\ & &  
 +\frac{7-108 z + 240 z^2 -128 z^3}{12}  \log^2 (z) + \frac{3 -44 z + 56 z^2 -32 z^3 }{4} \log^2 (1-z)   \nonumber \\ & &   - \frac{11 -156 z +  240 z^2 - 128 z^3}{6} \log (1-z)\log(z)
 \nonumber \\ && 
-\frac{1}{6} \left( 11 -22z+ 28 z^2\right) \log^3 z + \frac{2}{3} \left( z^2 + (1-z)^2 \right) \log^3 (1-z)  \nonumber \\ &&  -  \frac{1}{2}( 13 - 26 z+ 32 z^2) \log^2 (1-z) \log z
 + 7  (z^2 + (1-z)^2) \log (1-z) \log^2 (z)
  \nonumber \\  & & 
 +  \left( \frac{ -7 + 60 z - 72 z^2 + 32 z^3 }{6} - 3 \left( 1 - 2 z + 4 z^2 \right)\log(1-z) + (7-14z + 16 z^2) \log(z) \right) \Li_2(z) \nonumber \\
 && + \left( \frac{8}{3} (1-2z)^3  + 8 (z^2 + (1-z)^2) \left(\log(1-z)-\log(z)\right)\right) \Li_2\left(\frac{2z-1}{z}\right)
\nonumber \\ & & 
  -  ( 7 - 14 z + 20 z^2   ) \Li_3(1-z) - 2 ( 5 -10 z + 12 z^2) \Li_3(z)   + 2 (13 - 26 z + 28 z^2 ) \zeta(3) 
  \nonumber \\ & & 
  - 16 (z^2 + (1-z)^2 ) \left( \Li_3\left(\frac{1-2z}{1-z}\right) + \Li_3\left(\frac{2z-1}{z}\right)  \right)
 \nonumber \\ & & 
- 2 \pi^2  \left( (z^2 + (1-z)^2) \left|\log(1-z)-\log(z)\right| - \frac{1}{3} \left|1-2z\right|^3\right)  .  \label{gCFTR}
  \end{eqnarray}
The last line of Eq.~\eqref{gCFTR} causes the non-smooth behavior of $\mathcal G_g^{C_F T_R}$ at $z=1/2$, and it is canceled in the matching by an identical term in $D_g^{Q(2)}$.

The functions $g^{C_F T_R}$ and $\mathcal I^{C_F T_R}$ are related by the matching condition \eqref{rematch}. 
Focusing on the $C_F T_R$ color structure,  from the matching condition Eq.~\eqref{rematch}, and the tree level and one loop matching coefficients \eqref{match.2}--\eqref{match.g2}, one derives 
\begin{eqnarray}\label{relation2}
C_F T_R \mathcal I^{C_F T_R}(z) &=&  C_F T_R \left(\left. g^{C_F  T_R}(z) - F_g^{(C_F  T_R)}(z)  \right|_{\mu_0 = m_Q}\right) - \int \frac{\df \xi}{\xi} p_{g Q} (\xi) \left. D_Q^{Q (1)} \left(\frac{z}{\xi}\right) \right|_{\mu_0 = m_Q} 
\,.
\end{eqnarray}
$F_g^{(C_F T_R)}$ is the $C_F T_R$ component of $D_g^{Q(2)}$, in the notation of  Ref.~\cite{Mitov:2004du}, and it is given in Eq. (20) of that paper. 
$D_Q^{Q (1)}$ is given in Eq.~\eqref{qFF.2}.
Setting $\mu_0 = m_Q$ eliminates the logarithmic terms in the fragmentation functions, leaving only the finite pieces.
$p_{g Q}$ is the piece of  $\mathcal I_{g Q}^{(1)}$ proportional to $\delta(\omega r)$,
\begin{equation}
p_{g Q} (z) = T_R \Big( (z^2 + (1-z)^2) \log(z (1-z)) + 2 z (1-z) \Big).
\end{equation}

\end{document}